\documentclass[12pt,reqno]{amsart}
\usepackage[pdfborder={0 0 0.5 [3 2]}, plainpages=false]{hyperref}%
\usepackage[left=1in,right=1in,top=1in,bottom=1in]{geometry}%
\usepackage[citation-order]{amsrefs}%
\usepackage{amsmath}
\usepackage{enumerate}
\usepackage{amssymb}                
\usepackage{amsfonts}
\usepackage{amsthm}
\usepackage{bbm}
\usepackage[table,xcdraw]{xcolor}
\usepackage{float}
\usepackage{mathtools}
\usepackage{cool}
\usepackage{graphicx,epsfig}
\usepackage{booktabs}

\usepackage[capitalize,nameinlink]{cleveref}
\crefname{section}{section}{sections}
\crefname{subsection}{subsection}{subsections}
\Crefname{section}{Section}{Sections}
\Crefname{subsection}{Subsection}{Subsections}

\Crefname{figure}{Figure}{Figures}

\crefformat{equation}{\textup{#2(#1)#3}}
\crefrangeformat{equation}{\textup{#3(#1)#4--#5(#2)#6}}
\crefmultiformat{equation}{\textup{#2(#1)#3}}{ and \textup{#2(#1)#3}}
{, \textup{#2(#1)#3}}{, and \textup{#2(#1)#3}}
\crefrangemultiformat{equation}{\textup{#3(#1)#4--#5(#2)#6}}%
{ and \textup{#3(#1)#4--#5(#2)#6}}{, \textup{#3(#1)#4--#5(#2)#6}}{, and \textup{#3(#1)#4--#5(#2)#6}}

\Crefformat{equation}{#2Equation~\textup{(#1)}#3}
\Crefrangeformat{equation}{Equations~\textup{#3(#1)#4--#5(#2)#6}}
\Crefmultiformat{equation}{Equations~\textup{#2(#1)#3}}{ and \textup{#2(#1)#3}}
{, \textup{#2(#1)#3}}{, and \textup{#2(#1)#3}}
\Crefrangemultiformat{equation}{Equations~\textup{#3(#1)#4--#5(#2)#6}}%
{ and \textup{#3(#1)#4--#5(#2)#6}}{, \textup{#3(#1)#4--#5(#2)#6}}{, and \textup{#3(#1)#4--#5(#2)#6}}

\crefdefaultlabelformat{#2\textup{#1}#3}

\def\C{{\mathbb C}}

\def\Mod{\mathop\mathrm{mod}\nolimits}
\def\Re{\mathop\mathrm{Re}\nolimits}
\def\Im{\mathop\mathrm{Im}\nolimits}

\graphicspath{ {images/} }

\begin{document}

\title[Standing and Traveling Waves in a Nonlinearly Dispersive Lattice Model]{Standing and Traveling Waves in a Minimal Nonlinearly Dispersive Lattice Model}

\author[R Parker]{Ross Parker}
\address{IDA Center for Communications Research - Princeton, Princeton, NJ, 08540, USA}
\email{r.parker@ccr-princeton.org}

\author[P Germain]{Pierre Germain}
\address{Department of Mathematics, Imperial College,
London SW7 2AZ, UK}
\email{p.germain@imperial.ac.uk}


\author[J Cuevas-Maraver]{Jes\'us Cuevas-Maraver}
\address{Grupo de F\'{\i}sica No Lineal, Departamento de F\'{\i}sica Aplicada I,
Universidad de Sevilla. Escuela Polit\'{e}cnica Superior, C/ Virgen de Africa, 7, 41011-Sevilla, Spain}
\address{Instituto de Matem\'{a}ticas de la Universidad de Sevilla (IMUS). Edificio
Celestino Mutis. Avda. Reina Mercedes s/n, 41012-Sevilla, Spain}
\email{jcuevas@us.es}

\author[A Aceves]{Alejandro Aceves}
\address{Department of Mathematics, Southern Methodist University, 
Dallas, TX 75275, USA}
\email{aaceves@smu.edu}

\author[P G Kevrekidis]{P.\,G. Kevrekidis} 
\address{Department of Mathematics and Statistics, University of Massachusetts, Amherst MA 01003, USA}
\email{kevrekid@math.umass.edu}

\begin{abstract}
In the work of~\cite{Colliander2010} a minimal lattice model
was constructed describing the transfer of energy to high
frequencies in the defocusing nonlinear Schr{\"o}dinger equation.
In the present work, we present a systematic study of 
the coherent structures, both standing and traveling, that arise in the context of this
model. We find that the
nonlinearly dispersive nature of the model is responsible
for standing waves in the form of discrete compactons.
On the other hand, analysis of the dynamical features of
the simplest nontrivial variant of the model, namely the dimer case, yields both solutions where the intensity is trapped in a single site and solutions where the intensity moves between the two sites, which suggests the possibility of moving excitations in larger lattices.
Such excitations are also suggested by the dynamical evolution
associated with modulational instability.
Our numerical computations confirm this
expectation, and we systematically construct such traveling states
as exact solutions in lattices of varying size, as well as explore
their stability. A remarkable feature of these traveling lattice
waves is that they are of ``antidark'' type, i.e., 
they are mounted on top of a non-vanishing background.
These studies shed light on the existence, stability and
dynamics of such standing and traveling
states in $1+1$ dimensions, and pave
the way for exploration of corresponding configurations in higher dimensions.
\end{abstract}

\maketitle

\section{Introduction}\label{sec:intro}

Lattice nonlinear dynamical systems are of wide interest
in a diverse array of physical 
applications~\cites{Aubry2006,Flach2008,kev09}.
Some typical recent examples include, but are not
limited to, the evolution of light beams in arrays
of optical waveguides~\cite{LEDERER20081}, the study of
mean-field atomic Bose-Einstein condensates (BECs) in 
the presence of optical lattice external potentials~\cite{RevModPhys.78.179}, and the 
propagation of traveling, breathing or shock
waves in nonlinear metamaterials such as granular 
crystals~\cites{Nester2001,yuli_book,Chong2018}.
Similar structures have been analyzed
in models and experiments of electrical circuits~\cite{remoissenet},
in micromechanical cantilever arrays~\cites{cantilevers}, and
in superconducting Josephson junction lattices~\cites{alex,alex2},
as well as argued to be present during the denaturation of
the DNA double strand~\cite{Peybi}.

Arguably, one of the most prototypical models that has arisen
in the context of the interplay of dispersion (diffraction) on a lattice
 and nonlinearity is the discrete nonlinear
Schr{\"o}dinger (DNLS) equation~\cites{kev09,chriseil}. This model
has been central in the theoretical analysis
and significant experimental developments associated with
discrete solitons in optics~\cite{sukho}. Moreover, it has played 
a role in unveiling instabilities (both theoretically~\cite{PhysRevLett.89.170402}
and experimentally~\cite{Cataliotti_2003}), as well
as intriguing dynamical behavior (such as
coherent perfect absorption~\cite{doi:10.1126/sciadv.aat6539})
in atomic BECs. Finally, its role cannot be understated
as a quintessential model within mathematical physics~\cite{cole},
at the intersection of integrable and non-integrable
variants of the continuum NLS equation~\cite{AblowitzPrinariTrubatch}. 

While the DNLS equation is characterized by linear dispersion and explores its interplay
with nonlinearity, there are reasons to examine the scenario 
where dispersion is purely nonlinear (and does not have
a linear component). For instance, in the work of~\cite{vvk1},
motivated by the complicated nonlinearities associated 
with Frenkel excitons in~\cite{vvk2}, bright
discrete compactons were studied, and the results were subsequently extended
to encompass some exact results, including ones regarding moving
discrete states in such models~\cite{vvk3}. However, the focus
and motivation of the present work is different. It stems instead
from a fundamental study regarding energy cascades in models
of turbulence, which arise from considerations in the context
of the defocusing NLS equation~\cite{Colliander2010}.
The latter  considers a suitably modified notion of a ``lattice node''
as representing a group of wavenumbers in the Fourier
space formulation of the original problem. In this
setting, a minimal
model of lattice dynamics was developed in order to offer
insights regarding the transfer of energy to high frequencies.

The minimal model of~\cite{Colliander2010} has spurred considerable
further activity in its own right, including dynamical simulations
illustrating the existence of cascades in the model~\cite{jeremy1},
the exploration of the connection with Burgers equation (notably
towards the study of rarefaction waves~\cite{Her}), a consideration
of the continuum limit of the model~\cite{jeremy2}, as well as a comparative
study of integrators of such a model~\cite{gideon2}.
A notable associated question, however, remains in identifying the principal ``vehicle'' enabling the cascades within this
model. 

In the present work, motivated by all of the above interconnected
factors --- namely the broad interest in nonlinear lattice models,
the special features of this model such as its lack of linear
dispersion (and hence potential for compactly supported states),
and its nontrivial appeal as a minimal model for 
transfer of energy across wavenumbers --- we revisit this 
prototypical nonlinearly dispersive setting.
After setting the stage and reviewing
some basic properties of the model in \cref{sec:model}, we proceed
to briefly examine its modulational instability in \cref{sec:MI}, identifying
already at that level the potential for both localized and propagating
states. We then corroborate this expectation through the identification
of stationary compactly supported states in \cref{sec:compacton}, 
accompanied by the study of their spectral stability.
In \cref{sec:dimer}, we start to explore the dynamics of the system via the simplest
nontrivial case thereof, namely that of two lattice nodes, i.e., the
nonlinearly dispersive dimer. We revisit the important
``slider'' states earlier identified in~\cite{Colliander2010},
but importantly we showcase their sensitivity as separatrices in the full system dynamics which we are able to completely characterize with exact, analytical solutions and illustrate with a two-dimensional phase portrait involving relevant dynamical variables.  
Finally, this complete understanding of the
dimer case, and, in particular, the presence of states wherein the intensity is transferred between the two sites,
prompts us to explore genuinely traveling states in progressively
larger lattices in \cref{sec:moving}.
We also examine the stability of the associated waveforms. \Cref{sec:conclusions} summarizes 
our findings and presents a number of directions for future studies. We briefly comment on the continuum limit of the model in an appendix.

\section{Model}\label{sec:model}

\begin{figure}
\begin{center}
\includegraphics[width=8cm]{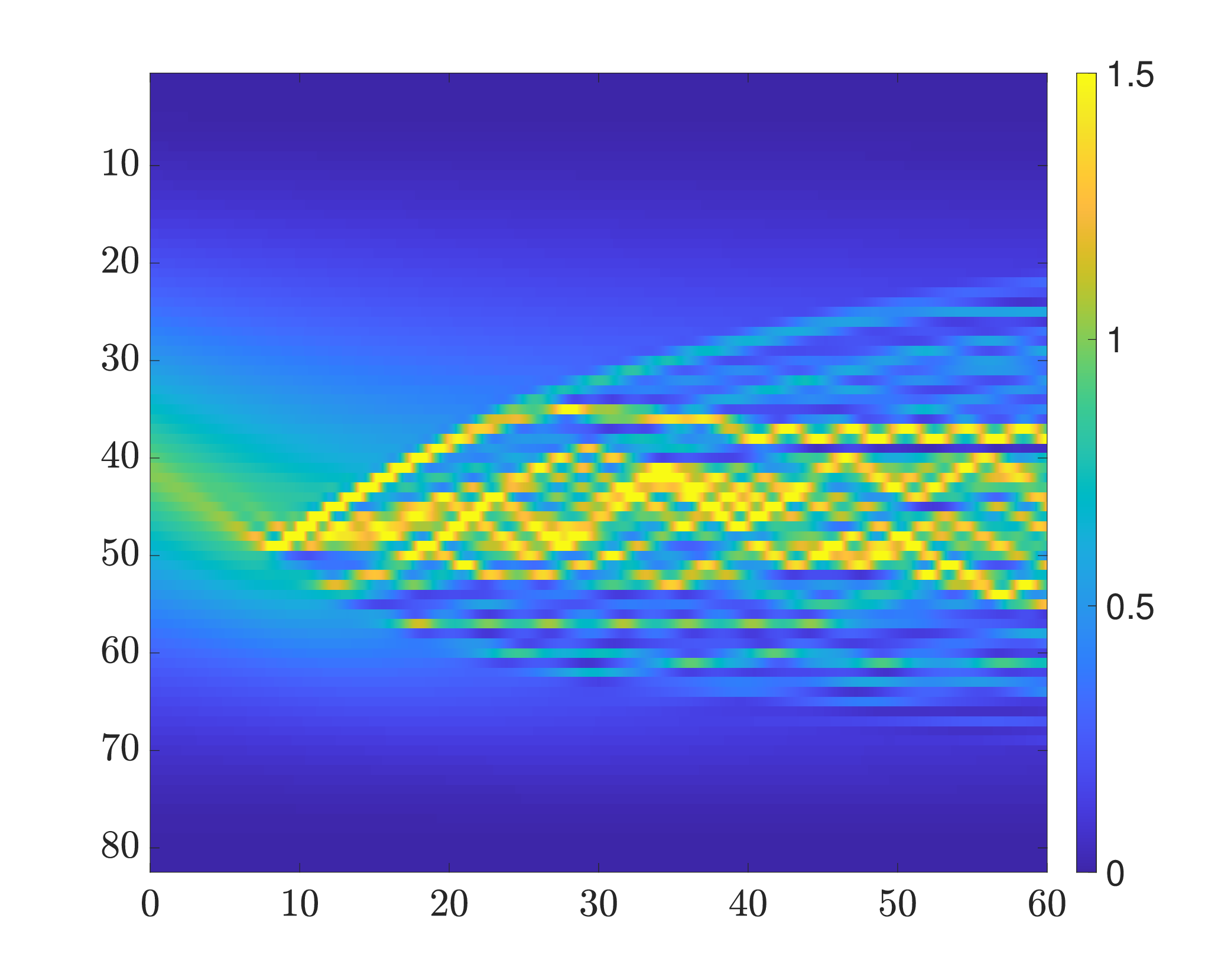}
\includegraphics[width=8cm]{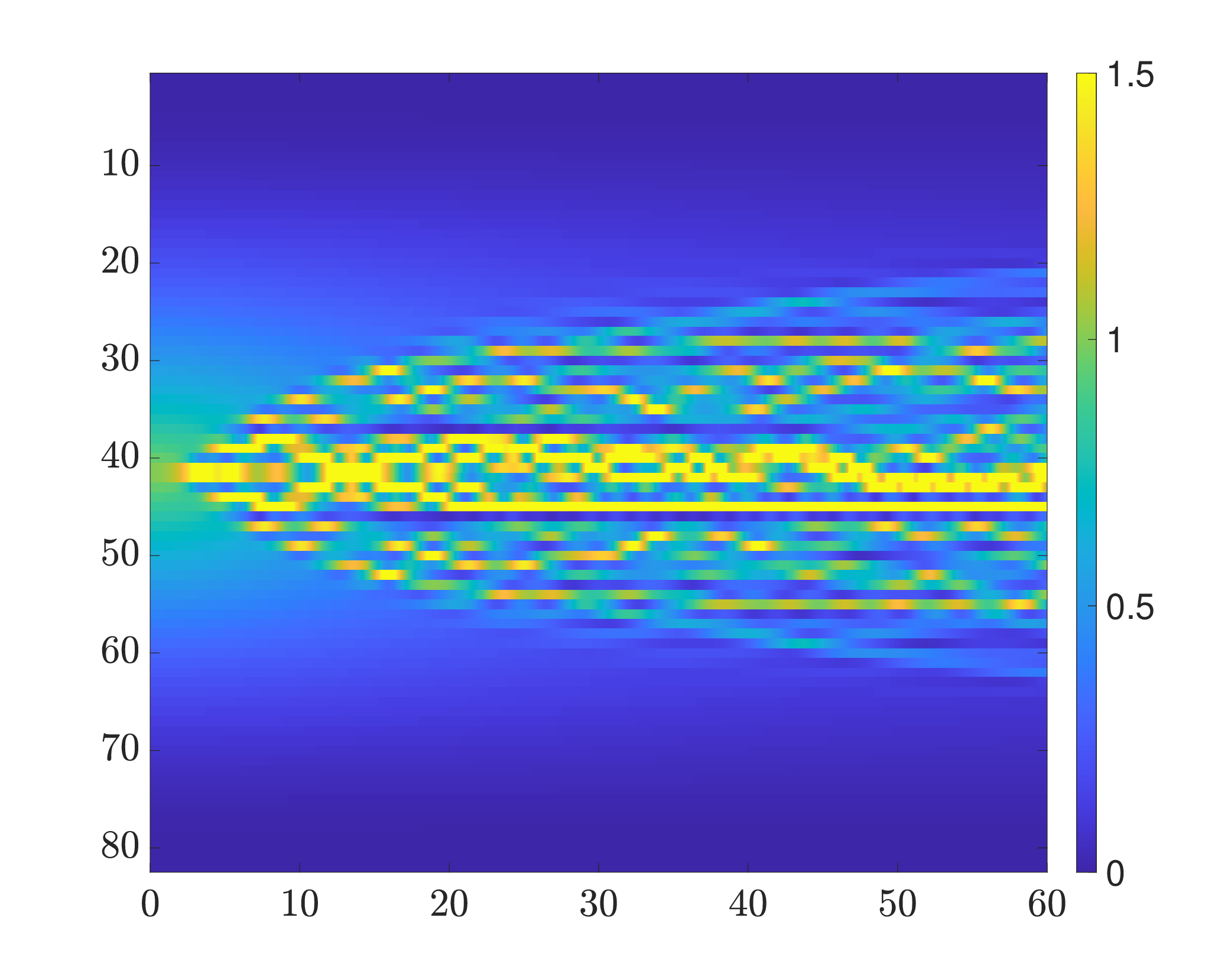}
\end{center}
\caption{Colormap of intensity $|u_j|^2$ for evolution of \cref{eq:model} in $t$. Horizontal axis is $t$, vertical axis is lattice index $j$. Initial condition is $u_j = c_j e^{i j \phi}$, where $c_j$ is a piecewise linear ramp from 0 to 1 and back, defined by $u_j = 2j/N$ for $0 \leq j \leq N/2$ and $u_j = 2 - 2j/N$ for $N/2 \leq j \leq N$. $\phi = \pi/4$ (left) and $\phi = \pi/2$
 (right). $N=80$ lattice sites, $d=0.25$. The time evolution is performed using the Dormand-Prince integrator, implemented in Matlab by means of the \texttt{ode45} function.}
\label{fig:evol1}
\end{figure}

The model we will be considering here is the fully nonlinear lattice differential equation
\begin{equation}\label{eq:model}
i \dot{u}_j + d (u_{j-1}^2 + u_{j+1}^2) \overline{u}_j - |u_j|^2 u_j = 0,
\end{equation}
where $u_j \in \C$ and $d>0$ quantifies the nonlinear nearest neighbor coupling. (See section 2 of \cite{Colliander2010} for a derivation of this model, which is equation (2.15) in that reference with $d=2$.)
The present exploration of solutions to \cref{eq:model} is motivated by timestepping experiments showing the appearance and breakdown of a diverse array of coherent structures which exist in different parts of the lattice; see~\cref{fig:evol1}
for a pertinent illustration. Examples suggested by the figure
include traveling solutions (\cref{fig:evol1}, left, starting around $t=8$ and $j=50 $, the intensity moves to lower $j$), ``breather'' solutions (\cref{fig:evol1}, left, the intensity alternates regularly between sites 41 and 42 starting around $t=40$) and stationary solutions (\cref{fig:evol1}, right, the intensity is constant at site 46 starting around $t=20$).

Equation \cref{eq:model} is Hamiltonian, with conserved energy given by
\begin{equation}\label{eq:H}
H(u) = \sum_j \left( \frac{1}{4}|u_j|^4 - 
\frac{d}{4}\left( \overline{u}_j^2 u_{j-1}^2
+ u_j^2 \overline{u}_{j-1}^2 \right) \right),
\end{equation}
which follows from translation symmetry of \cref{eq:model} in $t$. By the Cauchy-Schwarz inequality,
$$
\frac{1-2d}{4} \sum_j |u_j|^4 \leq H(u) \leq \frac{1+2d}{4} \sum_j |u_j|^4,
$$
which implies, in particular, that the Hamiltonian is coercive if $d \in (0,1/2)$ (it is then equivalent to the $\ell^4$ norm); we will see that it is useful to think of this case as defocusing. 
The power of the solution (squared $\ell^2$ norm)
\[{}
P(u) = \| u \|_{\ell^2}^2 = \sum_j |u_j|^2
\]
is also conserved, which follows from the gauge symmetry $u \mapsto e^{i\theta} u$ of \cref{eq:model}. In addition, the model is invariant under the transformation $u \mapsto a u$, $t \mapsto a^3 t$, for a real constant $a$. As a consequence, scaling the amplitude of the solution does not qualitatively affect the solution but merely speeds up or slows down its time evolution. Finally, some ``staggering transforms'' act in an interesting way on the equation. The transform $u_j \mapsto \epsilon_j u_j$, where $\epsilon_j = \pm 1$, leaves the equation invariant. The transform $u_j \mapsto i^j u_j$ amounts to flipping the sign of $d$, which shows that the case $d<0$ is included in our analysis, thus we can take $d>0$ without loss of generality.

Defining the density matrix elements by 
\begin{equation}\label{eq:rho}
\rho_{jk} = u_j \overline{u_k},
\end{equation}
the evolution equation for $\rho_{jk}$ is given by
\begin{equation}\label{eq:evolrho}
\begin{aligned}
\frac{d}{dt}\rho_{jk} = i &\Big[ 
d\left( \rho_{j-1,j}\rho_{j-1,k} + \rho_{j+1,j}\rho_{j+1,k}
- \rho_{j,k-1}\rho_{k,k-1} - \rho_{j,k+1}\rho_{j,k+1} \right) \\
&\quad + \left( \rho_{kk} - \rho_{jj} \right) \rho_{jk}
\Big].
\end{aligned}
\end{equation}
The intensity at lattice site $j$ is given by $\rho_{jj} = |u_j|^2$, which has evolution 
\begin{equation}\label{eq:evolint}
\begin{aligned}
\frac{d}{dt}\rho_{jj} &= i d\left( \rho_{j-1,j}^2 + \rho_{j+1,j}^2 - \rho_{j,j-1}^2 - \rho_{j,j+1}^2 \right) \\
&= -2 d \Im \left( \rho_{j-1,j}^2 +  \rho_{j+1,j}^2 \right),
\end{aligned}
\end{equation}
where we used the fact that $\rho_{jk} = \overline{\rho_{kj}}$. We can also separate real and imaginary parts by writing $u_j = a_j + i b_j$ for real $a_j$ and $b_j$. Equation \cref{eq:model} can then be written as
\[
\frac{d}{dt}\begin{pmatrix} a_j \\ b_j \end{pmatrix} = 
\begin{pmatrix}
(a_j^2 + b_j^2) b_j - 2 d a_j ( a_{j-1}b_{j-1} + a_{j+1}b_{j+1}) 
+ d b_j ( a_{j-1}^2 + a_{j+1}^2 - b_{j-1}^2 - b_{j+1}^2) \\
 -(a_j^2 + b_j^2) a_j + 2 d b_j ( a_{j-1}b_{j-1} + a_{j+1}b_{j+1} ) 
+ d a_j (a_{j-1}^2 + a_{j+1}^2 - b_{j-1}^2 - b_{j+1}^2 )
\end{pmatrix}.
\]
This form of the equation is useful for numerical analysis, as well as for the linear stability analysis in \cref{sec:compactonlinear} below.

The system \cref{eq:model} can be posed either on the full integer lattice or on a finite lattice comprising $N$ nodes. Since equation \cref{eq:model} can be written as  
\begin{equation}\label{eq:model2}
\dot{u}_j = i \left[ d (u_{j-1}^2 + u_{j+1}^2)  - u_j^2 \right] \overline{u_j},
\end{equation}
it follows that if $u_j(0) = 0$, then $u_j(t) = 0$ for all $t > 0$. If the initial data on the full integer lattice is nonzero only at a finite number of lattice sites, the system is equivalent to one on a finite lattice. In other words, intensity cannot spread to sites which are initialized to 0 (or bypass these sites), which is a feature fundamentally different from the linear dispersion case.

\section{Modulational instability}\label{sec:MI}

We now turn to an analysis of modulational
instability (MI) in the model, in order to further motivate the wave features which we will subsequently explore. Plane wave solutions of \cref{eq:model} 
can be found of the form
$$
u_j(t) = B e^{i( kj-\omega t)}, 
$$
with $k \in [-\pi,\pi]$. Substituting this into equation \cref{eq:model}, these plane waves satisfy the dispersion relation
$$
\omega = |B|^2 (1 - 2d \cos(2 k)).
$$
To understand the stability of such plane waves, we 
perturb according to
$$
u_j(t) = B e^{i(kj-\omega t)} (1 + a_j(t)).
$$
Linearizing in $a_j$ (and using the dispersion relation) leads to the equation
$$
- i \partial_t a_j = \omega (a_j - \overline{a_j}) + |B|^2 (-2 a_j + 2 d e^{2  ik} a_{j+1} + 2d e^{-2 i k} a_{j-1}).
$$
Taking the Fourier transform normalized as
$$
\widehat{a}(\theta) = \sum a_j e^{- i \theta j},
$$
with $\theta \in [0,2\pi]$ being the wavenumber
of the perturbation, this becomes
$$
- i \partial_t \widehat{a}(\theta) = \omega  (\widehat{a}(\theta) -  \overline{\widehat{a}(-\theta)}) +  |B|^2 (-2 + 4 d \cos(2k+\theta))\widehat{a}(\theta).
$$
This can be written as the vector equation
$$
-i \partial_t A = M A,
$$
with $A = \widehat{a}(\theta)/\,\overline{\widehat{a}(-\theta)}$ and
\begin{equation}\label{eq:MImatrix}
\begin{aligned}
&M = \begin{pmatrix} |B|^2(-2 + 4d \cos(2k + \theta)) + \omega & -\omega \\  \omega & -|B|^2(-2 + 4d \cos(2k - \theta)) + \omega \end{pmatrix}.
\end{aligned}
\end{equation}
Stability is then equivalent to the matrix $M$ having real eigenvalues, or, in other words,
\begin{equation}\label{eq:MIcrit}
(\cos(2k+\theta) + \cos(2k - \theta) - 2\cos(2k))(-1  + d \cos(2k+\theta) + d \cos(2k - \theta)) > 0.
\end{equation}
Using standard trigonometric identities, this criterion simplifies to
\begin{equation}\label{eq:MIcrit2}
h(\theta) = 2 \cos(2 k) (\cos \theta - 1)(2d \cos(2k) \cos\theta - 1) > 0,
\end{equation}
which is quadratic in $\cos \theta$ for fixed $d$ and $k$. Equation \cref{eq:MIcrit2} always has a root at $\theta = 0$; for $\qquad |2 d \cos(2k)| \geq 1$, it has an additional pair of roots at 
\[
\theta = \pm \arccos\left( \frac{1}{2d \cos(2k)} \right).
\]
Given the dependence of these expressions on $\cos \theta$, we can restrict the
discussion (by mirror symmetry) to $\theta>0$ hereafter.

\begin{figure}
\begin{center}
\includegraphics[width=5cm]{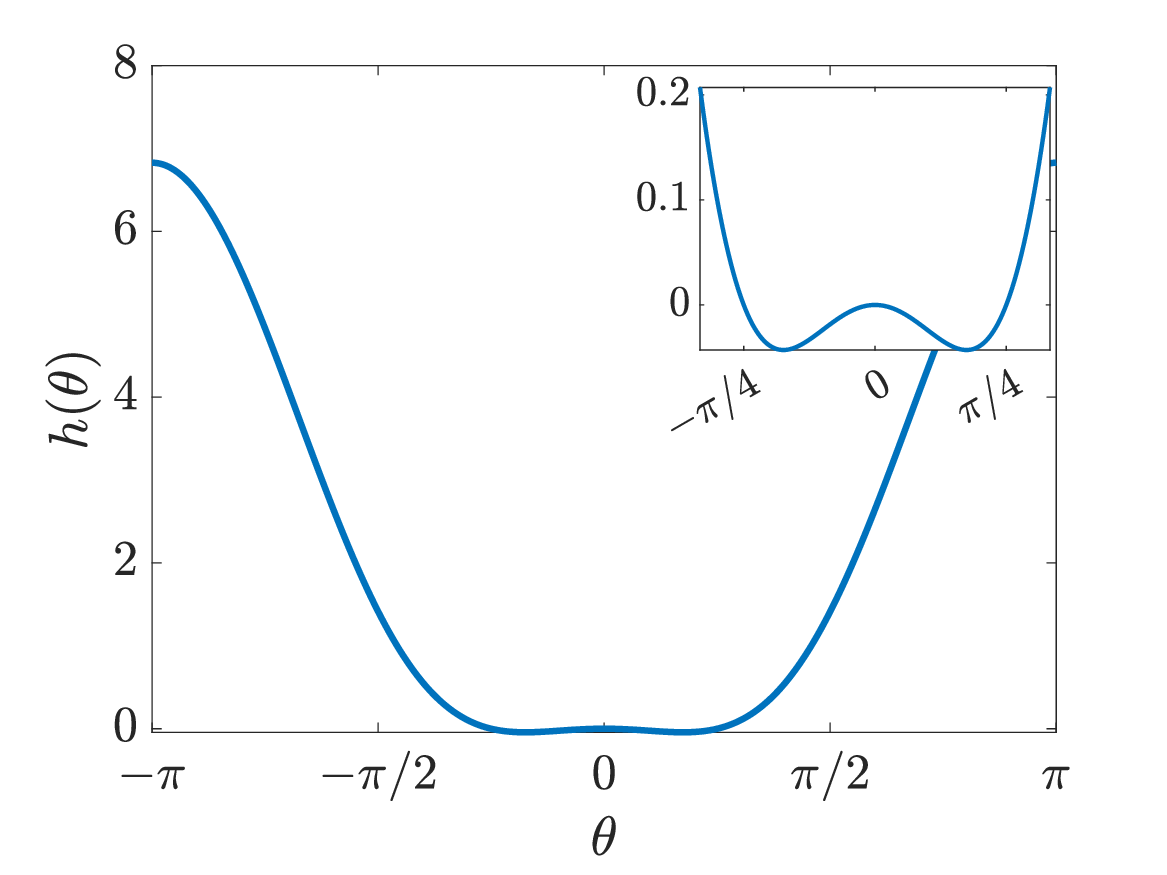}
\includegraphics[width=5cm]{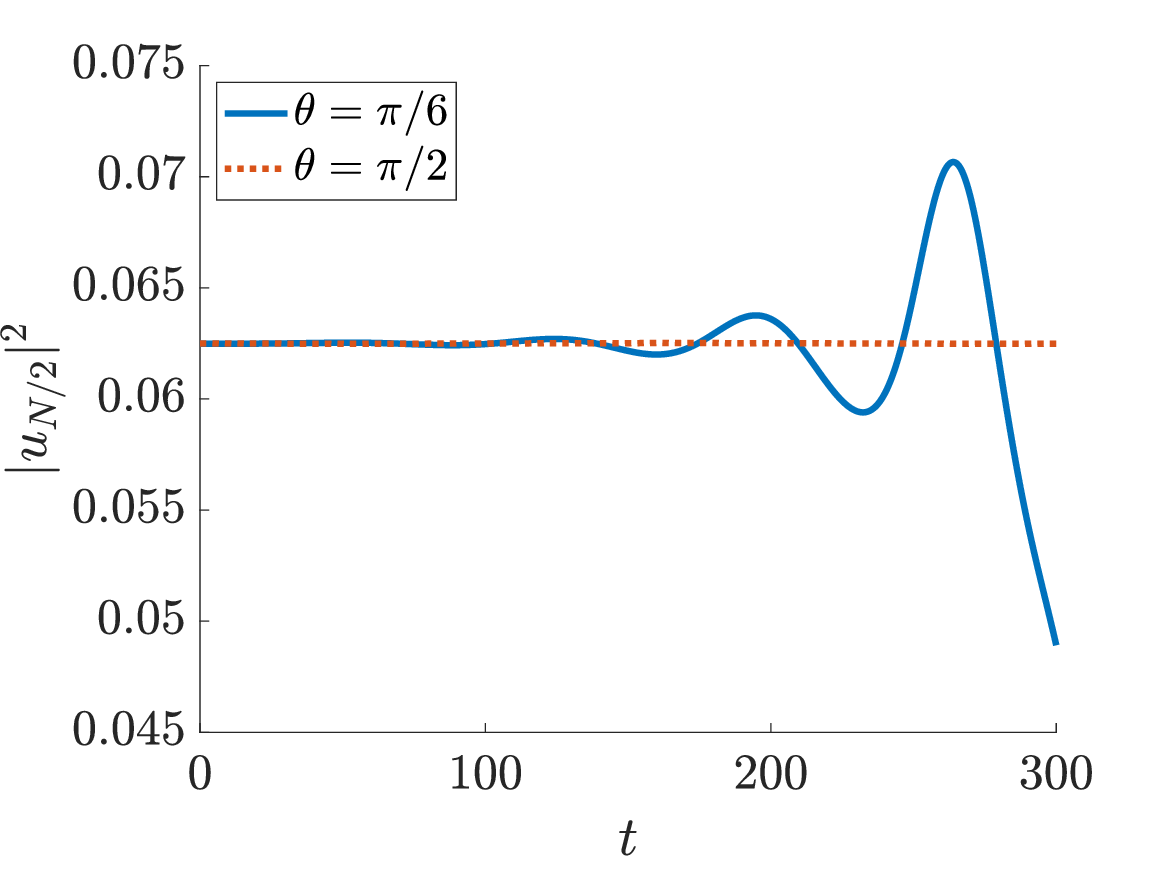}
\includegraphics[width=5cm]{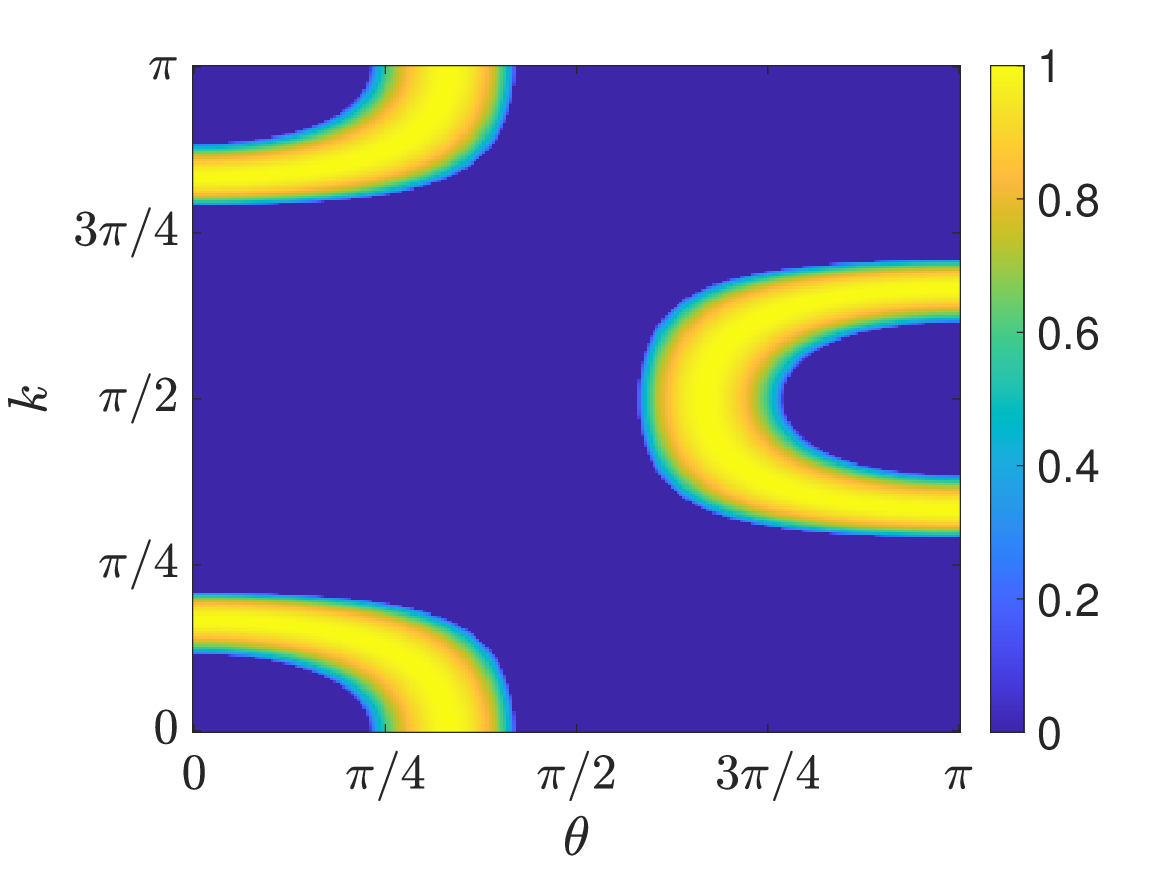}
\end{center}
\caption{Left: plot of $h(\theta)$ from \cref{eq:MIcrit2} vs. $\theta$ for $k=\pi/8$. 
Center: Evolution of the perturbation of plane wave using the initial condition $u_j(0) = B e^{i k j}(1 + \epsilon e^{-i\theta j})$ with $\epsilon = 0.0001$ and $B = 1/4$ for $k=\pi/8$. The time evolution is performed using the Dormand-Prince integrator, implemented in Matlab by means of the \texttt{ode45} function.
Right: Regions of MI in the $(\theta,k)$ plane; intensity of colormap is maximum imaginary part of matrix $M$ from \cref{eq:MImatrix}. The plot can be extended to negative $\theta$ and $k$ by symmetry. $d=1$ for all plots.}
\label{fig:MI1}
\end{figure}

As an example, the left panel of \cref{fig:MI1} plots $h(\theta)$ from \cref{eq:MIcrit2} vs. $\theta$ for $k=\pi/8$. (The specific wavenumber is chosen so that the periodic boundary conditions on a lattice of size $N=256$ nodes are satisfied.) The roots of $h(\theta)$ are at 0 and $\pm \pi/4$, thus $h(\theta)$ is negative for $\pi/4 < \theta < 0$ and $0 < \theta < \pi/4$, which is the MI region. 
Compare the evolution of the two perturbations in the center panel of \cref{fig:MI1}; the perturbation with $\theta = \pi/6$ (solid blue line) is within the MI region and grows exponentially, in contrast to the perturbation with $\theta = \pi/2$ (dotted orange line), which is outside the MI region and thus does not grow. 
A colormap showing regions of MI in the $(\theta, k)$ plane is shown in the right panel of \cref{fig:MI1}; the color indicates the growth rate of MI, as given by the maximal imaginary part of the eigenvalues of \cref{eq:MImatrix}.
Two interesting observations here are as follows.
First, MI is not always (and in particular is not for $k=0$) a long-wavelength instability with a band starting at
$\theta=0$, as is typically the case in NLS models. Second, there are regions of modulationally stable
wavenumbers $k$.

\begin{figure}
\begin{center}
\includegraphics[width=7cm]{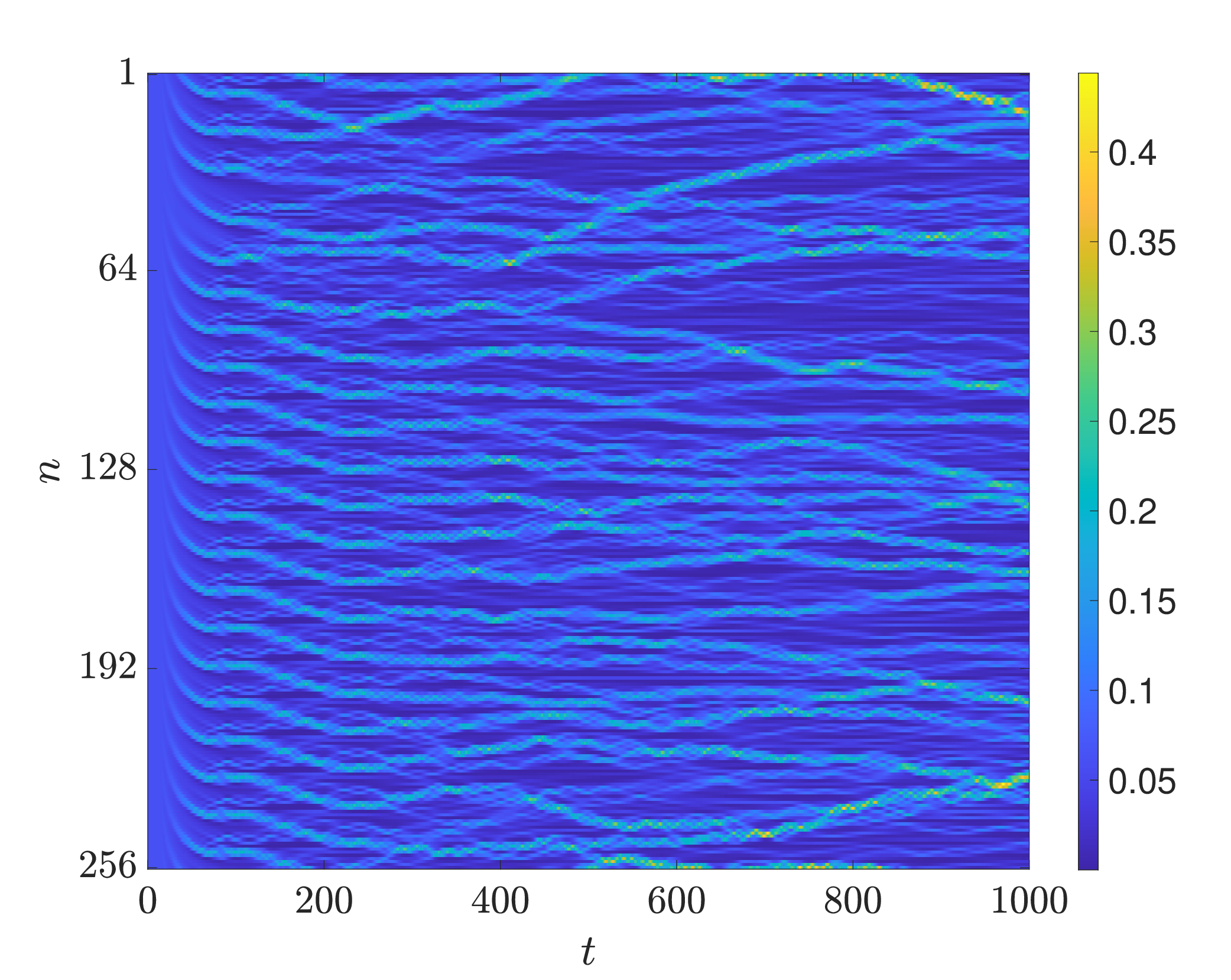}
\includegraphics[width=7cm]{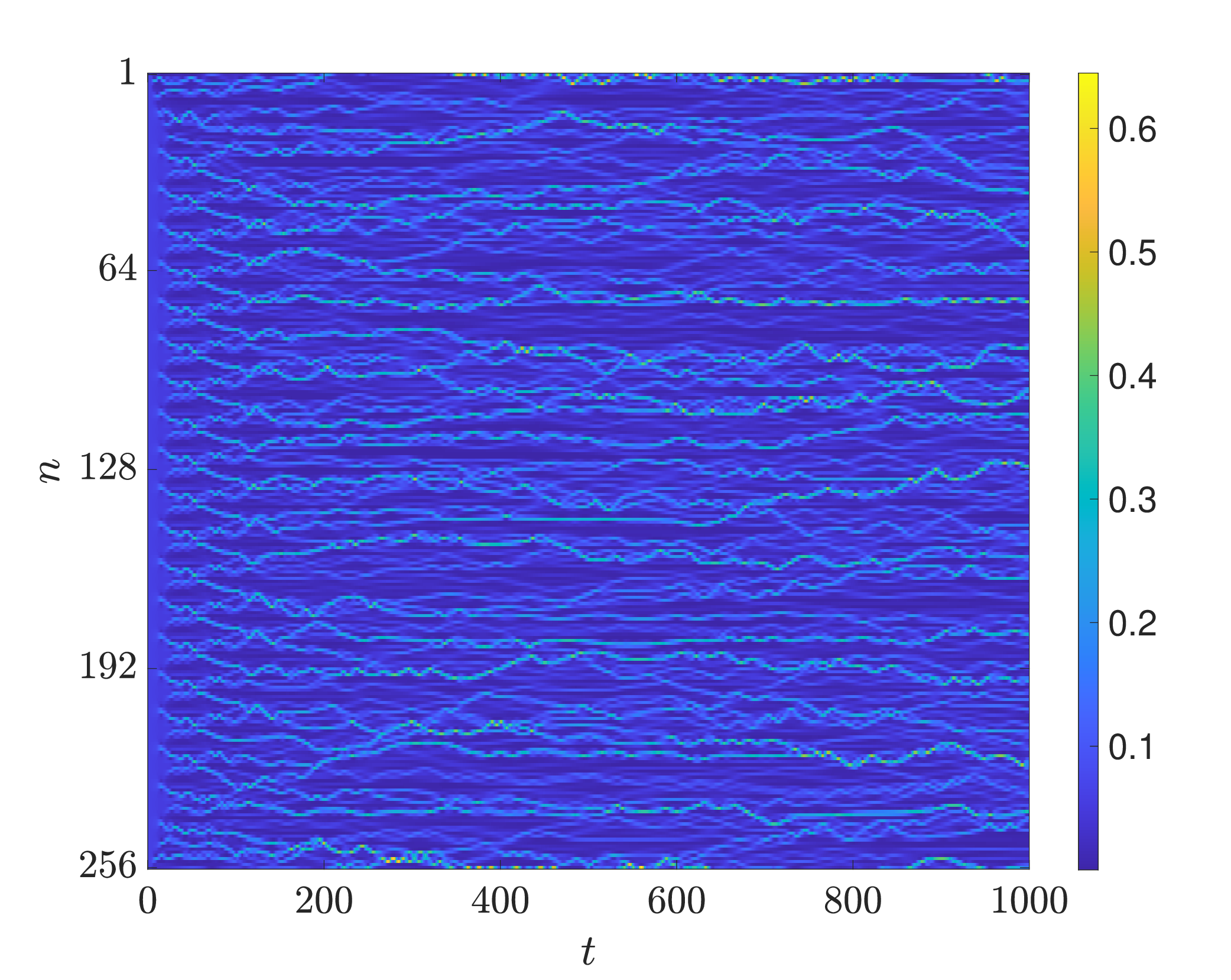}
\end{center}
\caption{Evolution of perturbation of a plane wave using initial condition $u_j(0) = B e^{i k j}(1 + \epsilon e^{-i\theta j})$ with $\epsilon = 0.0001$ and $B = 1/4$ 
for $k=\pi/8$, $\theta = \pi/6$ (left) and $k=\pi/2$, $\theta = \pi/6$ (right). For all plots, $N=256$ lattice nodes with periodic boundary conditions and $d=1$. The time evolution is performed using the Dormand-Prince integrator, implemented in Matlab by means of the \texttt{ode45} function.}
\label{fig:MI2}
\end{figure}

Colormaps showing the evolution of MI for all lattice sites are shown in \cref{fig:MI2}; comparison of these to the evolution plots in \cref{fig:evol1} suggests that MI plays a significant role in the dynamics of this system.
Importantly, the astute reader can discern a number
of both standing and moving waves in the pattern
that results from the MI. It is to these coherent
structures
that we now turn in more detail in what follows.

\section{Standing Waves: Compactons}\label{sec:compacton}

The first nonlinear structures of interest are compactons, which are standing waves supported on a finite set of $N$ adjacent lattice sites. (These also appear in different nonlinearly dispersive DNLS variants in, e.g.,~\cites{vvk1,vvk2,vvk3}, as discussed earlier.) Standing waves are solutions of the form 
\begin{equation}\label{eq:standingansatz}
u_j = c_j e^{-i \omega t}, 
\end{equation}
with frequency $\omega$ and amplitudes $c_j$. Although these amplitudes are traditionally real (as in, for example, the DNLS equation), we will see below that there is a class of solutions (the staggered compactons) where this is not the case. 

\subsection{Real compactons}\label{sec:realcompactons}

\begin{figure}
\begin{center}
\includegraphics[width=12cm]{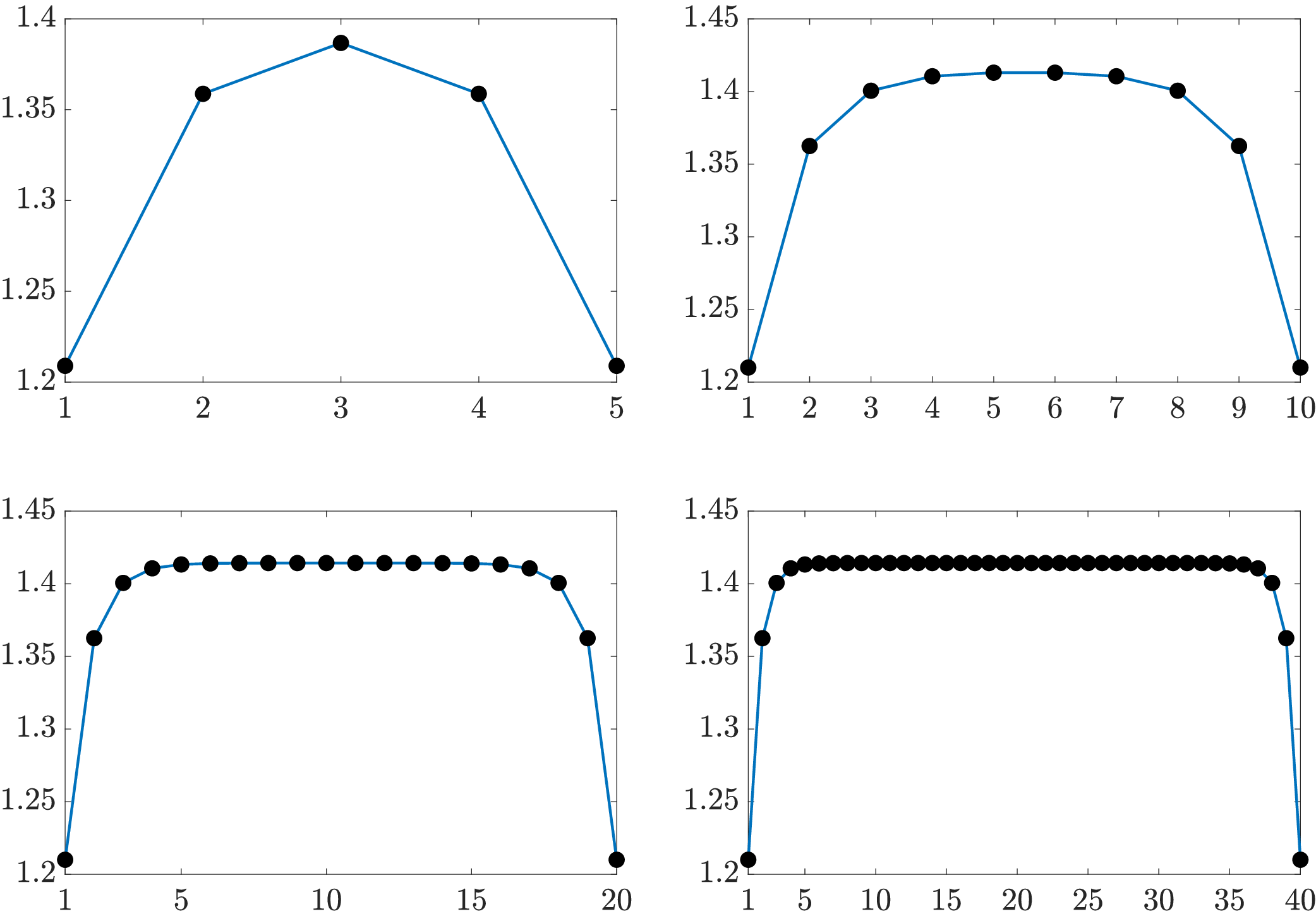}
\end{center}
\caption{Compacton solutions obtained by solving equation \cref{eq:cjeq} for $N=5, 10, 20$, and $40$. Positive amplitude $c_j$ is chosen at each site, $d=0.25$, $\omega = 1$.}
\label{fig:compactons}
\end{figure}

Real compactons are solutions of the form \cref{eq:standingansatz}, where all the $c_j$ are taken to be real. In this case, substituting \cref{eq:standingansatz} into \cref{eq:model} and simplifying, we obtain the standing wave equation
\begin{equation}\label{eq:st}
d (c_{j-1}^2 + c_{j+1}^2) c_j - c_j^3 + \omega c_j = 0.
\end{equation}
For a compacton comprising $N$ sites labeled $c_1$ to $c_N$, since $c_j \neq 0$ for all $j$, we can divide equation \cref{eq:st} by $c_j$ to obtain
\begin{equation}\label{eq:cjeq}
\begin{pmatrix}
1 & -d \\
-d & 1 & -d \\
& -d & 1 & -d \\
& & \ddots & \ddots & \ddots \\
& & & -d & 1 & -d \\
& & & & -d & 1
\end{pmatrix}
\begin{pmatrix}
c_1^2 \\ c_2^2 \\ c_3^2 \\ \vdots \\ c_{N-1}^2 \\ c_N^2
\end{pmatrix}
=
\begin{pmatrix}
\omega \\ \omega \\ \omega \\ \vdots \\ \omega \\ \omega
\end{pmatrix},
\end{equation}
which is linear in the square amplitudes $c_j^2$ and can be solved by row reduction. 
Although \cref{eq:cjeq} has a unique solution whenever the matrix is nonsingular, this solution is only valid if $c_j^2 > 0$ for all $j$, since we are taking the amplitudes $c_j$ to be real.
See \cref{fig:compactons} for representative compacton solutions. We note that we can take either the positive or the negative root for each amplitude $c_j$.

For $N=1$, the compacton is a single-site standing wave $u_1 = c_1 e^{-i \omega t}$, with $c_1 = \pm \sqrt{\omega}$. For $N=2,3,4$, and $5$, solving this linear system yields the solutions from \cref{table:compactons}. For fixed $\omega>0$, the norm of these solutions blows up as $d$ approaches $1$, $1/\sqrt{2}$, $(\sqrt{5}-1)/2$, and $1/\sqrt{3}$ (respectively) from below; the matrix in \cref{eq:cjeq} is singular at these values of $d$.
For a given $\omega$, a compacton solution exists only if all of the square amplitudes $c_j^2$ are positive (see the intervals of existence in \cref{table:compactons}); this depends on whether $\omega > 0$ or $\omega < 0$. For example, for $N=2$, a real compacton exists on $(0, 1)$ for $\omega > 0$ and on $(1, \infty)$ for $\omega < 0$. Interestingly, the 2-site compacton is spectrally stable on both of these intervals. (See \cref{sec:linearrealcompacton} below for further discussion of stability; we note here that spectral stability does not change at the existence thresholds in \cref{table:compactons} where the norm of the solution blows up.)

For $N=2$ and $N=3$, the matrix in \cref{eq:cjeq} is only singular at the values of $d$ already discussed. For $N \geq 4$, however, the matrix is singular at other values of $d$. For example, when $N=4$, the matrix is singular when $d=(\sqrt{5}+1)/2$. At this value of $d$, the solution in \cref{table:compactons} exists but is not unique; we can add any multiple of the kernel vector $(d, 1, -1, -d)$ to obtain another solution. The case when $N=5$ is similar. The matrix in \cref{eq:cjeq} is singular when $d=1$, in which case we can add any multiple of the kernel vector $(1, 1, 0, -1, -1)$ to get another solution. (See \cref{sec:linearrealcompacton} below for a discussion on how spectral stability changes at these singular points.)
In addition, for $N=5$, real compactons do not exist when $d > (1+\sqrt{5})/2$, since, in that case, $c_1^2$ and $c_2^2$ have opposite signs for all $\omega$. For larger $N$, analytic computation of exact solutions is less straightforward. Numerical computations strongly suggest that real compactons of all sizes $N$ exist for $0 < d < 1/2$ (see, in addition, the discussion below). Existence results for compactons for $d$ outside this interval are more complicated due the requirement that all the $c_j^2 > 0$. (The blue filled circles in \cref{fig:compactonH} indicate which compactons exist for a few values of $d > 1/2$.)
Finally, we note that the real compacton solutions are characterized by a plateau in the center of the solution. For $0 < d < 1/2$, the height of this plateau approaches 
\begin{equation}\label{eq:compactonplateau}
c^2=\frac{\omega}{1-2d}
\end{equation}
for large $N$, which is found by taking $c_j=c_{j-1}=c_{j+1}=c$ in \cref{eq:st} and solving for $c$.

\begin{table}
\begin{tabular}{@{}llllll@{}}
\toprule
$N$ & real compacton solution &&& $\omega > 0$ & $\omega < 0$ \\ \midrule
2 &  $c_1^2,\,c_2^2 = \frac{\omega}{1-d}$ & & & $(0,1)$ & $(1,\infty)$ \\
3 &  $c_1^2,\,c_3^2 = \frac{\omega(1 + d)}{1 - 2 d^2}$ & $c_2^2 = \frac{\omega(1 + 2 d)}{1 - 2 d^2}$ & & $\left(0,\frac{1}{\sqrt{2}}\right)$ & $\left(\frac{1}{\sqrt{2}},\infty\right)$ \\
4 &  $c_{1}^2,\,c_4^2 = \frac{\omega}{1 - d - d^2}$ &
$c_{2}^2,\,c_{3}^2 = \frac{\omega(1+d)}{1 - d - d^2}$ & & $\left(0,\frac{\sqrt{5}-1}{2}\right)$ & $\left(\frac{\sqrt{5}-1}{2},\infty\right)$ \\
5 &  $c_{1}^2,\,c_5^2 = \frac{\omega(1+d-d^2)}{1-3d^2}$ &
$c_{2}^2,\,c_{4}^2 = \frac{\omega(1+2d)}{1-3d^2}$ & $c_3^2 = \frac{\omega(1 + d)^2}{1 - 3 d^2}$ & $\left(0,\frac{1}{\sqrt{3}}\right)$ & $\left(\frac{1}{\sqrt{3}},\frac{\sqrt{5}+1}{2}\right)$ \\ \bottomrule
\end{tabular}
\caption{Square amplitudes for real compacton solutions for $N=2,3,4$ and $5$, together with intervals of existence for these solutions for $\omega > 0$ and $\omega < 0$.}
\label{table:compactons}
\end{table}

To better understand the solutions of the above linear problem, we observe first that it suffices to consider the case $\omega=1$. Denoting $M$ for the matrix in \cref{eq:cjeq}, $\mathbf{x}$ for the vector $(c_j^2)$ and $\mathbf{1}$ for the vector $(1,\dots,1)^\top$, equation \cref{eq:cjeq} becomes
$$
M \mathbf{x} = \mathbf{1}.
$$
Since $M$ is a tridiagonal Toeplitz matrix, its eigenvalues $\lambda_k$ and eigenvectors $\mathbf{e}_k$ can be computed explicitly (see, for instance~\cite{Smith1985}, page 154):
$$
\lambda_k = 1 - 2d \cos \left( \frac{k\pi}{N+1} \right),\qquad \mathbf{e}_k = \left(\sin \left( \frac{kj \pi}{N+1} \right) \right)_{j=1,\dots,N}
\qquad \mbox{$k=1 \dots N$.} 
$$
For $d \in (0,1/2)$, the eigenvalues $\lambda_k$ are positive, and the matrix $M$ is an $M$-matrix; in particular its inverse has positive entries, so that $\mathbf{x}$ has positive entries.

Using trigonometric identities, we can then compute
\begin{align*}
& \| \mathbf{e}_k \|^2 = \sum_{j=1}^N \sin \left( \frac{kj \pi}{N+1} \right)^2 = \frac{N}{2} - \frac{1}{2} \sum_{j=1}^N \cos \left( \frac{2kj \pi}{N+1} \right) = \frac{N-1}{2} \\
& \langle \mathbf{1} , \mathbf{e}_k \rangle = \sum_{j=1}^N \sin \left( \frac{kj \pi}{N+1} \right) =
\left\{
\begin{array}{ll}
0 &  \mbox{if $k$ even} \\
\mbox{cotan} \left( \frac{k \pi}{2(N+1)} \right) & \mbox{if $k$ odd}.
\end{array}
\right.
\end{align*}
Since $M$ is self-adjoint, we can now invert the linear equation through the formula
$$
\mathbf{x} = \sum_{k=1}^N \frac{1}{\| \mathbf{e}_k \|^2} \langle  \mathbf{1} , \mathbf{e}_k \rangle \mathbf{e}_k.
$$
In other words, the coordinates of $\mathbf{x}$ are given by
$$
x^\ell = \frac{2}{N-1} \sum_{\substack{k \in 2 \mathbb{N}-1}} \frac{1}{1 - 2d \cos \left( \frac{k\pi}{N+1} \right)} \mbox{cotan} \left( \frac{k\pi}{2(N+1)} \right) \sin \left( \frac{k \ell \pi}{N+1} \right),
$$

We now want to find the limit as $N \to \infty$ of this expression, when $\ell$ is away from the extremities; we will assume that $\frac{\ell}{N} \to \alpha \in (0,1)$. In the above sum, the leading contribution is given by small values of $k$, due to the singularity at zero of the $\mbox{cotan}$ function. Therefore, it is legitimate to expand in $\frac{k}{N}$, which gives
$$
x^\ell \sim \frac{2}{N-1} \sum_{\substack{k \in 2 \mathbb{N}-1}} \frac{1}{1-2d} \frac{2(N+1)}{k \pi} \sin\left( \alpha k \pi \right) \sim \frac{4}{\pi(1-2d)} \sum_{\substack{k \in 2 \mathbb{N}-1}} \frac{\sin\left( \alpha k \pi \right)}{k}.
$$
By the formula for the Fourier series of the sawtooth wave
$$\sum_{k =1}^\infty \frac{\sin(kx)}{k} = \frac{\pi}{2} - \frac{x}{2}
$$ 
for $x \in (0,2\pi)$, we obtain for $\alpha \in (0,\pi)$
\begin{align*}
x^\ell & \sim \frac{4}{\pi(1-2d)} \left[ \sum_{k \in \mathbb{N}} \frac{\sin\left( \alpha k \pi \right)}{k} - \sum_{k \in \mathbb{N}} \frac{\sin\left(2 \alpha k \pi \right)}{2k} \right] = \frac{1}{1-2d}.
\end{align*}
This result suggests, in close correspondence
with \cref{fig:compactons}, that the compacton solution
becomes nearly flat in its center for sufficiently large $N$.

\subsection{The staggered compacton} \label{sec:stagcompactons}

For solutions supported on $N$ sites, where $N \geq 2$, another standing wave solution is obtained by using the ansatz
$$
u_j = i^j c_j e^{-i\omega t},
$$
where the $c_j$ are again real. We call this a staggered compacton, since there is a phase difference of $\pi/2$ between each pair of adjacent sites. Substituting this ansatz into \cref{eq:model} and simplifying, the amplitudes $c_j$ solve the equation
\begin{equation}\label{eq:ststag}
- d (c_{j-1}^2 + c_{j+1}^2) c_j - c_j^3 + \omega c_j = 0,
\end{equation}
which leads to the linear system
\begin{equation}\label{eq:cjeqstag}
\begin{pmatrix}
1 & d \\
d & 1 & d \\
& d & 1 & d \\
& & \ddots & \ddots & \ddots \\
& & & d & 1 & d \\
& & & & d & 1
\end{pmatrix}
\begin{pmatrix}
c_1^2 \\ c_2^2 \\ c_3^2 \\ \vdots \\ c_{N-1}^2 \\ c_N^2
\end{pmatrix}
=
\begin{pmatrix}
\omega \\ \omega \\ \omega \\ \vdots \\ \omega \\ \omega
\end{pmatrix}.
\end{equation}
We note that these are the same equations as those satisfied by the real compacton, except that $d$ has been changed to $-d$. As with the real compacton, a solution to \cref{eq:cjeqstag} is only valid if $c_j^2 > 0$ for all $j$.
Staggered compacton solutions for small $N$ are shown in \cref{table:stagcompactons}. Although the solutions from this table are obtained from those in \cref{table:compactons} by replacing $d$ with $-d$, their intervals of existence are very different. Of note, for $\omega > 0$, the 2-site compacton exists for all $d$; in particular, its norm does not blow up for any $d>0$. As in the real compacton case, numerical computations suggest that staggered compactons of all sizes $N$ exist for $0 < d < 1/2$. Existence results are similarly more complicated for $d > 1/2$ (see the orange unfilled circles in \cref{fig:compactonH}, which indicate compactons that exist for a few values of $d > 1/2$).

\begin{table}
\begin{tabular}{@{}llllll@{}}
\toprule
$N$ & staggered compacton solution &&& $\omega > 0$  \\ \midrule
2 &  $c_1^2,\,c_2^2 = \frac{\omega}{1+d}$ & & & $(0,\infty)$  \\
3 &  $c_1^2,\,c_3^2 = \frac{\omega(1 - d)}{1 - 2 d^2}$ & $c_2^2 = \frac{\omega(1 - 2 d)}{1 - 2 d^2}$ & & $\left(0,\frac{1}{2} \right)$, $\left(1,\infty\right)$ \\
4 &  $c_{1}^2,\,c_4^2 = \frac{\omega}{1 + d - d^2}$ &
$c_{2}^2,\,c_{3}^2 = \frac{\omega(1-d)}{1 + d - d^2}$ & & $\left(0,1\right)$ \\
5 &  $c_{1}^2,\,c_5^2 = \frac{\omega(1-d-d^2)}{1-3d^2}$ &
$c_{2}^2,\,c_{4}^2 = \frac{\omega(1-2d)}{1-3d^2}$ & $c_3^2 = \frac{\omega(1 - d)^2}{1 - 3 d^2}$ & $\left(0,\frac{1}{2}\right)$ \\ \bottomrule
\end{tabular}
\caption{Square amplitudes for staggered compacton solutions for $N=2,3,4$ and $5$, together with intervals of existence for these solutions for $\omega > 0$ . These four solutions do not exist for $\omega < 0$.}
\label{table:stagcompactons}
\end{table}

\begin{figure}
\begin{center}
\includegraphics[width=12cm]{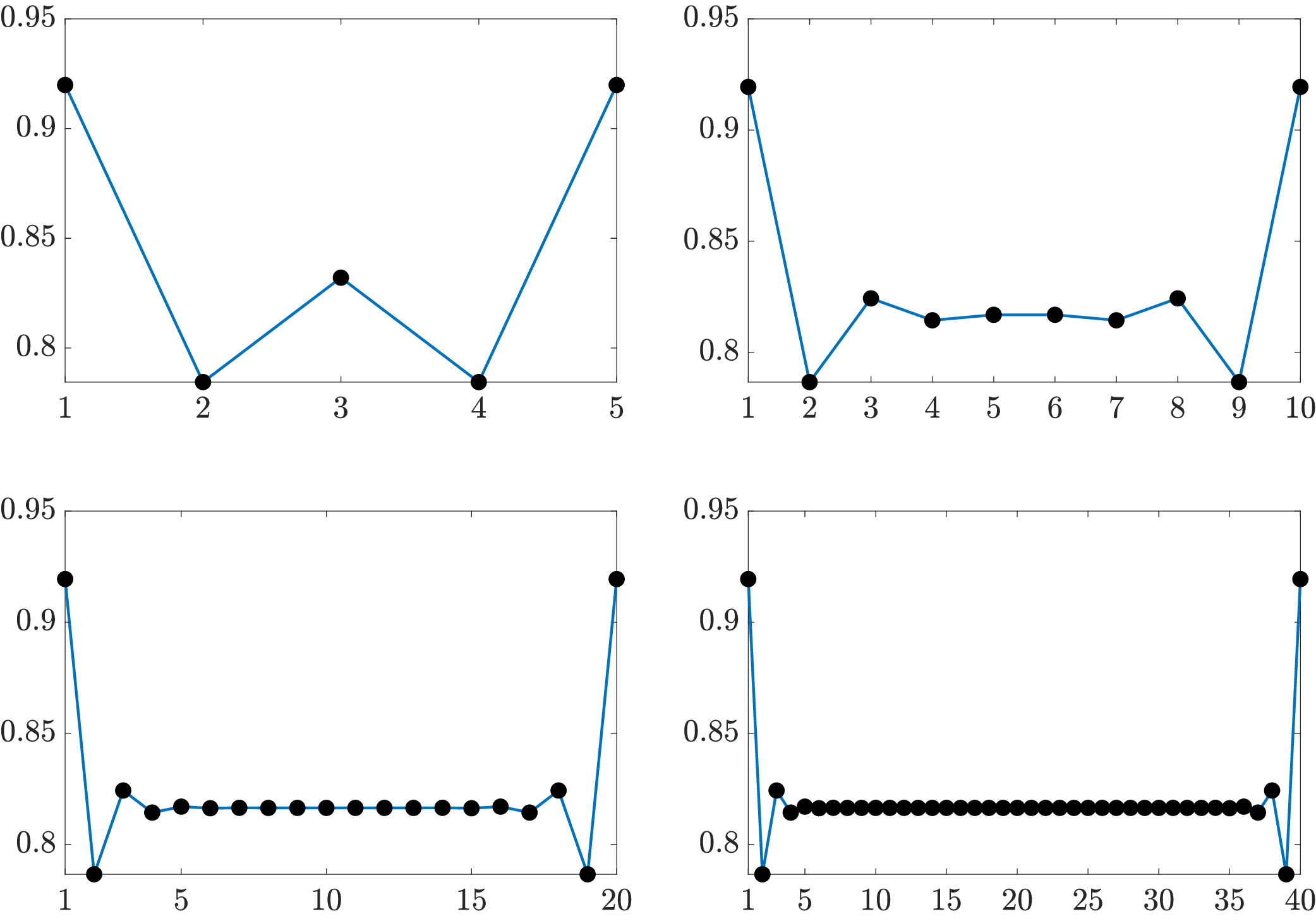}
\end{center}
\caption{Staggered compacton solutions obtained by solving equation \cref{eq:cjeqstag} for $N=5, 10, 20, 40$. Positive amplitude at each site, $d=0.25$, $\omega = 1$.}
\label{fig:compactons_stag}
\end{figure}

Plots of staggered compactons for the same values of $N$ as the real compactons are shown in \cref{fig:compactons_stag}. There is an intensity plateau in the middle of the solution; for large $N$, this plateau approaches 
\begin{equation}\label{eq:stagplateau}
c^2=\frac{\omega}{1+2d}
\end{equation}
for $0 < d < 1/2$.
As in the case of the real compacton, this corresponds to a solution $\mathbf{x}$ such that
$$
x^\ell \to \frac{1}{1+2d} \qquad \mbox{as $N \to \infty$}.
$$

\subsection{Mixed compactons}

The phase differences between adjacent lattice sites are 0 (or $\pi$, if negative roots are taken for the $c_j$) for real compactons and $\pi/2$ for staggered compactons. It is possible to construct compactons which have ``mixed'' phase differences. For example, for a 3-site compacton, we can take the ansatz
\[
u_1 = c_1, \quad u_2 = c_2, \quad u_3 = i c_3,
\]
where $c_1$, $c_2$, and $c_3$ are real. A compacton solution is then given by
\[
c_1^2 = \frac{\omega(1+d-2d^2)}{1-2d^2}, \quad
c_2^2 = \frac{\omega}{1-2d^2}, \quad
c_3^2 = \frac{\omega(1-d-2d^2)}{1-2d^2},
\]
for $0 < d < 1/2$ (the square amplitudes $c_j^2$ are all positive on this interval). This compacton is unstable. Indeed, numerical computations suggest that all such compactons are unstable, hence 
we will not consider such ``mixed-phase'' solutions hereafter.

\subsection{Energy considerations}

\begin{table}
\begin{tabular}{@{}ll@{}}
\toprule
$N$ & energy ($H$) \\ \midrule
2 & $\frac{\omega^2}{2(1-d)} = \frac{(1-d)P^2}{8}$ \\
3 & $\frac{(3+4d)\omega^2}{4(1-2d^2)} = \frac{(1-2d^2)P^2}{4(3+4d)}$  \\
4 & $\frac{(2+d)\omega^2}{2(1- d - d^2)} = \frac{(1-d-d^2)P^2}{8(2+d)}$ \\
5 & $\frac{(5+8d-d^2)\omega^2}{4(1-3d^2)} = \frac{(1-3d^2)P^2}{4(5+8d-d^2)}$\\ \bottomrule
\end{tabular}
\caption{Energy for real compacton solutions for $N=2,3,4$ and $5$ as a function of frequency $\omega$ or power of solution $P$. Energy for staggered compactons is found by replacing $d$ with $-d$. Single-site solution has energy $H=P^2/4$.}
\label{table:compactonH}
\end{table}

\begin{figure}
\begin{center}
\includegraphics[width=7cm]{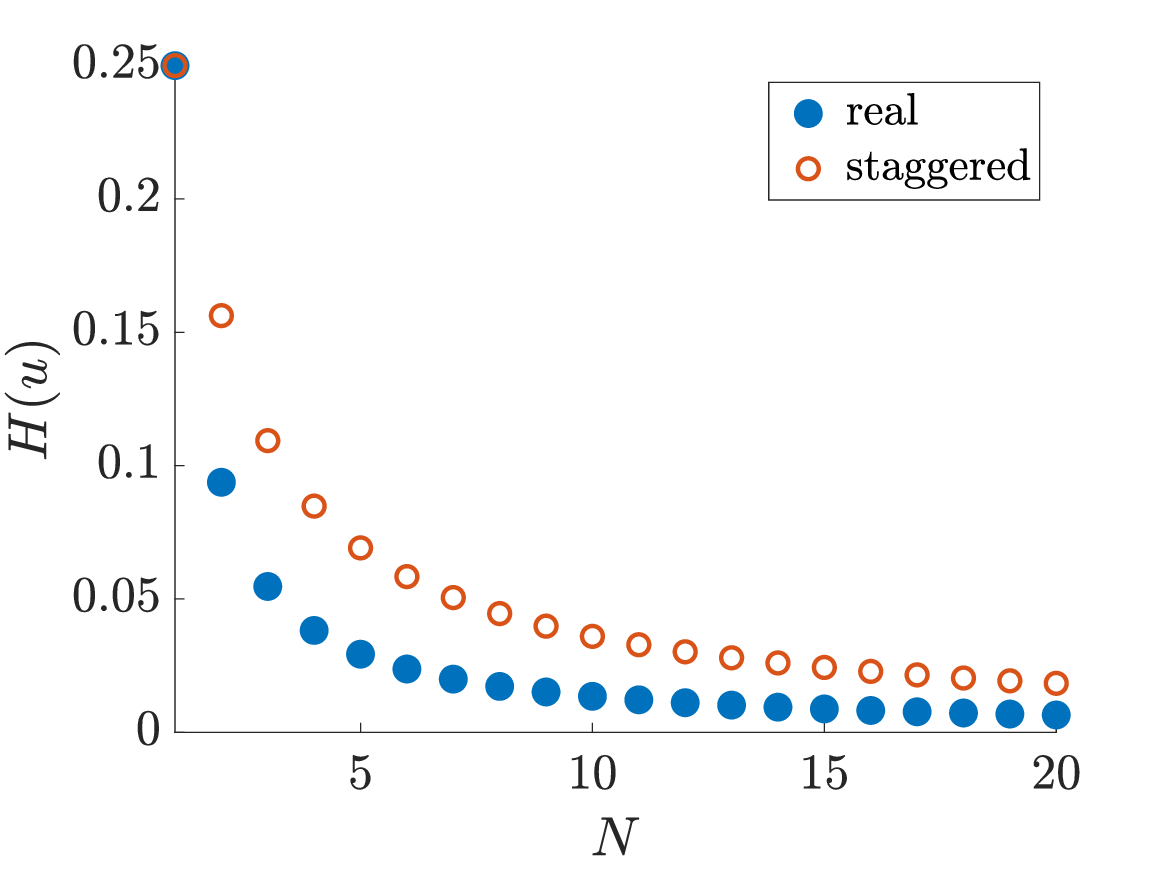}
\includegraphics[width=7cm]{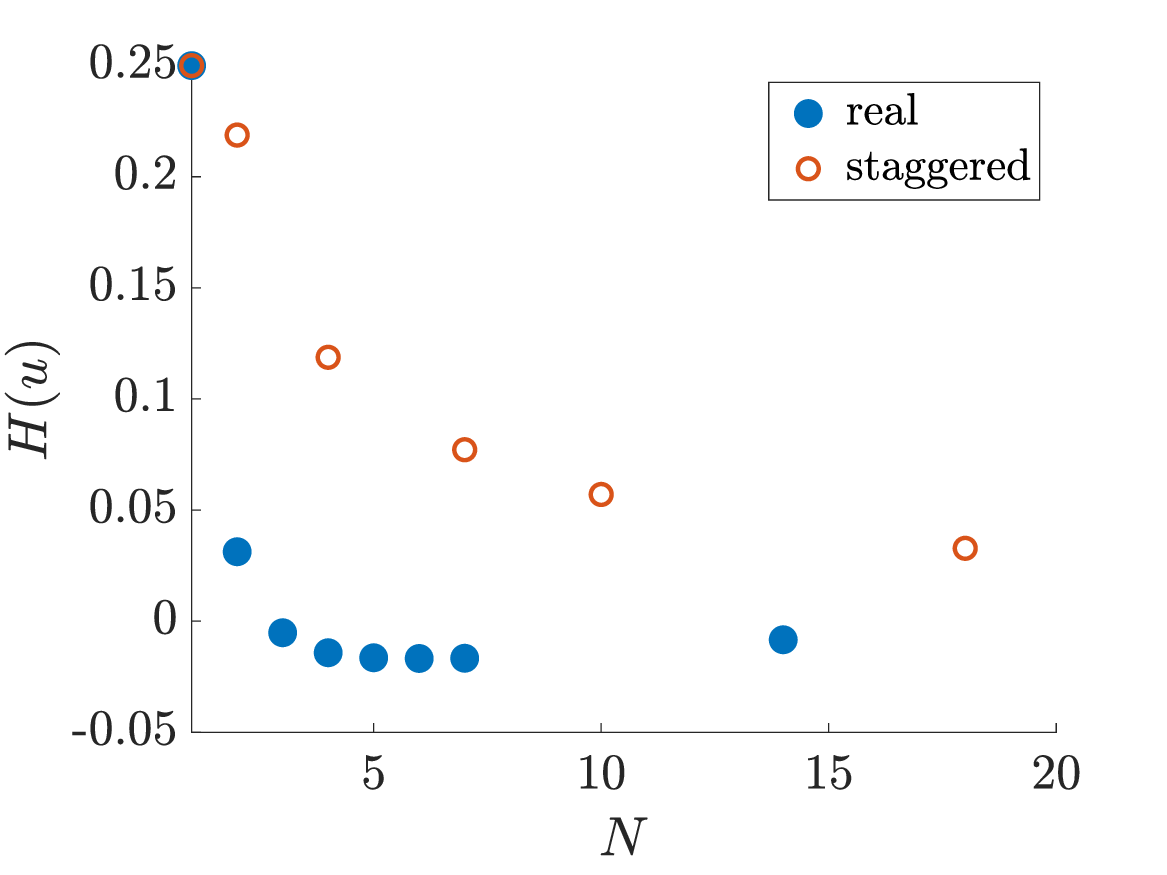}
\includegraphics[width=7cm]{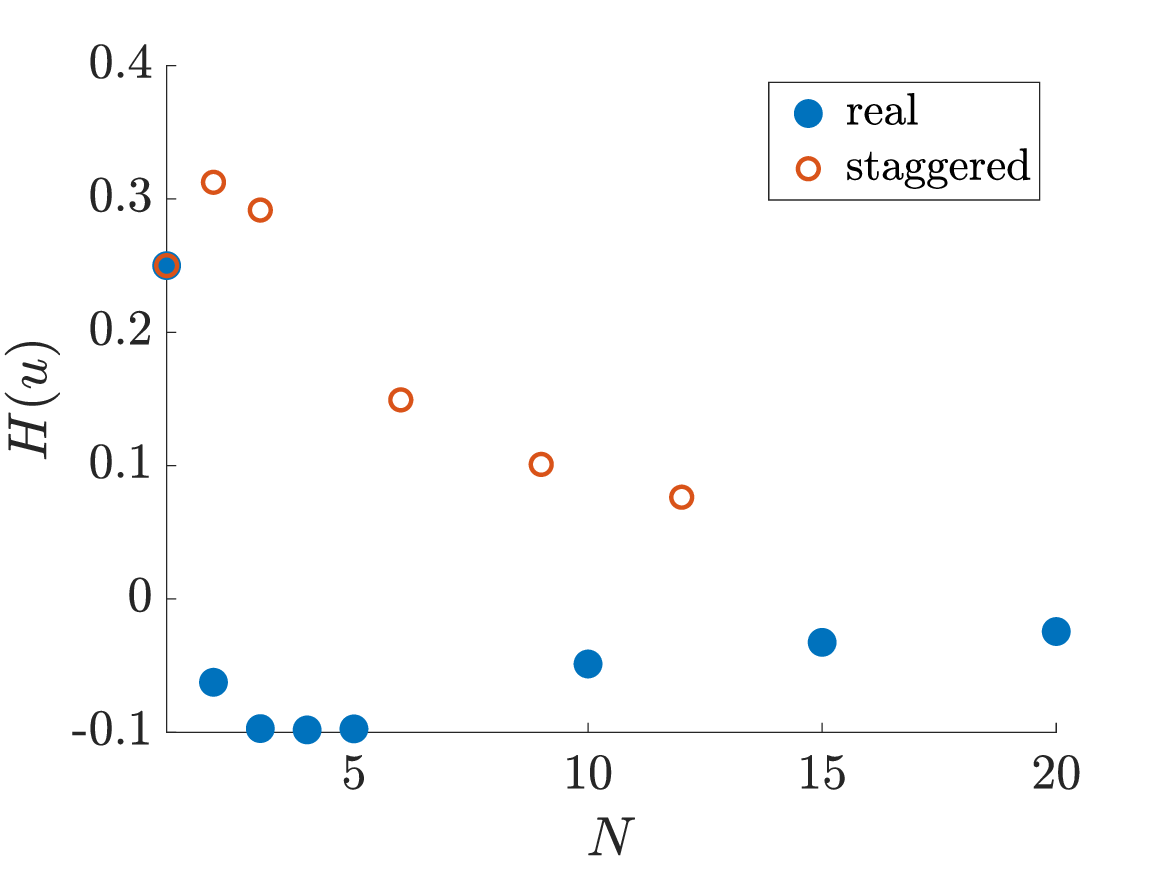}
\includegraphics[width=7cm]{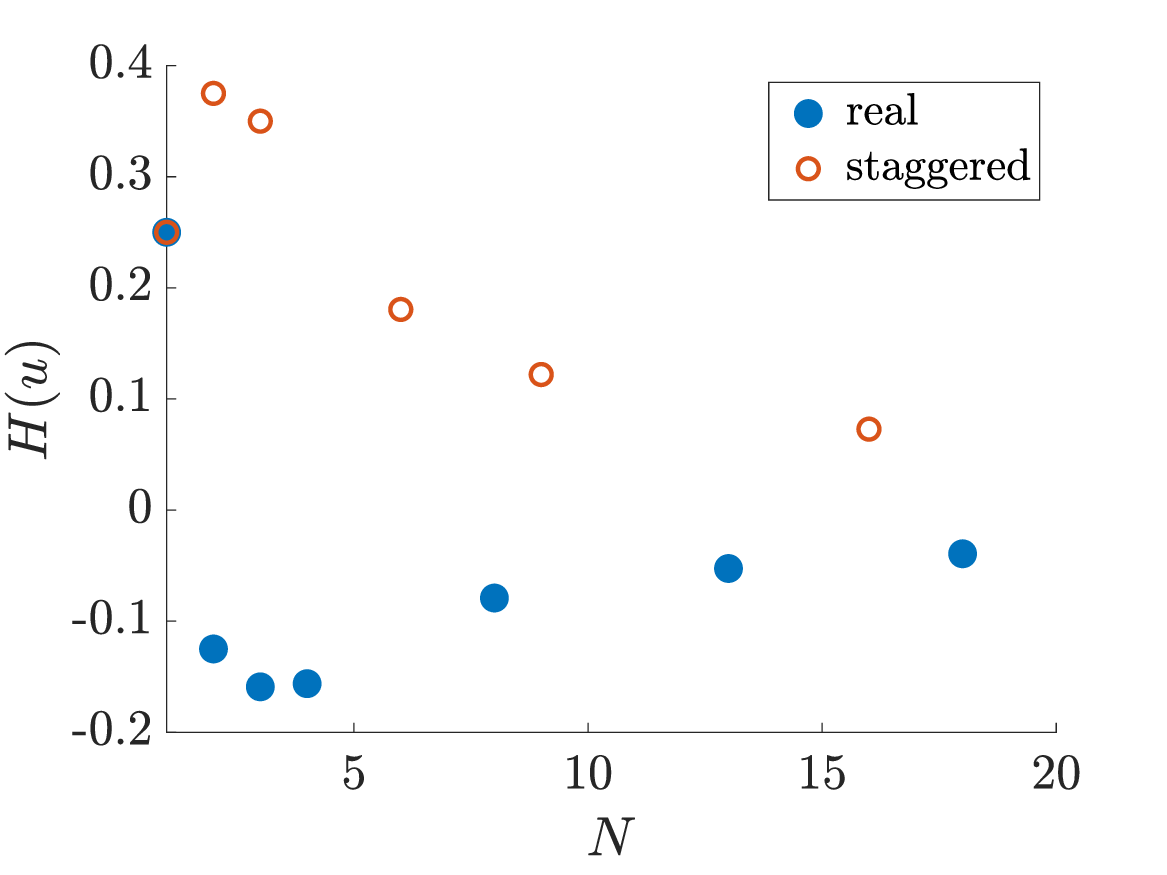}
\end{center}
\caption{Energy $H$ for real and staggered compacton solutions scaled so that the power $P=1$ for all solutions. Coupling parameter $d=0.25, 0.75, 1.5$, and $2$ (top to bottom, left to right). For $d>1/2$, solutions for a particular $N$ do not exist if a marker is not shown. Real and staggered compacton solutions obtained by solving equations \cref{eq:cjeq} and \cref{eq:cjeqstag}, respectively, together with the condition that the square amplitudes $c_j^2 > 0$ for all $j$.
}
\label{fig:compactonH}
\end{figure}

\begin{figure}
\begin{center}
\includegraphics[width=7cm]{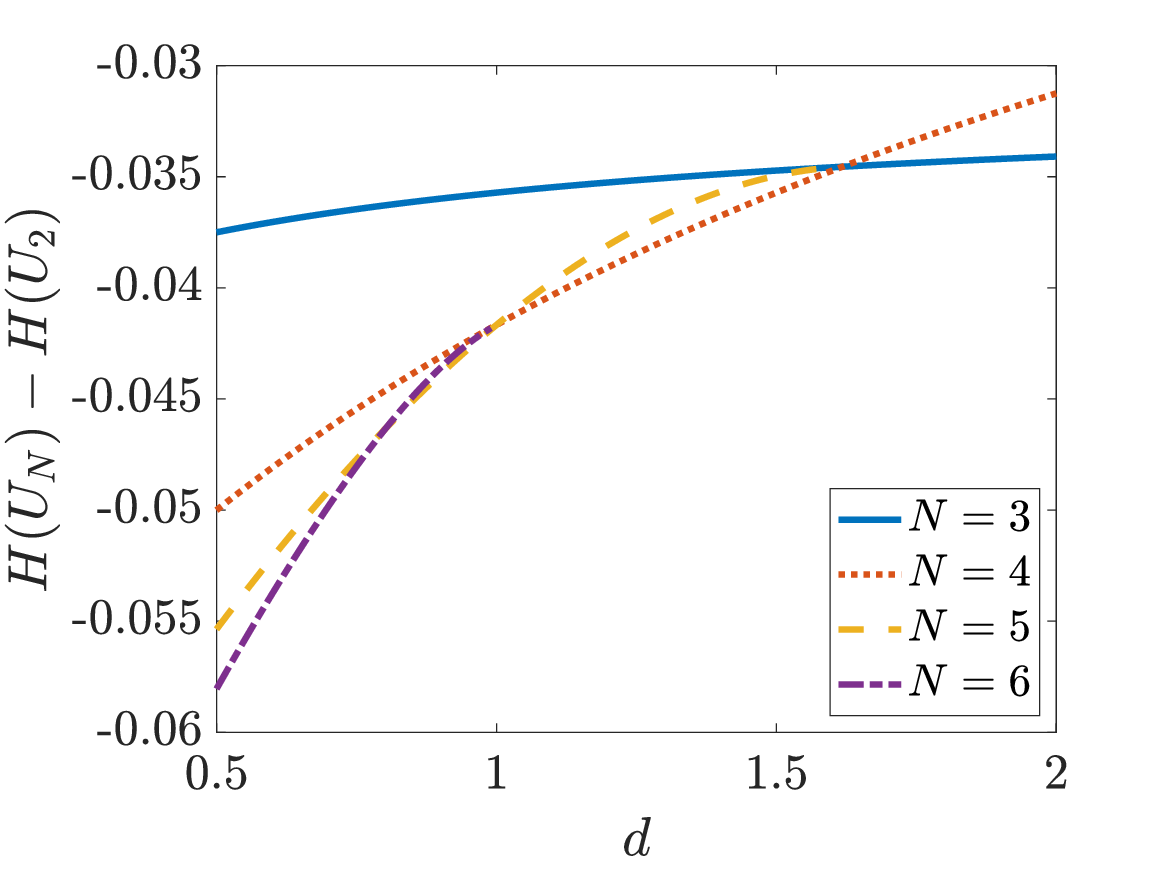}
\includegraphics[width=7cm]{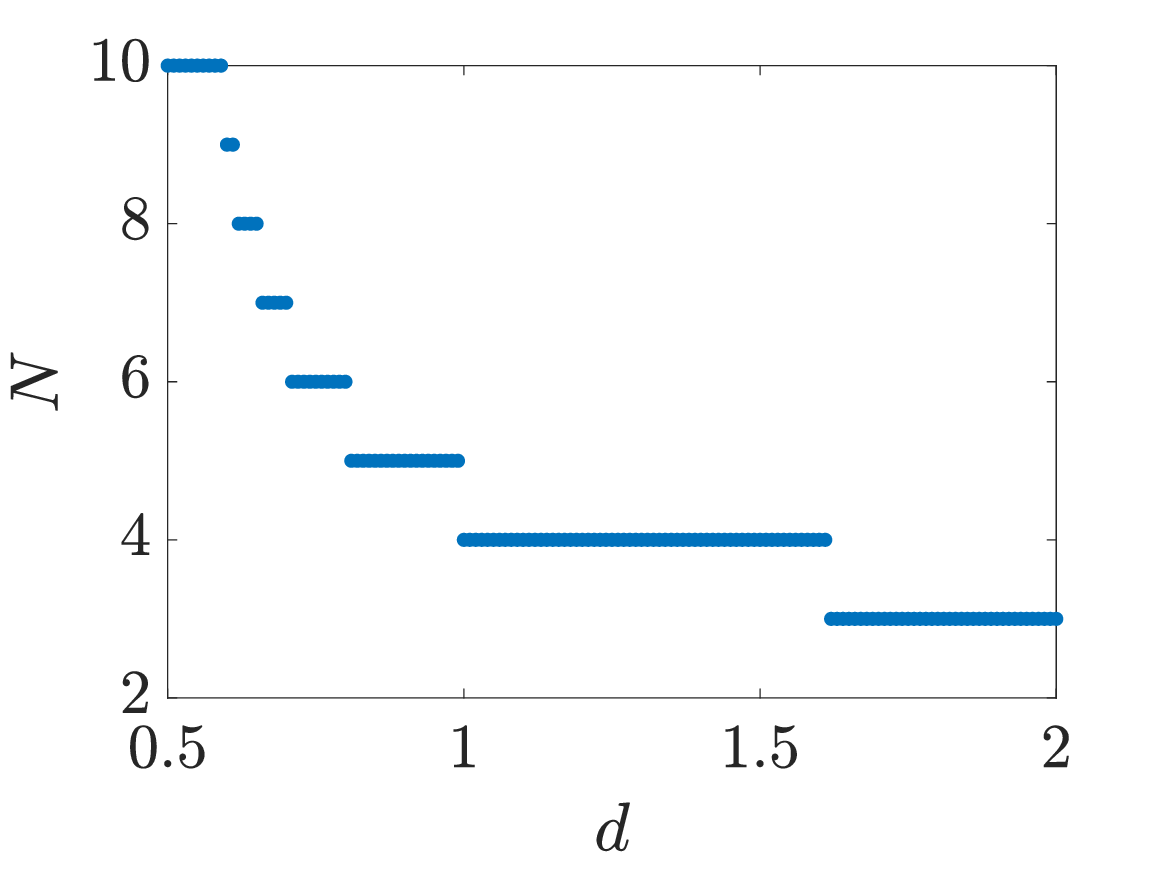}
\end{center}
\caption{Left: Energy difference between real compacton of size $N$ and real compacton of size 2 vs. $d$. Right: size of real compacton of minimum energy (selected only from among compactons of sizes from $N=1$ to $N=10$) vs. $d$. All solutions scaled so that power $P=1$. The value of $N$ selected for the given $d$ corresponds to the spatial extent of the ground
state of the system.}
\label{fig:compactonHd}
\end{figure}

The energy \cref{eq:H} of real and staggered compactons as a function of $N$ and for various $d$ are plotted in \cref{fig:compactonH} (the power of the solution is scaled to $1$ for all $N$). Formulas for the energy of small compactons are also given in \cref{table:compactonH}.
For $0 < d < 1/2$ and fixed power (exemplified by \cref{fig:compactonH}, top left), the energy decreases monotonically with increasing $N$, both for real and staggered compactons. The staggered compacton has higher energy than the real compacton, although this difference becomes smaller with increasing $N$; the latter
is natural to expect, as the profile of both compactons asymptotes to a 
constant near the center of the respective structure. For all $d$, the single-site solution (compacton with $N=1$) of power $P$ has energy $H = P^2/4$, which is independent of $d$. For $0 < d < 1$, this single-site solution is the energy maximum. The energy of the two-site staggered compacton is $(1+d)P^2/8$, which increases with increasing $d$, and surpasses the energy of the single site solution at $d=1$. For $d > 1$, the two-site staggered compacton is the energy maximum, implying that it is stable for $d >1$ (see \cref{sec:linearcompactonstag} below). Numerical computations indicate that these solutions (single site solution if $d<1$ and two-site staggered compacton if $d>1$) maximize the Hamiltonian $H$ over all vectors in $\ell^2$, under the constraint that the power $P$ is fixed. Since $H$ and $P$ are conserved quantities of the system, this implies that both are stable in the sense of Lyapunov (for the appropriate value of $d$). Note that the growth mechanism exhibited in~\cite{Colliander2010} exploits the instability of the single site solution, which follows from the value $d=2$.

For $d>1/2$, numerical computations suggest that the energy minimizer is the real compacton of a finite size (which depends on $d$). See the left panel of \cref{fig:compactonHd} for a plot of the energies of real compactons of sizes $N=3,4,5$ and $6$ vs. $d$ (for ease of visualization, the vertical axis actually plots the energy difference with the 2-site real compacton). This is further confirmed by performing a constrained minimization using Matlab's \texttt{fmincon} function, with fixed power as the sole constraint; we note that we do not restrict ourselves to standing wave solutions. For $d = 0.75$, 1.5, and 2, the energy minimizer is the real compacton comprising $N=6$, 4, and 3 sites, respectively\footnote{We learned from Jeremy Marzuola that the minimality of the 3-site compacton can be established rigorously if $d=2$. This result will be published in a forthcoming article.}. Numerical computations suggest that as $d$ is increased, the size of the compacton that minimizes the energy also decreases. Specifically, the energy minimizer becomes the $N=4$ real compacton at $d=1$ and then finally the $N=3$ real compacton at $d=(1+\sqrt{5})/2$ (see \cref{fig:compactonHd}, right panel). 
Conversely, as $d$ approaches 1/2 from above, the size of the compacton with minimum energy increases. We note that the numerical minimization is performed on a finite lattice ($N=10$ in the right panel of \cref{fig:compactonHd}). As $d$ approaches 1/2 from above, the size of the compacton with minimum energy monotonically approaches 10, which is the maximum allowable size. Repeating the experiment with a lattice size of 20, the compacton with minimum energy approaches 20 as $d$ decreases to 1/2. 
We hypothesize that if this restriction were removed, i.e., if we were considering the full integer lattice, the size of the compacton with minimum energy would approach infinity as $d$ decreases to 1/2.


\subsection{Linearization and stability}\label{sec:compactonlinear}

Linearizing about a standing wave of the form $(a_j + i b_j)e^{-i\omega t}$, where $a_j$ and $b_j$ are real, we obtain the eigenvalue problem
\begin{equation}\label{eq:stwaveEVP}
\begin{pmatrix}
\widetilde{L}^- & L^- \\ -L^+ & -\widetilde{L}^+ 
\end{pmatrix}\begin{pmatrix} v_j \\ w_j \end{pmatrix}
 = \lambda \begin{pmatrix} v_j \\ w_j \end{pmatrix},
\end{equation}
where
\begin{equation}\label{eq:stwaveEVPops}
\begin{aligned}
\widetilde{L}^- v_j &= 2[a_j b_j - d(a_{j-1}b_{j-1} + a_{j+1}b_{j+1})]v_j \\
&\qquad + 2 d [ (a_{j-1}b_j - a_j b_{j-1}) v_{j-1} + (a_{j+1}b_j - a_j b_{j+1}) v_{j+1}] \\
L^- w_j &= [ (a_j^2 + 3 b_j^2) + d( a_{j-1}^2 + a_{j+1}^2 - b_{j-1}^2 - b_{j+1}^2) - \omega] w_j \\
&\qquad - 2 d [ (a_{j-1} a_j + b_{j-1} b_j)  w_{j-1} + (a_{j+1} a_j + b_{j+1} b_j)  w_{j+1} ] \\
L^+ v_j &= [ (3 a_j^2 + b_j^2) - d( a_{j-1}^2 + a_{j+1}^2 - b_{j-1}^2 - b_{j+1}^2) - \omega] v_j \\
&\qquad - 2 d [ (a_{j-1} a_j + b_{j-1} b_j)  v_{j-1} + (a_{j+1} a_j + b_{j+1} b_j)  v_{j+1} ] \\
\widetilde{L}^+ w_j &= 2[a_j b_j - d(a_{j-1}b_{j-1} + a_{j+1}b_{j+1})]w_j \\
&\qquad - 2 d [ (a_{j-1}b_j - a_j b_{j-1}) w_{j-1} + (a_{j+1}b_j - a_j b_{j+1}) w_{j+1}].
\end{aligned}
\end{equation}
For both the real and the staggered compacton, these linear operators simplify significantly. We treat these two cases separately below.

\subsubsection{Real compactons}\label{sec:linearrealcompacton}

For a compacton solution $c_j e^{-i\omega t}$ where the amplitudes
$c_j$ are real, the linear operators $\widetilde{L}^\pm$ are 0, thus the eigenvalue problem becomes
\begin{align}\label{eq:realcompactonEVP}
&\begin{pmatrix}
0 & L^- \\ -L^+ & 0 
\end{pmatrix}\begin{pmatrix} v_j \\ w_j \end{pmatrix}
 = \lambda \begin{pmatrix} v_j \\ w_j \end{pmatrix},
\end{align}
where
\begin{align*}
L^- w_j &= (c_j^2 + d( c_{j+1}^2 + c_{j-1}^2) - \omega) w_j - 2 d c_j (c_{j+1} w_{j+1} + c_{j-1} w_{j-1}) \\
L^+ v_j &= ( 3 c_j^2 - d( c_{j+1}^2 + c_{j-1}^2) - \omega) v_j - 2 d c_j (c_{j+1} v_{j+1} + c_{j-1} v_{j-1}).
\end{align*}
Since $d(c_{j-1}^2 + c_{j+1}^2) = c_j^2 - \omega$ from \cref{eq:cjeq}, we can rewrite $L^-$ and $L^+$ as
\begin{align*}
&L^- w_j = 2 (c_j^2  - \omega) w_j - 2 d c_j (c_{j+1} w_{j+1} + c_{j-1} w_{j-1}) \\
&L^+ v_j = 2 c_j^2 v_j - 2 d c_j (c_{j+1} v_{j+1} + c_{j-1} v_{j-1}),
\end{align*}
from which it follows that $L^- = L^+ - 2 \omega I$. Furthermore, if we let $M$ be the matrix in \cref{eq:cjeq}, $L^+ = 2\,\text{diag}(c)\,M\, \text{diag}(c)$, where $\text{diag}(c)$ is the diagonal matrix with the amplitudes $c_j$ on the diagonal.
The eigenvalues $\lambda$ do not depend on whether we take the positive or negative root for $c_j$. To see this, if $L^+$ is the matrix associated with a compacton with all positive amplitudes, and $L_j^+$ is the matrix associated with the same compacton, except the amplitude $c_j$ of site $j$ is negative, then $L_j^+ = A L^+ A$, where $A$ is the self-invertible matrix formed by changing the $j$th diagonal element of the identity matrix to $-1$. 

Since the eigenvalue problem \cref{eq:realcompactonEVP} can be written as $L^- L^+ v = -\lambda^2 v$, and the matrix $L^- L^+ = (L^+ - 2 \omega I)L^+$ is symmetric, the eigenvalues of $L^- L^+$ are real, which implies that the eigenvalues $\lambda$ come in pairs which are either real or purely imaginary. For all $N$, there is an eigenvalue of algebraic multiplicity 2 and geometric multiplicity 1 at the origin due to the gauge symmetry of the system. 
For $N=1$, this double eigenvalue at $0$ is the only eigenvalue, thus the single-site compacton solution is spectrally neutrally stable. For $N=2$, there is an additional pair of eigenvalues on the imaginary axis at
\[
\lambda = \pm 2 \omega \frac{\sqrt{2d(1+d)}}{1-d} i.
\]
Since these are imaginary for both $0<d<1$ and $d>1$ (and for all $\omega$), the 2-site real compacton is spectrally stable. Perturbations of the 2-site real compacton yield oscillatory states which remain close to the unperturbed compacton (see \cref{sec:dimerdyn}, in particular \cref{fig:dimerdynamic}). 

Exact formulas for eigenvalues are less straightforward to obtain (and present) for $N \geq 3$. 
For $0 < d < 1/2$, numerical computations strongly suggest that for a compacton of size $N$, all of the nonzero eigenvalues are purely imaginary. This implies that real compactons of all sizes are spectrally stable in the parameter range $0 < d < 1/2$. For values of $d$ outside that range, we provide 
further details in the case of small compactons in what follows.
Numerical computations suggest that the 3-site real compacton is spectrally stable for all $d>0$ and all $\omega$ for which it exists (see \cref{table:compactons}). Similarly, computations suggest that the 4-site real compacton is spectrally stable for $d \in \left( 0, \frac{\sqrt{5}+1}{2}\right)$, and the 5-site real compacton is spectrally stable for $d \in (0,1)$. Stability is lost at the right endpoints of these intervals as a pair of imaginary eigenvalues collides at the origin and becomes real. We note that these endpoints coincide precisely with the additional values of $d$ at which the matrix in \cref{eq:cjeq} is singular (see \cref{sec:realcompactons} above). 

\subsubsection{Staggered compactons}\label{sec:linearcompactonstag}

The staggered compacton alternates between sites which are real and sites which are purely imaginary, thus all terms of the form $a_j b_j$, $a_j a_{j-1}$, $a_j a_{j+1}$, $b_j b_{j-1}$, and $b_j b_{j+1}$ are 0, which reduces the four linear operators in \cref{eq:stwaveEVPops} to
\begin{equation}
\begin{aligned}
\widetilde{L}^- v_j &= 2 d [ (a_{j-1}b_j - a_j b_{j-1}) v_{j-1} + (a_{j+1}b_j - a_j b_{j+1}) v_{j+1}] \\
L^- w_j &= [ (a_j^2 + 3 b_j^2) + d( a_{j-1}^2 + a_{j+1}^2 - b_{j-1}^2 - b_{j+1}^2) - \omega] w_j  \\
L^+ v_j &= [ (3 a_j^2 + b_j^2) - d( a_{j-1}^2 + a_{j+1}^2 - b_{j-1}^2 - b_{j+1}^2) - \omega] v_j  \\
\widetilde{L}^+ w_j &= - 2 d [ (a_{j-1}b_j - a_j b_{j-1}) w_{j-1} + (a_{j+1}b_j - a_j b_{j+1}) w_{j+1}].
\end{aligned}
\end{equation}
We note that for the staggered compacton, $\widetilde{L}^+ = -\widetilde{L}^-$, and that both $L^+$ and $L^-$ are diagonal. As with the real compactons, there is an eigenvalue of algebraic multiplicity 2 and geometric multiplicity 1 at the origin due to the gauge symmetry of the system. For $N=2$, there is an additional pair of eigenvalues at
\begin{equation}
\lambda = \pm 2 \omega \frac{\sqrt{2d(1-d)}}{1+d},
\end{equation}
which are real when $0 < d < 1$, implying instability, and purely imaginary when $d > 1$, implying spectral stability. 
A bifurcation occurs at $d=1$, when the pair of real eigenvalues collides at the origin and moves onto the imaginary axis.
The nature of this bifuration and the behavior of perturbations to the 2-site staggered compacton can be fully understood using the phase plane analysis in \cref{sec:dimerdyn}, 
noting that the bifurcation there occurs for $d=1/2$, rather than for $d=1$, due to different boundary conditions.
The staggered compacton corresponds to the fixed point at $(p,\phi) = (0,\pi/2)$ in \cref{fig:dimerdynamic}, where $p = |u_2|^2 - |u_1|^2$ and $\phi$ is the phase difference between $u_2$ and $u_1$ (see \cref{sec:dimer} below for details).

\begin{figure}
\begin{center}
\includegraphics[width=7cm]{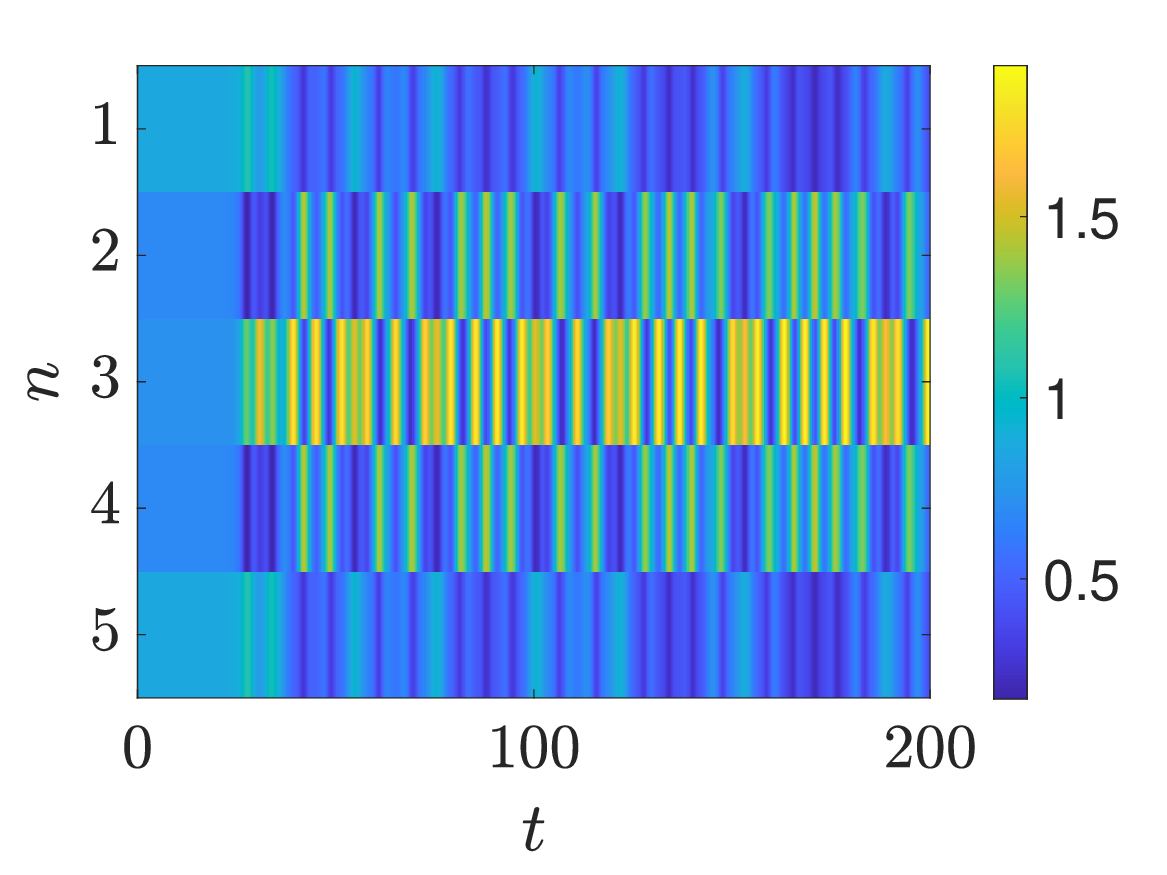}
\includegraphics[width=7cm]{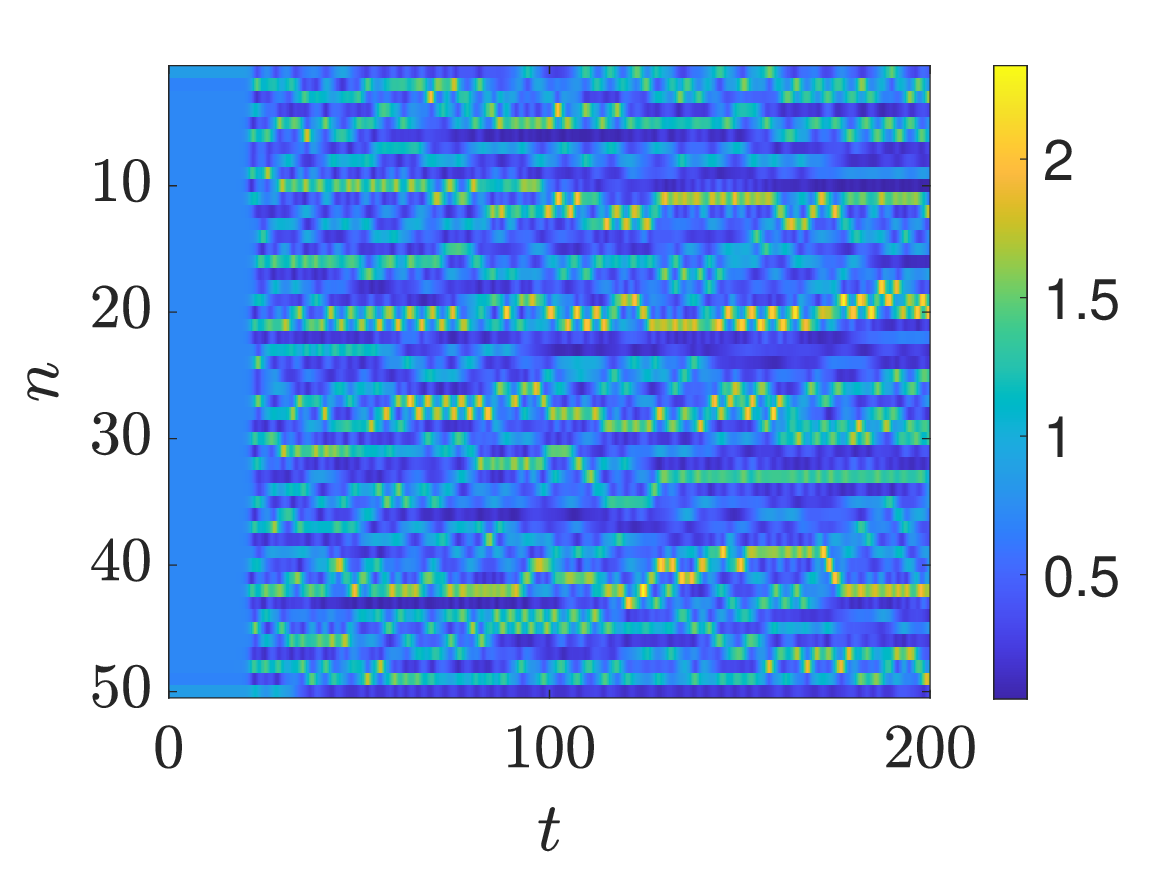}
\end{center}
\caption{Colormap of square intensity $|u_n|^2$ of unperturbed staggered compactons with $N=5$ (left) and $N=50$ (right). $\omega = 1$, $d = 0.2$. Vertical axis is lattice site. The time evolution is performed using the Dormand-Prince 
integrator, implemented in Matlab by means of the \texttt{ode45} function.}
\label{fig:stagpert}
\end{figure}

\begin{figure}
\begin{center}
\includegraphics[width=7cm]{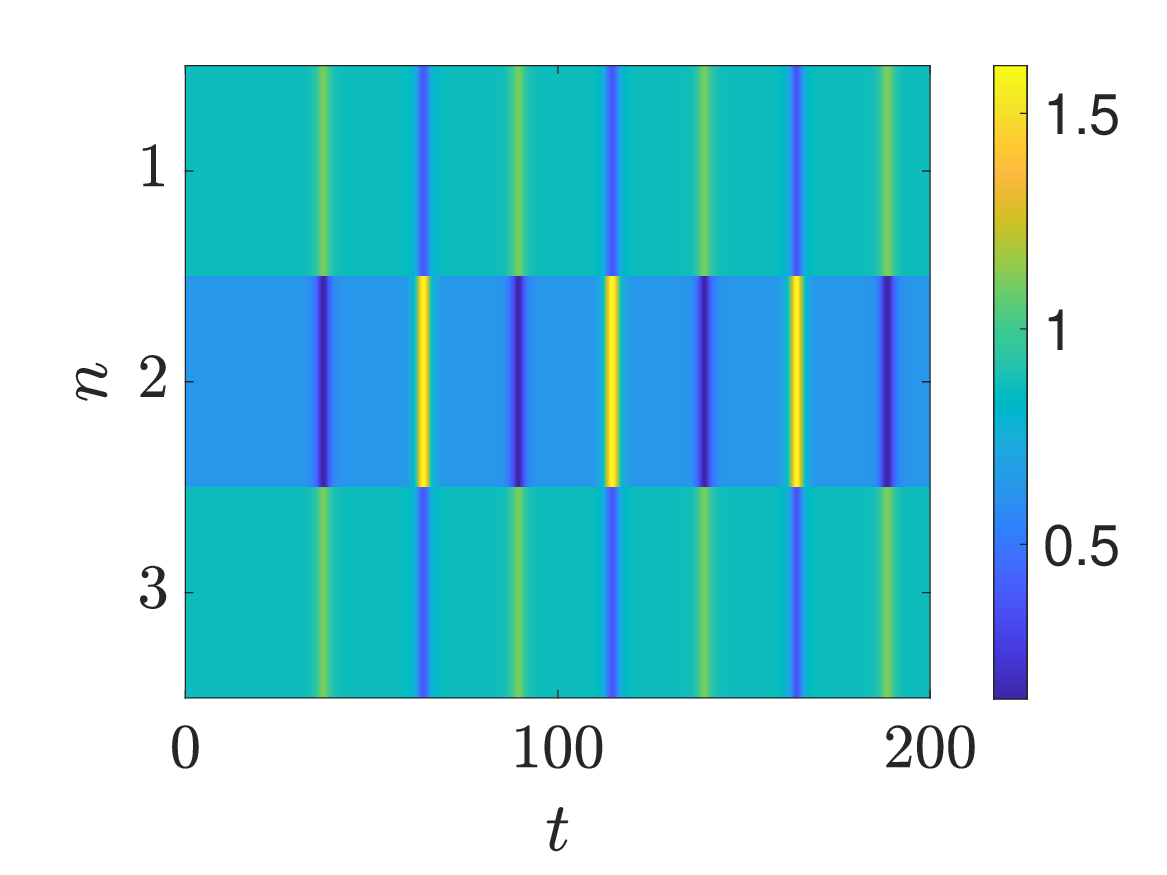}
\includegraphics[width=7cm]{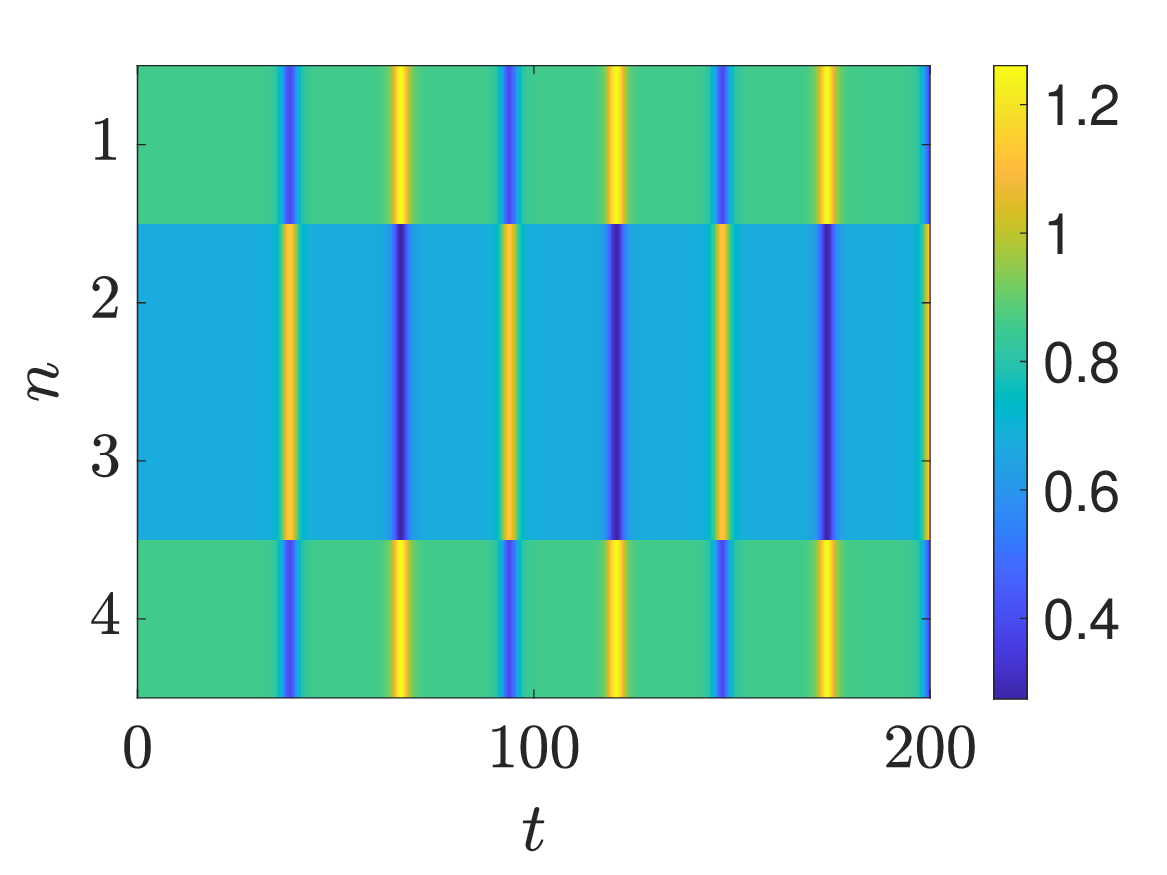}
\includegraphics[width=7cm]{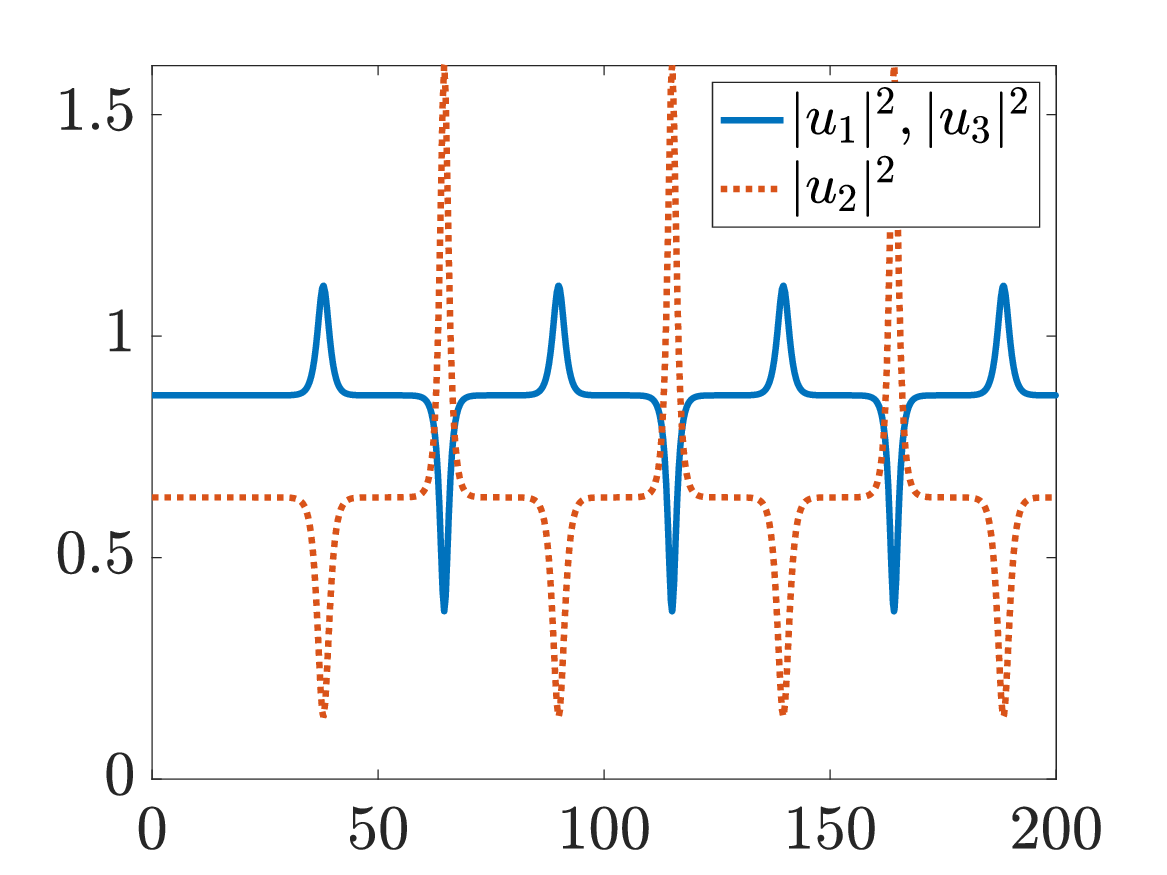}
\includegraphics[width=7cm]{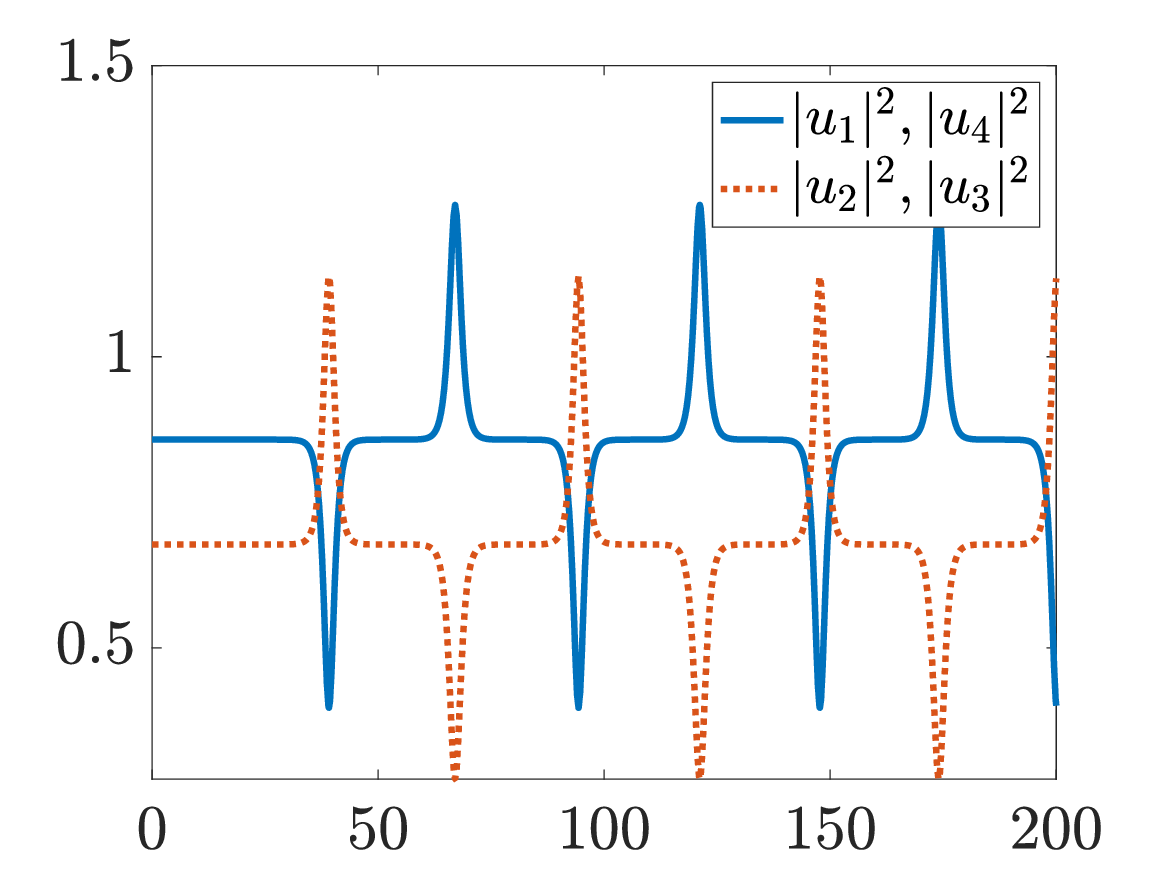}
\includegraphics[width=7cm]{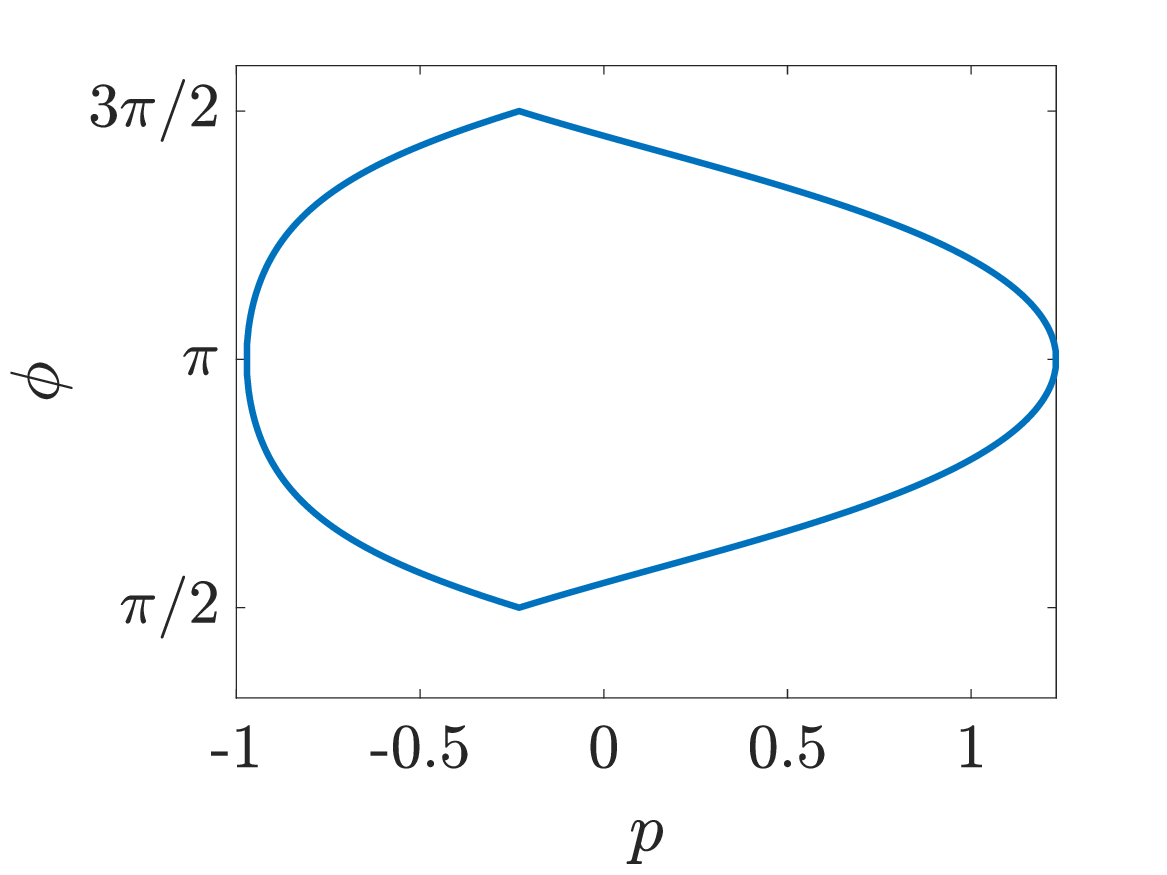}
\includegraphics[width=7cm]{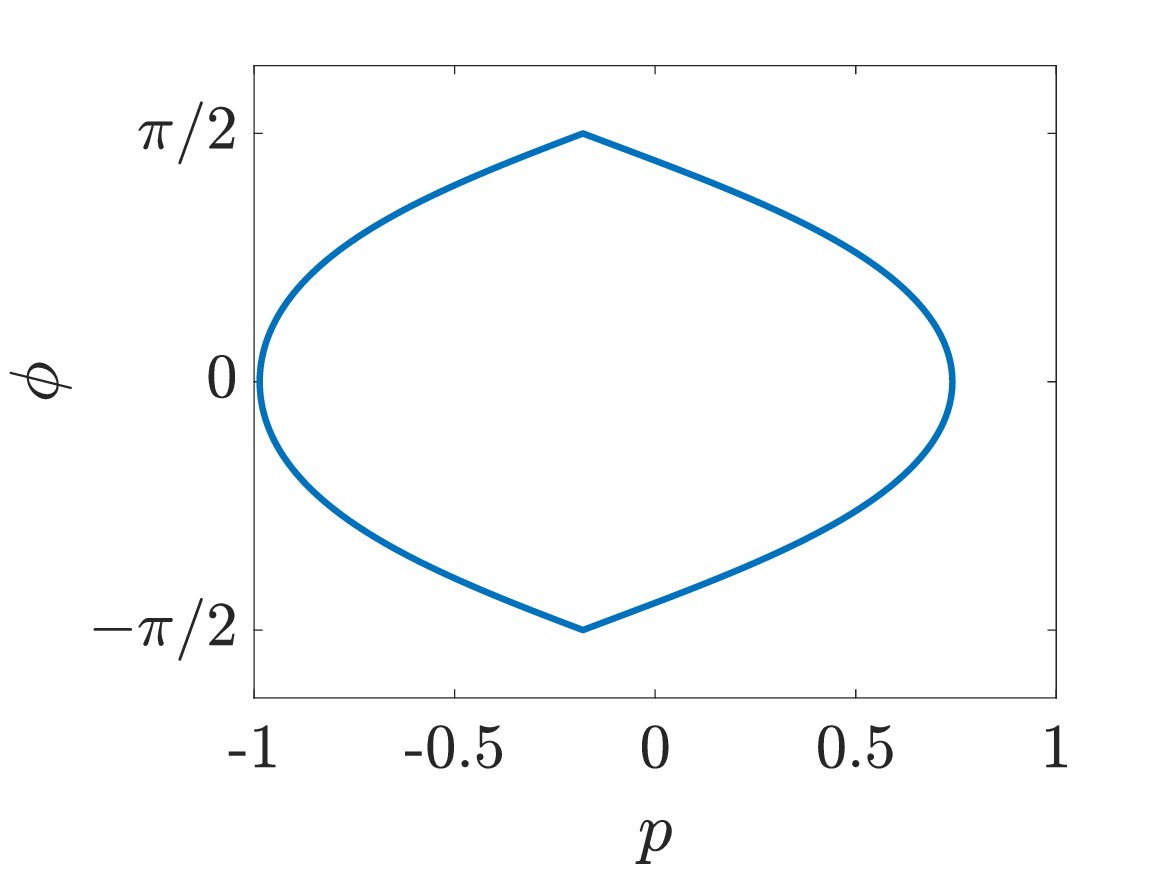}
\end{center}
\caption{Top: colormap of square intensity $|u_n|^2$ of unperturbed staggered compactons. Vertical axis is lattice site. Middle: square intensity vs. $t$ of these solutions. Bottom: phase portraits corresponding to periodic orbits from the middle plots; $p = |u_2|^2 - |u_1|^2$, and $\phi$ is the phase difference between $u_2$ and $u_1$. A ``near-corner'' occurs at the top and bottom of these periodic orbits as they pass very close to a saddle point equilibrium (compare to left panel of \cref{fig:dimerdynamic}).
Left: $N=3$, $u_3 = -u_1$ in periodic orbit. Right: $N=4$, $u_3 = i u_2$ and $u_4 = -i u_1$ in periodic orbit. $d=0.21$, $\omega = 1$. The time evolution is performed using the Dormand-Prince integrator, implemented in Matlab by means of the \texttt{ode45} function.}
\label{fig:stagpert2}
\end{figure}

For $N\geq 3$, numerical computations strongly suggest that for staggered compactons with coupling parameter $0 < d < 1/2$, all of the nonzero eigenvalues are real, thus all staggered compactons are all unstable in that parameter regime. This instability is sufficiently strong that it can be demonstrated using numerical evolution experiments from unperturbed initial conditions (\cref{fig:stagpert}). In general, these structures break down and do not tend towards or oscillate about any stable coherent structure. For example, for large $N$ (right panel of \cref{fig:stagpert}), the staggered compacton solution breaks down into smaller structures similar to those seen in \cref{fig:evol1}. For small staggered compactons ($N=3$ and $N=4$) and particular values of $d$, however, the staggered compacton appears to decay into a coherent periodic orbit (\cref{fig:stagpert2}), but this behavior appears to be uncommon. For $N=3$, the periodic orbit in the left of \cref{fig:stagpert2} has the symmetry $u_3 = -u_1$, and the system \cref{eq:model} reduces to the pair of equations with asymmetric coupling terms
\begin{equation}
\begin{aligned}
&i \dot{u}_1 + d u_2^2 \overline{u_1} - |u_1|^2 u_1 = 0 \\
&i \dot{u}_2 + 2 d u_1^2 \overline{u_2} - |u_2|^2 u_2 = 0.
\end{aligned}
\end{equation}
For $N=4$, the periodic orbit in the right of \cref{fig:stagpert2} has the symmetry $u_3 = i u_2$ and $u_4 = -i u_1$, and the system reduces to the pair of equations with asymmetric nonlinear terms
\begin{equation}
\begin{aligned}
&i \dot{u}_1 + d u_2^2 \overline{u_1} - |u_1|^2 u_1 = 0 \\
&i \dot{u}_2 + d u_1^2 \overline{u_2} - (1+d)|u_2|^2 u_2 = 0.
\end{aligned}
\end{equation}
Compare both of these cases to equation \cref{eq:dimer} below for the symmetric dimer. Phase portraits of these periodic orbits are shown in the bottom panels of \cref{fig:stagpert2}, where $p = |u_2|^2 - |u_1|^2$ and $\phi$ is the phase difference between $u_2$ and $u_1$ (see also \cref{sec:dimerdyn} below). 

\section{Dynamical considerations: the dimer case}\label{sec:dimer}

\begin{figure}
\begin{center}
\includegraphics[width=12cm]{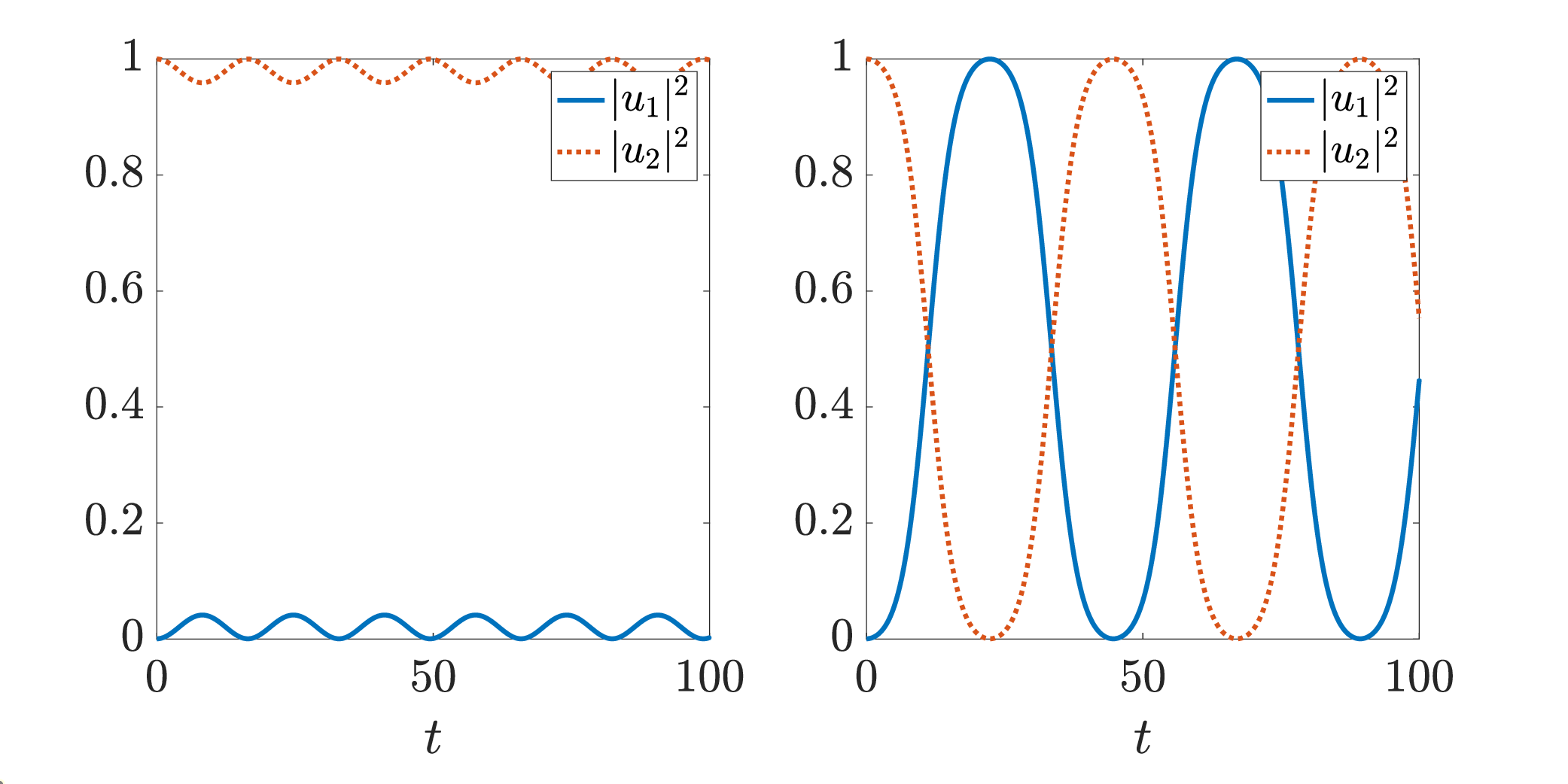}
\end{center}
\caption{Evolution of dimer equation \cref{eq:dimer} on periodic lattice with initial conditions $u_1(0) = 0.02$, $u_2(0) = 1$. Coupling parameter $d = 0.49$ (left), $d = 0.51$ (right). One can clearly discern the self-trapping transition
occurring at $d=1/2$. The time evolution is performed using the Dormand-Prince integrator, implemented in Matlab by means of the \texttt{ode45} function.}
\label{fig:dimer1}
\end{figure}

Next, we look at solutions in which the intensity moves along the lattice. It turns out that a useful starting
point is the dimer (two-site solution) on a periodic lattice:
\begin{equation}\label{eq:dimer}
\begin{aligned}
&i \dot{u}_1 + 2 d u_2^2 \overline{u_1} - |u_1|^2 u_1 = 0 \\
&i \dot{u}_2 + 2 d u_1^2 \overline{u_2} - |u_2|^2 u_2 = 0.
\end{aligned}
\end{equation}
Note that if we take Dirichlet instead of periodic boundary conditions on the lattice, we replace $d$ with $d/2$ in \cref{eq:dimer}. Numerical evolution experiments show that if one site is initialized to high intensity and the other to low intensity, optical intensity moves periodically between the two sites if $d>1/2$, but remains confined to the initial sites if $d<1/2$ (see \cref{fig:dimer1}). This suggests that a bifurcation occurs at $d=1/2$, which we will explore in detail in the following subsections.

\subsection{Phase plane analysis}\label{sec:dimerdyn}

We start by constructing a phase portrait for the dimer system.
Although the evolution of \cref{eq:dimer} occurs in a four-dimensional phase space comprising the real and imaginary parts of $u_1$ and $u_2$ (or, equivalently, the amplitude and phase of $u_1$ and $u_2$), we can reduce it to a two-dimensional dynamical system using the conservation of power and the gauge invariance of the system. To do this, we fix a power $P$, which will remain invariant as $t$ evolves. Writing $u_1 = r_1 e^{i\theta_1}$ and $u_2 = r_2 e^{i\theta_2}$, we recast the system in the two dynamical variables
\begin{equation}\label{eq:dimervar}
p = r_2^2 - r_1^2, \qquad 
\phi = \theta_2 - \theta_1,
\end{equation}
where $p$ and $\phi$ are the intensity difference and phase difference, respectively, between $u_1$ and $u_2$. 
Substituting the expressions for $u_1$ and $u_2$ into \cref{eq:dimer} and simplifying, we derive the following dynamical system for $p$ and $\phi$:
\begin{equation}\label{eq:dimerdyn}
\begin{aligned}
\dot{p} &= 2d(P^2 - p^2)\sin 2\phi \\
\dot{\phi} &= -p(1 + 2d \cos 2\phi),
\end{aligned}
\end{equation}
where $p \in [-P,P]$ and $\phi \in (-\pi, \pi)$. The system is Hamiltonian, with conserved quantity $H$ given by
\begin{equation}\label{eq:dimerH}
H(p, \phi) = \frac{1}{2}p^2 - d(P^2 - p^2) \cos 2\phi,
\end{equation}
and it can be written in standard Hamiltonian form as
\[
\frac{dp}{dt} = \frac{\partial H}{\partial \phi}, \qquad 
\frac{d\phi}{dt} = -\frac{\partial H}{\partial p}.
\]
The system \cref{eq:dimerdyn} has $\mathbb{Z}_2$ symmetry, i.e., is invariant under the transformation $(p, \phi) \mapsto (-p, -\phi)$. This can also be seen from the Hamiltonian \cref{eq:dimerH} by noting that $H(-p, -\phi) = H(p, \phi)$. Equation \cref{eq:dimerdyn} is also reversible, i.e., is infvariant under the transformation $t \mapsto -t$, $\phi \mapsto -\phi$. The sets $P^\pm = \{(p, \phi) : p = \pm P\}$ are invariant sets, since $\dot{\phi}=0$ on $P^{\pm}$. In terms of the original system \cref{eq:dimer}, the sets $P^\pm$ represent the case where the intensity is completely confined to one site and thus is zero at the other site; the phase difference $\phi$ in this situation is not physically meaningful.

\begin{figure}
\begin{center}
\includegraphics[width=5.25cm]{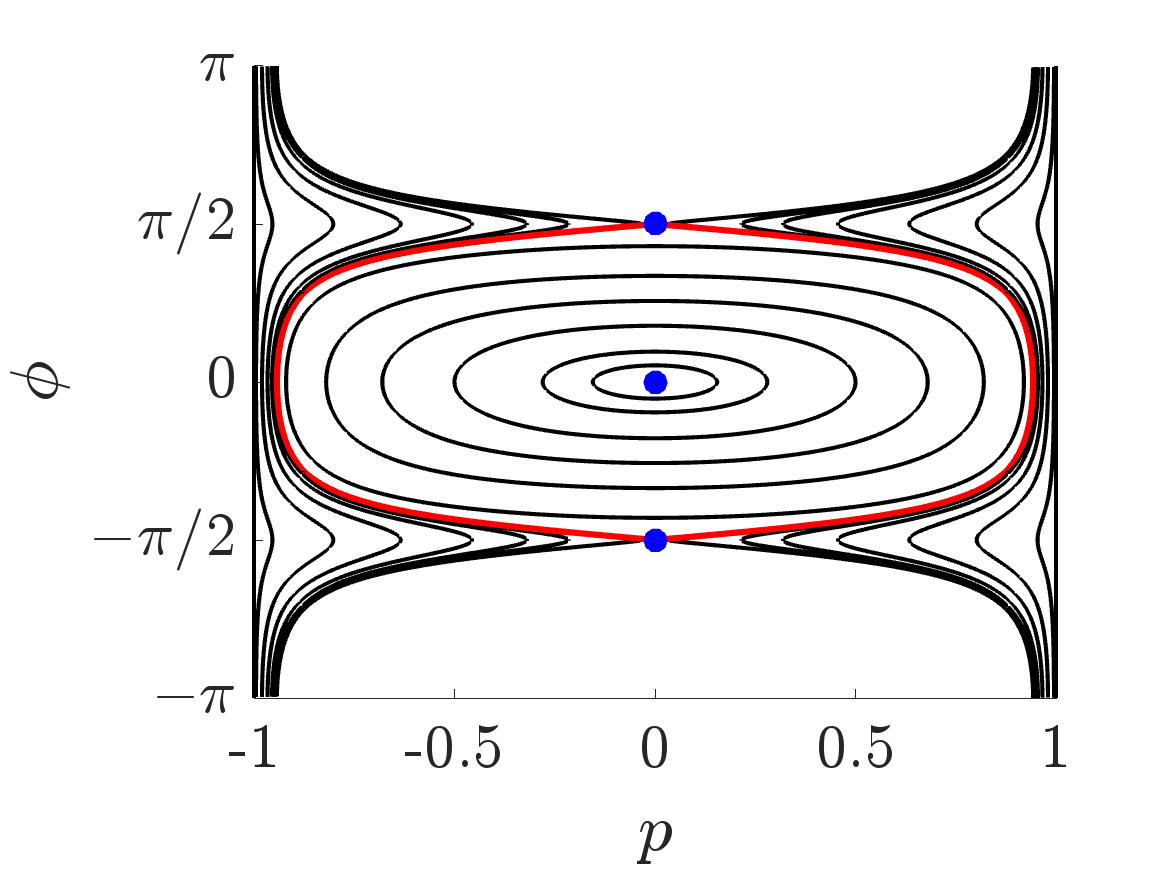}
\includegraphics[width=5.25cm]{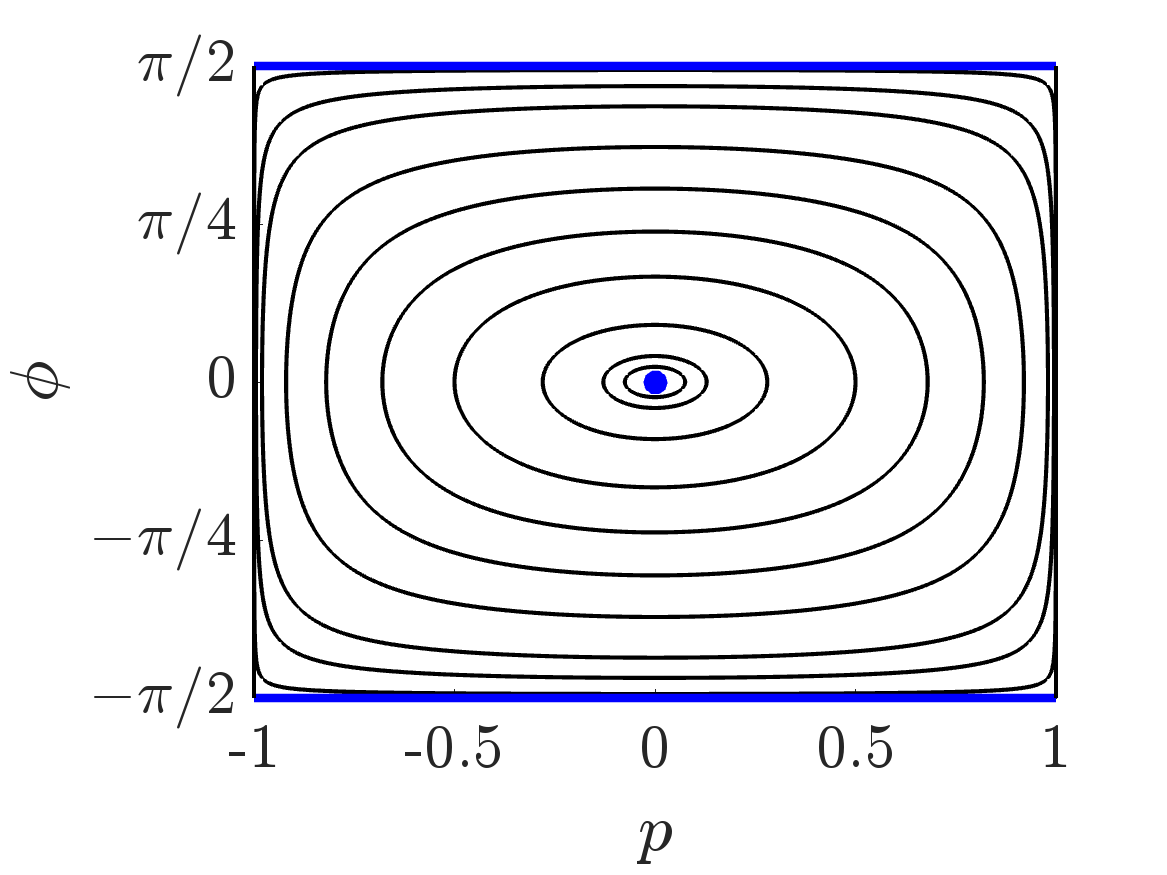}
\includegraphics[width=5.25cm]{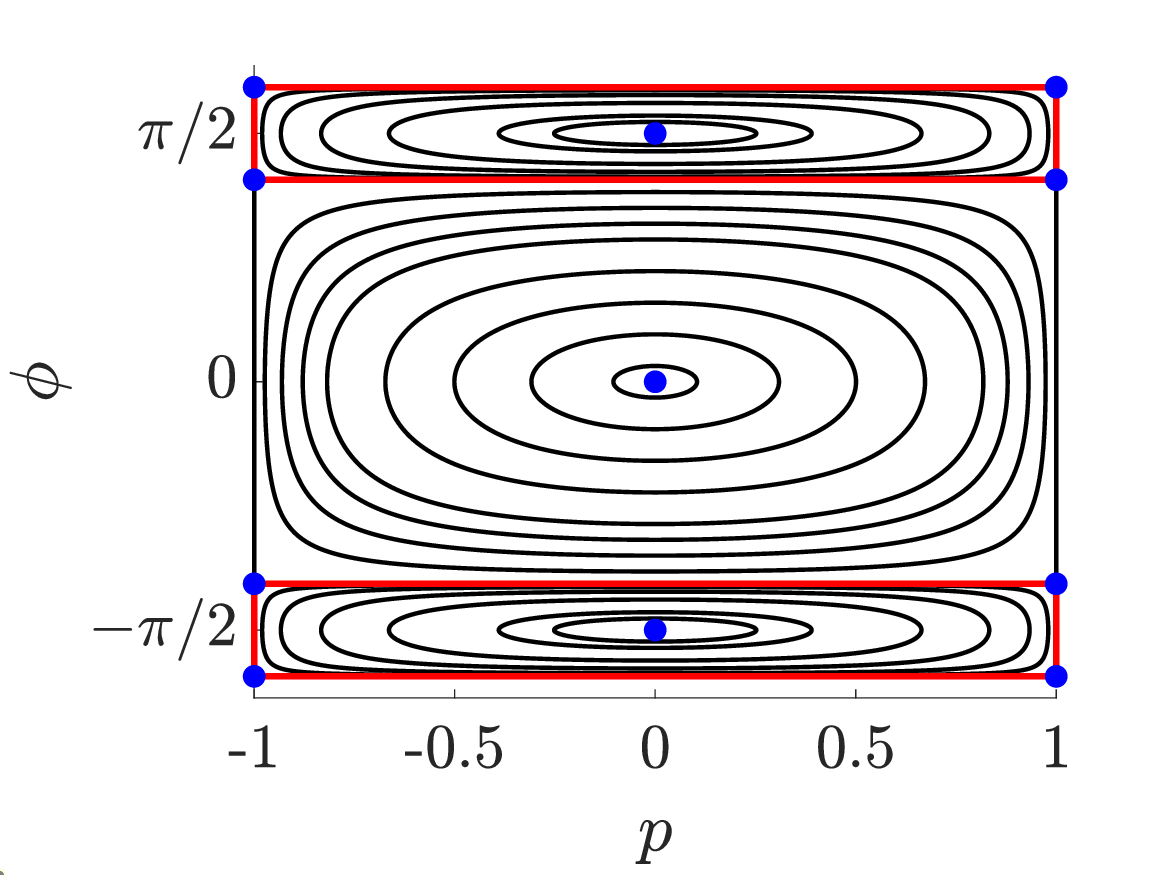}
\end{center}
\caption{Phase portrait of the dynamical system \cref{eq:dimerdyn} 
for $d < 1/2$ (left), $d=1/2$ (center), and $d > 1/2$ (right).
The red solid lines in the left panel are the limiting solutions \cref{eq:sechlimit}. The blue solid lines in the center panel are lines of equilibria which appear at $d=1/2$.
The red box in the right panel is the heteroclinic cycle produced by the degenerate Hamiltonian pitchfork bifurcation at $d=1/2$. The top and bottom of the red box are the limiting solutions \cref{eq:tanhlimit}. Equilibrium points in all panels are shown with blue dots. Power $P=1$, $d=0.4$, 0.5, and 0.6 (left to right). Trajectories computed using exact formulas derived below.}
\label{fig:dimerdynamic}
\end{figure}

\begin{figure}
\begin{center}
\includegraphics[width=15cm]{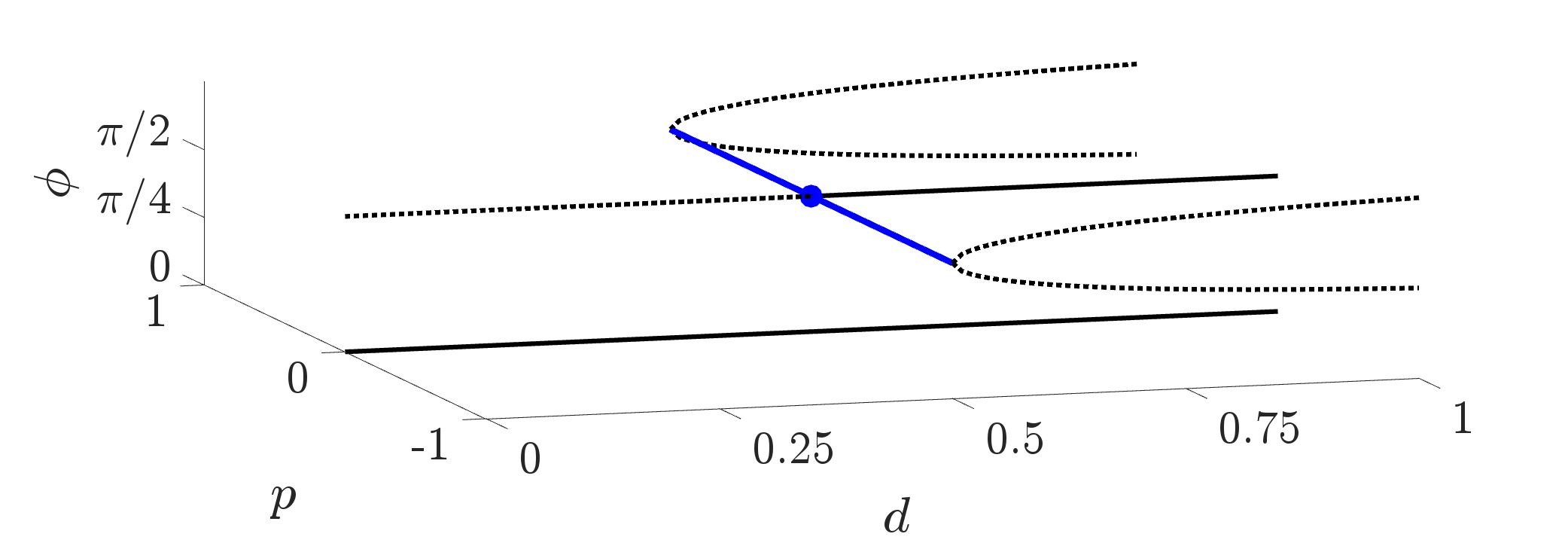}
\end{center}
\caption{Bifurcation diagram in $(d, p, \phi)$ space for dimer system, plotting the location of equilibrium points in $(p, \phi)$ plane as a function of coupling parameter $d$. Due to symmetries of system, diagram is only shown for $\phi \geq 0$.
Saddle points (unstable) indicated with dotted lines, centers (stable) indicated with solid lines. Blue solid line is line of equilibria which occurs at $d=1/2$. Blue dot at $(1/2, 0, \pi/2)$ is degenerate Hamiltonian pitchfork bifurcation.}
\label{fig:dimerBD}
\end{figure}

We will first describe the equilibria of the system, where the angular variable is considered modulo $\pi$. For stability analysis, linearization about an equilibrium point $(p,\phi)$ yields the $2\times 2$ Jacobian matrix
\[
\begin{pmatrix}
-4dp \sin 2 \phi & 4d(P^2 - p^2) \\
-1 - 2d \cos 2 \phi & 4dp \sin 2 \phi
\end{pmatrix}.
\]
Equation \cref{eq:dimerdyn} always has equilibria at $(0, 0)$ and $(0, \pm \pi/2)$ (blue dots in the left and right panels of \cref{fig:dimerdynamic}), which correspond to the real compacton and the staggered compacton, respectively. These are the only equilibria for $0 < d < 1/2$. 
The equilibrium at $(0,0)$ has a pair of eigenvalues $\lambda = \pm 2 i P \sqrt{d(1+2d)}$. Since these are always imaginary, this equilibrium is a linear center (and, in fact, is a nonlinear center, since the system is Hamiltonian). 
The equilibria at $(0, \pm \pi/2)$  have a pair of eigenvalues $\lambda = \pm 2 P \sqrt{d(1-2d)}$, which are real for $0 < d < 1/2$ and imaginary for $d>1/2$. These equilibria are saddle points for $0 < d < 1/2$ and linear centers for $d > 1/2$. 
Stability of these equilibria changes at $d=1/2$, when the pair of real eigenvalues collides at the origin and moves onto the imaginary axis. 
(We note that if we take Dirichlet boundary conditions, where $d$ is replaced with $d/2$, this bifurcation takes place at $d=1$, which is consistent with \cref{sec:compactonlinear}).
The bifurcation at $d=1/2$ is a degenerate Hamiltonian pitchfork bifurcation. At $d=1/2$, there are two continuous lines of equilibria (blue lines in \cref{fig:dimerdynamic}, center) which are given by $(p,\pm \pi/2)$ for $p \in [-P,P]$. The center points of these lines are the equilibria at $(0, \pm \pi/2)$. 
For $d<1/2$, the line of equilibria opens up into heteroclinic orbits (\cref{fig:dimerdynamic}, left)
For $d>1/2$, the line of equilibria opens up into a heteroclinic cycle in the shape of a long, thin box (red box in \cref{fig:dimerdynamic}, right). The left and right sides of the box arise from the invariant sets $P^\pm$, and the top and bottom arise from the line of equilibria. The corners of the box are saddle point equilibria located at $(\pm P, \phi_*)$, where
\begin{equation}\label{eq:phistar}
\cos 2\phi_* = -\frac{1}{2d}.
\end{equation}
A bifurcation diagram indicating the location of the equilibria in the $(p, \phi)$ plane as a function of $d$ is shown in \cref{fig:dimerBD}.

\begin{figure}
\begin{center}
\includegraphics[width=5.25cm]{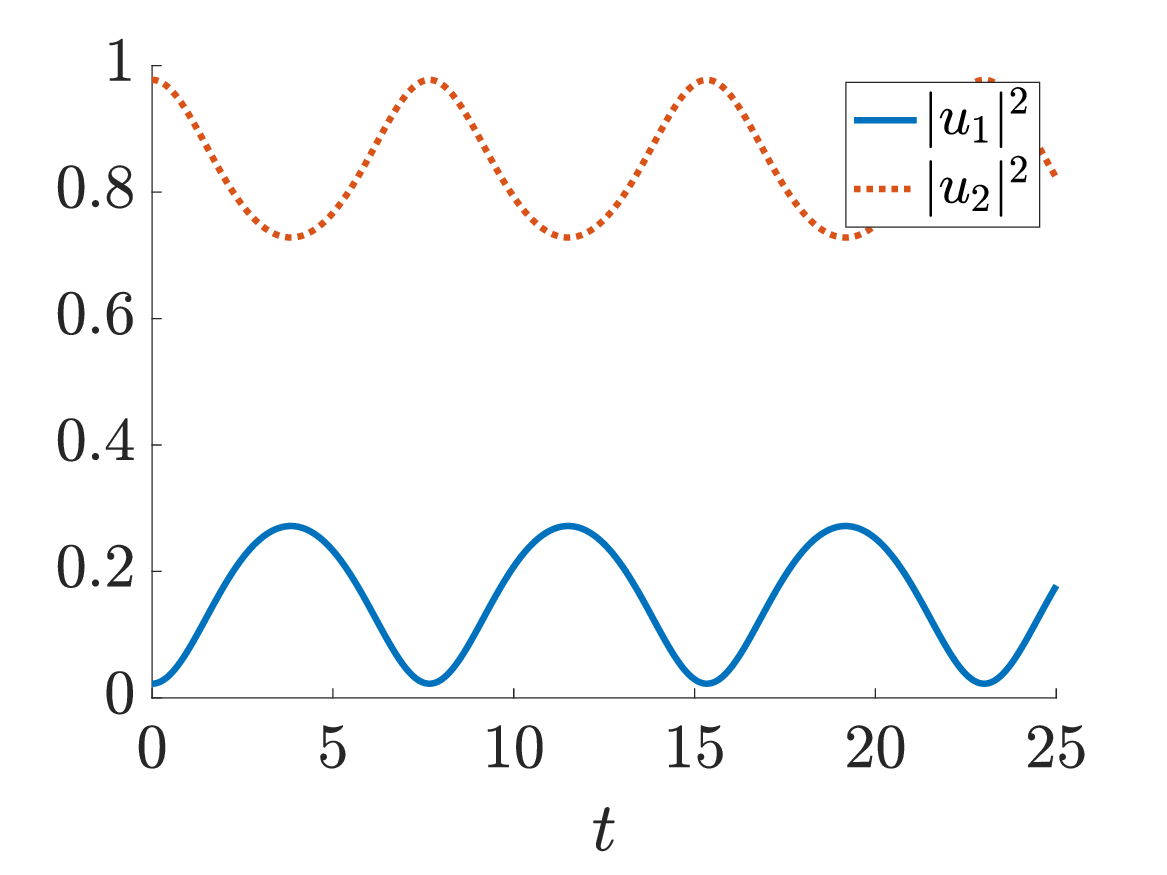}
\includegraphics[width=5.25cm]{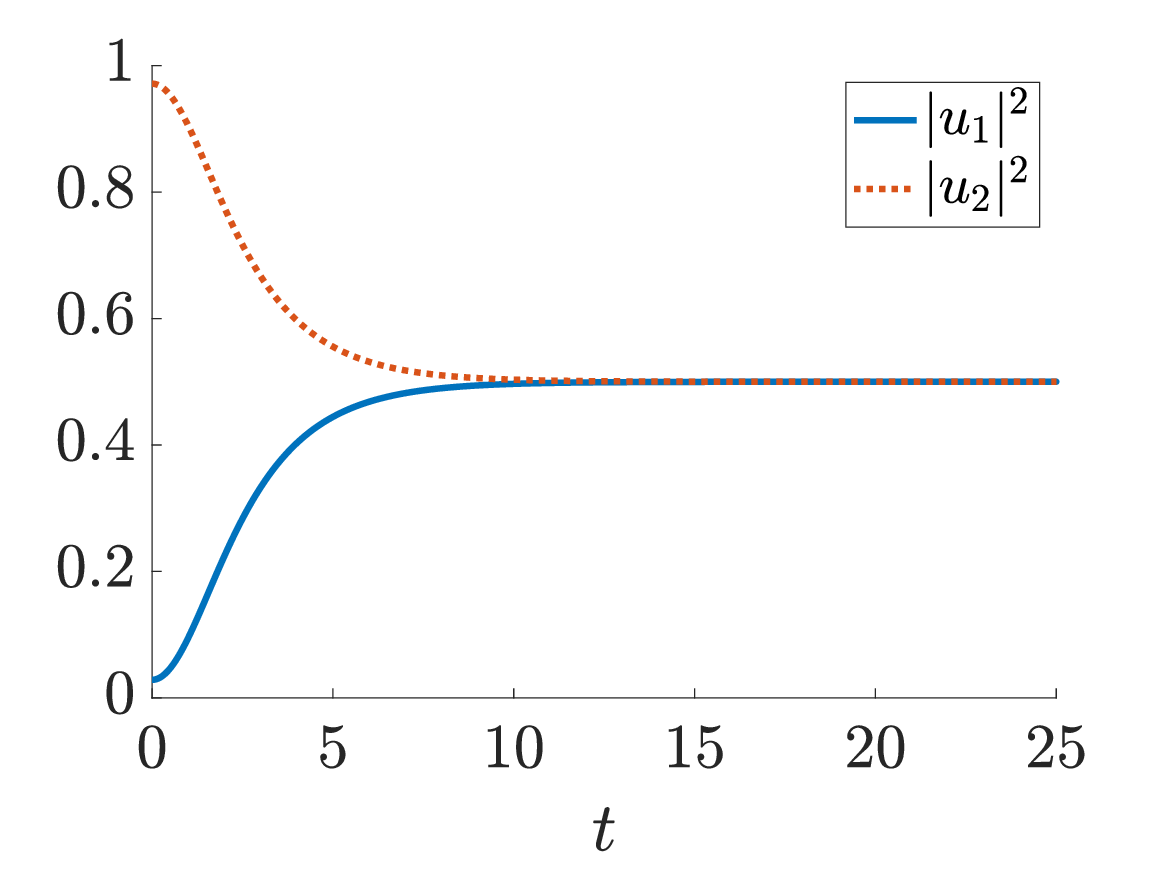}
\includegraphics[width=5.25cm]{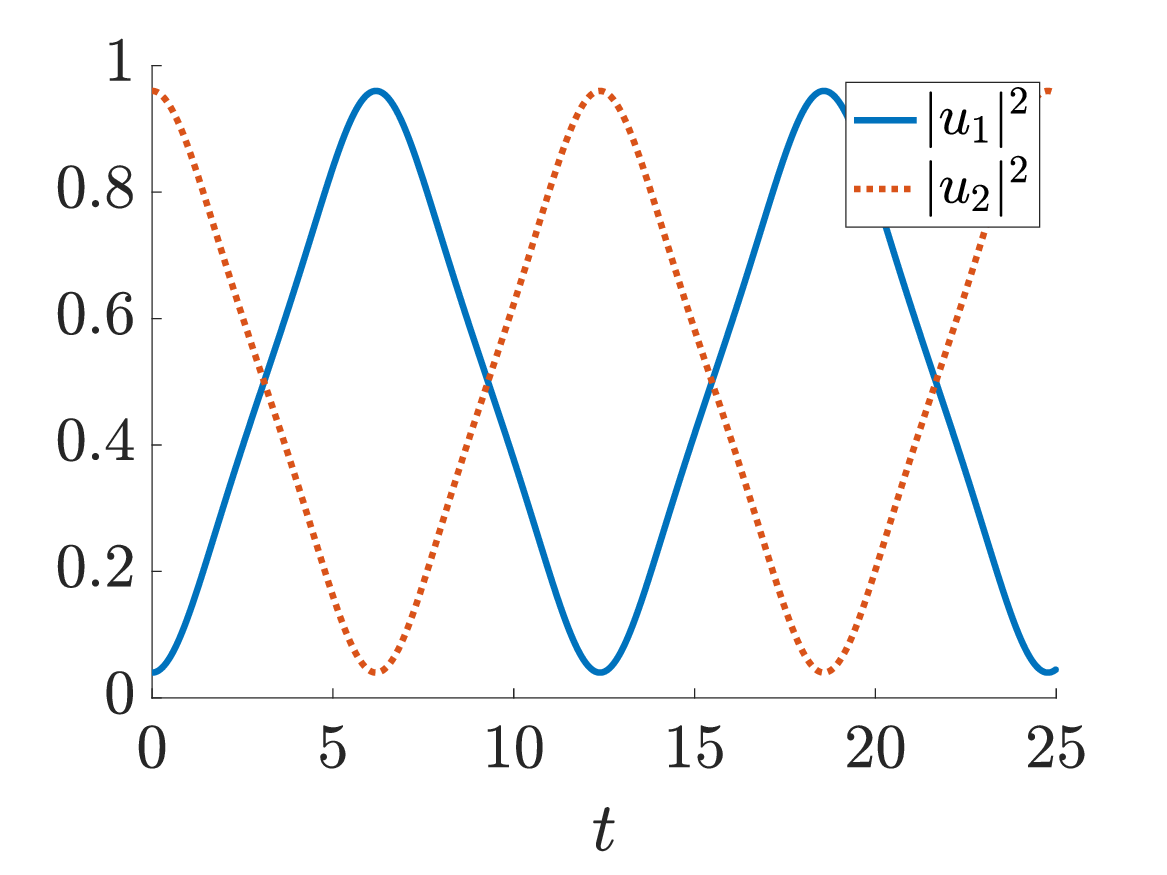}
\includegraphics[width=5.25cm]{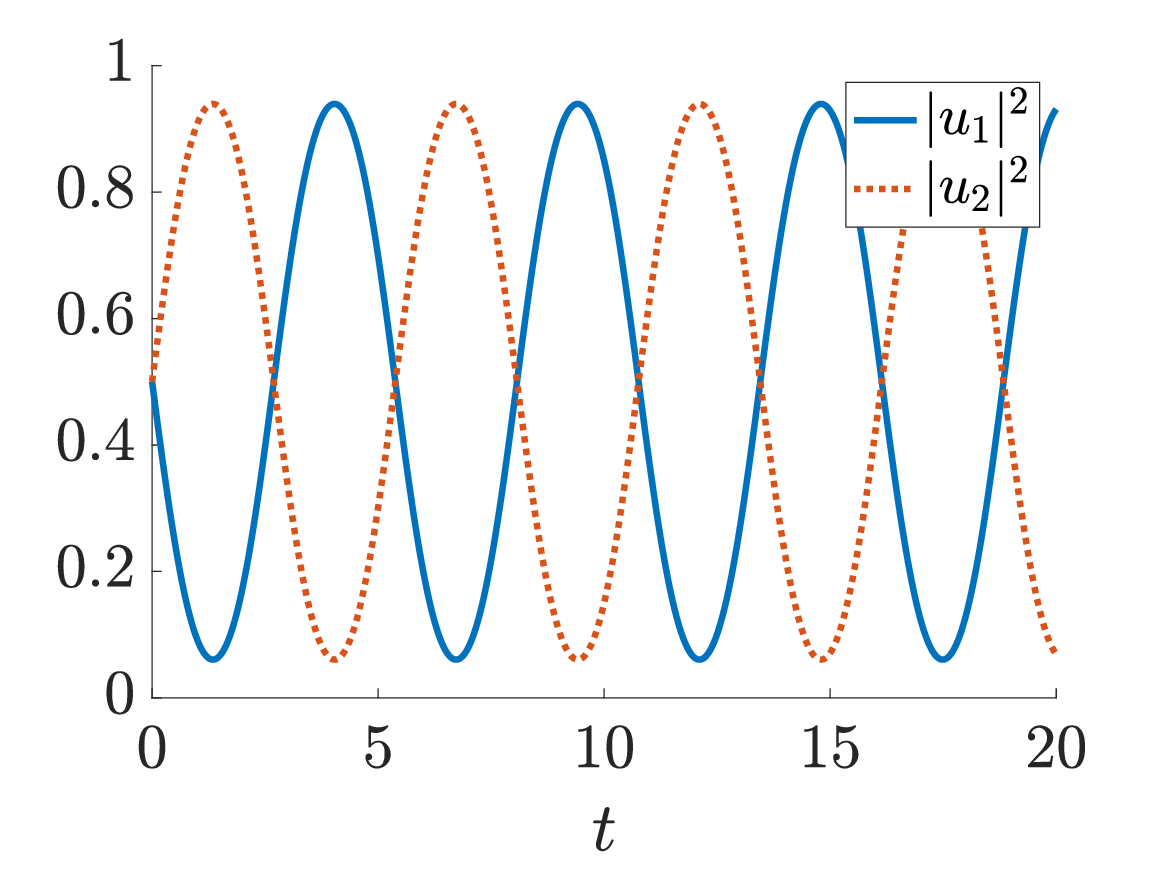}
\includegraphics[width=5.25cm]{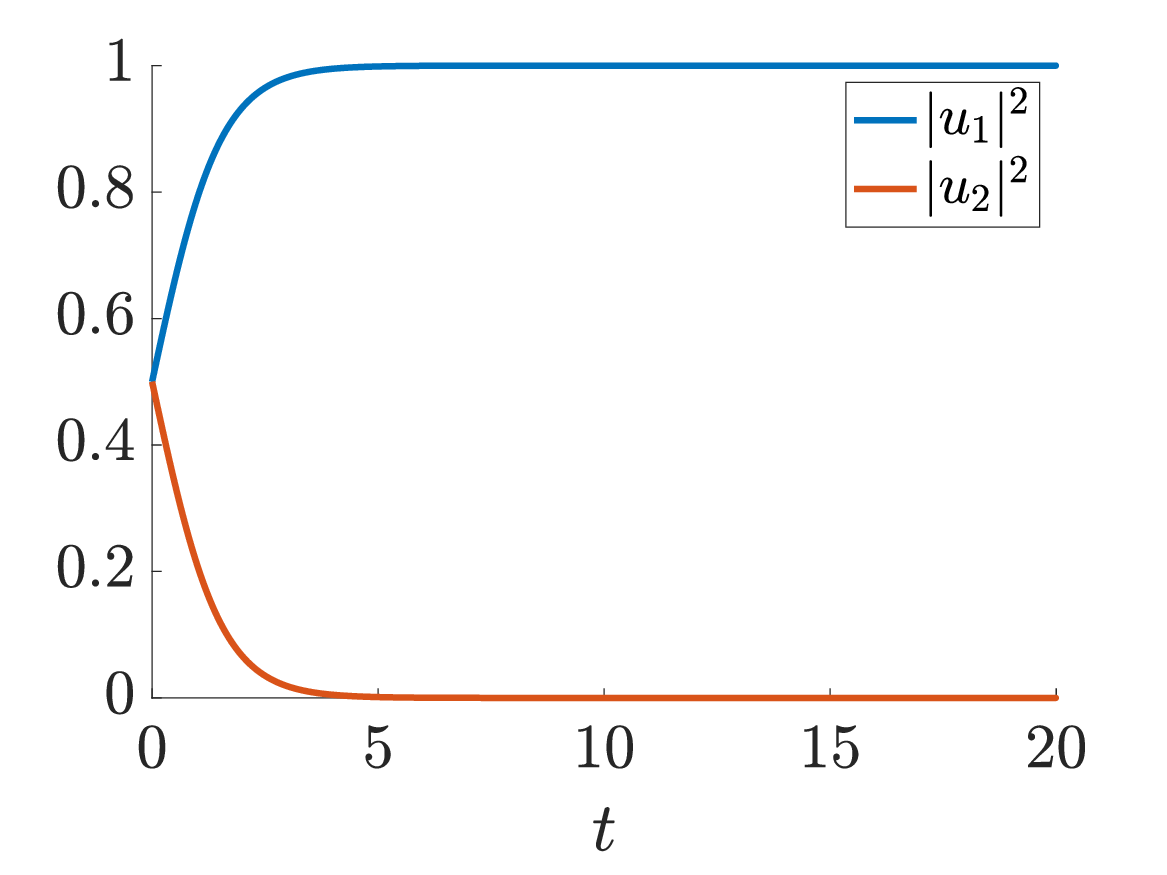}
\includegraphics[width=5.25cm]{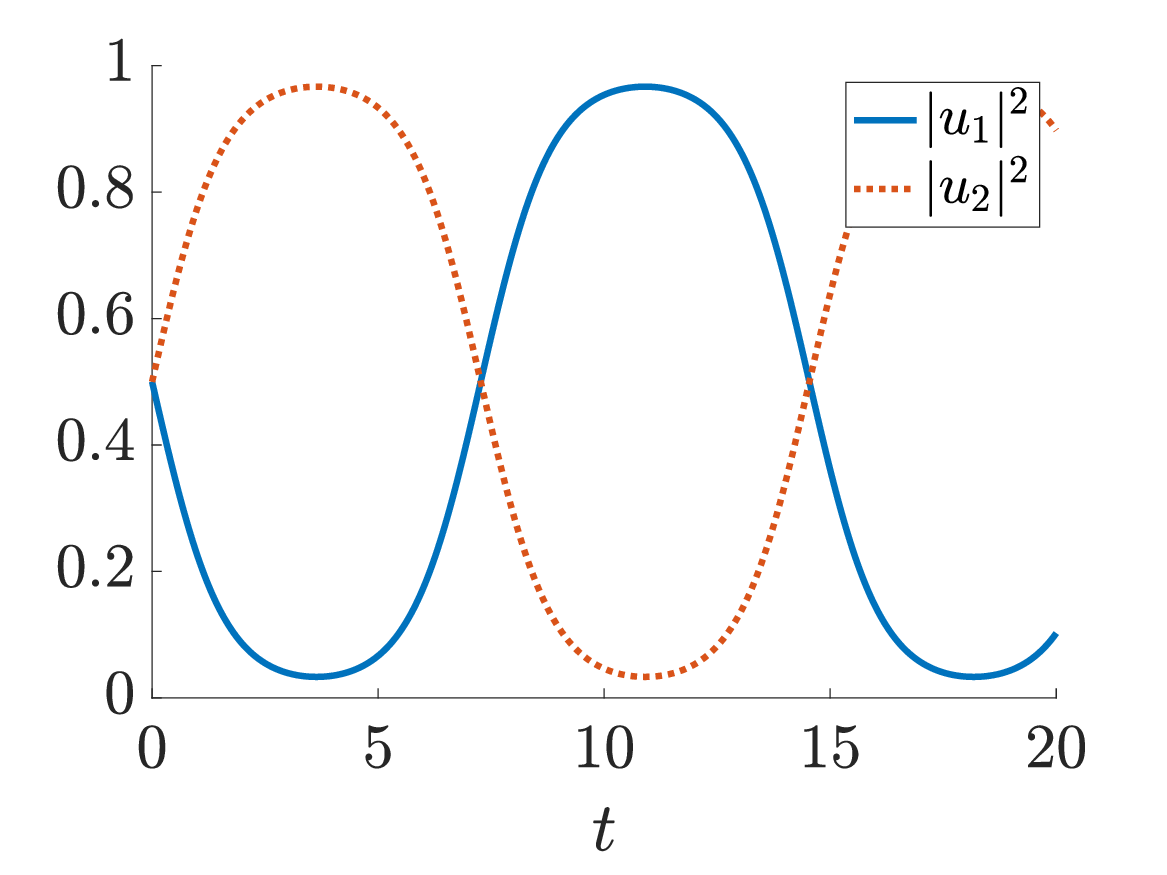}
\end{center}
\caption{Representative solutions of dimer. Top row: $d = 0.4 < 1/2$, corresponding to left panel of \cref{fig:dimerdynamic}. 
Initial intensity $|u_1(0)|^2 = 0.0225$, $0.0286$, and $0.04$ (increasing from left to right), and $|u_2(0)|^2 = 1 - |u_1(0)|^2$. Solution in the center panel corresponds to heteroclinic orbit in the left panel of \cref{fig:dimerdynamic}, as well as equation \cref{eq:sechlimit}.
Bottom row: $d = 0.6 > 1/2$, corresponding to right panel of \cref{fig:dimerdynamic}. Initial intensity $|u_1(0)|^2 = |u_2(0)|^2 = 0.5$.
Initial phase difference $\phi = 1$, $1.278$, and $1.298$ (increasing from left to right). Solution in the center panel is a slider solution from \cite{Colliander2010} and corresponds to the heteroclinic orbit (top and bottom of red box) in the right panel of \cref{fig:dimerdynamic}, as well as equation \cref{eq:tanhlimit}. Total power of solution is 1 in all cases. Solutions computed using exact formulas derived below.}
\label{fig:dimersols}
\end{figure}

Full phase portraits of this system for representative values of $d$ are shown in \cref{fig:dimerdynamic}, and 
plots of intensity vs. time for representative solutions are shown in \cref{fig:dimersols}. Exact solutions for all of the trajectories in the phase portrait are computed in the subsections that follow.
When $0 < d < 1/2$ (left panel of \cref{fig:dimerdynamic}), there is a family of concentric periodic orbits surrounding the equilibrium at the origin (see \cref{fig:dimersols}, top right, for a representative solution). As these periodic orbits move further from the origin, they approach a limiting solution, which is a pair of heteroclinic orbits connecting the saddle points at $(0, \pm \pi/2)$ (red lines in the left panel of \cref{fig:dimerdynamic}). Solutions on these trajectories (middle solution of the top row of \cref{fig:dimersols}) approach the saddle points, at which point the intensities of the two dimer sites are equal. Trajectories on the left and right sides of the heteroclinic orbits exhibit the self-trapped dynamics seen in \cref{fig:dimer1} (see \cref{fig:dimersols}, top left, for a representative solution). The manifolds of the saddle points at $(0, \pm \pi/2)$ prevent an orbit with strongly asymmetric initial data from oscillating with a changing sign of $p$.

When $d > 1/2$ (right panel of \cref{fig:dimerdynamic}), there are periodic orbits surrounding both the equilibrium at the origin and the equilibria at $(0, \pm \pi/2)$ (see \cref{fig:dimersols}, bottom right and bottom left, respectively, for representative solutions). Self-trapped dynamics is no longer possible, and it should be noted that the degenerate Hamiltonian pitchfork bifurcation at $d=1/2$ is responsible for the change from self-trapped dynamics (for $d<1/2$) to oscillatory behavior (for $d>1/2$) observed in \cref{fig:dimer1}. Both families of periodic orbits meet in limiting solutions, which are heteroclinic orbits connecting the saddle points at $(-P, \phi_*)$ and $(P, \phi_*)$, with $\phi_*$ defined in \cref{eq:phistar} (red horizontal lines in the right panel of \cref{fig:dimerdynamic}, which are the top and bottom of the heteroclinic cycle, and correspond to the middle solution in the bottom row of \cref{fig:dimersols}). We note that $\dot{\phi} = 0$, i.e., $\phi$ is constant, on these heteroclinic orbits. These solutions are the sliders in \cite{Colliander2010}. Their highly unstable nature that has been observed in our dynamics can be explained by the phase portrait, since any perturbation (no matter how small) moves the solution onto one of the nearby periodic orbits. Indeed, depending on the nature of the perturbation, the resulting orbit may be confined to completely different regions of phase space, corresponding 
to very different values of the relative phase $\phi$.

\subsection{Change of variables}

In the previous section, we characterized the qualitative behavior and bifurcations of the dimer system. We will now solve equation \cref{eq:dimer} exactly by using an appropriate change of variables. The system we obtain this way is less intuitive than the reduction \cref{eq:dimerdyn} from the previous section, but it will allow us to obtain an analytically tractable solution in terms of Jacobi elliptic functions. The analysis that follows is an adaptation of the method used in \cites{Kenkre1986,ParkerPerMod}.

We start by defining the four density matrix elements $\rho_{jk} = u_j \overline{u_k}$ for $j, k = 1, 2$. The time evolution of $\rho_{jk}$ is given by
\begin{equation}\label{eq:dimerrho}
\begin{aligned}
&\dot{\rho}_{11} = 2 i d \left( \rho_{21}^2 - \rho_{12}^2 \right) 
&&\quad\dot{\rho}_{12} = i\left(\rho_{22}-\rho_{11}\right)\left(2d \rho_{21} + \rho_{12}\right)
\\
&\dot{\rho}_{21} = -i\left(\rho_{22}-\rho_{11}\right)\left(2d \rho_{12} + \rho_{21}\right)
&&\quad\dot{\rho}_{22} = -2 i d \left( \rho_{21}^2 - \rho_{12}^2 \right).
\end{aligned}
\end{equation}
We note that in the degenerate case when one of the sites starts with zero intensity, e.g.,~$u_2(0) = 0$, then $\rho_{22}(0) = \rho_{12}(0) = \rho_{21}(0) = 0$, from which it follows that all four time derivatives in \cref{eq:dimerrho} are 0 for all $t\geq 0$. This implies that $u_2(t) = 0$ for all $t > 0$, thus we effectively have a single-site solution instead of a dimer. 
This is in line with our earlier comment regarding 
compactly supported initial data in this system.
From here on, we assume both initial conditions are nonzero.

Next, we define the variables
\begin{equation}\label{eq:pqrs}
\begin{aligned}
&p = \rho_{22}-\rho_{11},
\qquad q = i\left( \rho_{12} - \rho_{21} \right),
\qquad r = \rho_{12} + \rho_{21}, \\
&s = (1-2d)q^2 - (1+2d)r^2,
\end{aligned}
\end{equation}
where we note in particular that $p$ is the difference in intensity between the two sites in the dimer (as in the phase plane analysis above). Since
\begin{equation}\label{eq:qr2}
q = -2 \Im \rho_{12}, \qquad \quad
r = 2 \Re \rho_{12},
\end{equation}
all of these quantities are real. We can also write $s$ in terms of the density matrix elements as follows:
\begin{equation}\label{eq:s}
\begin{aligned}
s &= (q^2 - r^2) - 2d(q^2 + r^2)  \\
&=-\left(\rho_{12}^2 + \rho_{21}^2\right) - 4 d \rho_{12}\rho_{21} = -4 \Re \rho_{12}^2 - 8 d |\rho_{12}|^2.
\end{aligned}
\end{equation}
Letting
\[
P = |u_1|^2 + |u_2|^2 = \rho_{11} + \rho_{22}
\]
be the power of the solution, which is conserved in $t$, the intensities at the two lattice sites can be written in terms of $P$ and $p$ as
\begin{equation}\label{eq:pu12}
|u_1(t)|^2 = \frac{1}{2}\left( P - p(t) \right), \qquad |u_2(t)|^2 = \frac{1}{2}\left( P + p(t) \right).
\end{equation}
We note that $|p(t)| \leq P$, and since we are not considering the degenerate case, the inequality will always be strict.

The time derivatives of $p$, $q$, $r$, and $s$ are given by
\begin{equation}\label{eq:pqrsderiv}
\begin{aligned}
&\dot{p} = 4dqr &&\quad \dot{q} = -pr(1+2d) \\
&\dot{r} = pq(1-2d) && \quad \dot{s} = 4(4d^2-1)pqr.
\end{aligned}
\end{equation}
Since
\[
\frac{d}{dt}\left( p^2 \right) = 2 p \dot{p} = 8 d p q r,
\]
the equation for $\dot{s}$ becomes
\[
\dot{s} = \frac{4d^2-1}{2d}\frac{d}{dt}\left(p^2 \right).
\]
We can solve this to obtain
\begin{equation}
s = s_0 + \frac{4d^2-1}{2d}\left( p^2 - p_0^2 \right),
\end{equation}
where $p_0$ and $s_0$ are the initial conditions for $p$ and $s$. Since $\ddot{p} = 4d\left( \dot{q}r+q\dot{r}\right) = 4 d p s$, we obtain the second order differential equation for $p$
\begin{align*}
\ddot{p} &= 4dp\left[ s_0 + \frac{4d^2-1}{2d}\left( p^2 - p_0^2 \right) \right] \\
&= \left[ 4 d s_0  - 2(4d^2 - 1) p_0^2 \right] p + 2(4d^2-1)p^3,
\end{align*}
which we write as
\begin{equation}\label{eq:pdoteq}
\ddot{p} = (A - B p_0^2) p + B p^3,
\end{equation}
where
\[
A = 4 d s_0, \qquad B = 2(4d^2-1).
\]
The two initial conditions are 
\begin{equation}\label{eq:dimerICs}
\begin{aligned}
&p_0 = p(0) = \rho_{22}(0) - \rho_{11}(0) \\
&\dot{p}_0 = \dot{p}(0) = 4dq(0)r(0) = -8d 
\Re \rho_{12}(0) \Im \rho_{12}(0),
\end{aligned}
\end{equation}
where the second line follows from \cref{eq:pqrsderiv} and \cref{eq:qr2}.

The solutions for $p(t)$ will be in terms of Jacobi elliptic functions. To facilitate this, we look for a solution of the form
\begin{equation}\label{eq:pform}
p(t) = C y(Tt + \phi),
\end{equation}
where the function $y$ will be a Jacobi elliptic function. Making this substitution and simplifying, equation \cref{eq:pdoteq} becomes
\begin{equation}\label{eq:pdoteq2}
\ddot{y} = \frac{A - B p_0^2}{T^2} y + \frac{B C^2}{T^2} y^3.
\end{equation}

\subsection{Real initial conditions}

We first consider the case where the initial conditions $u_1(0)$ and $u_2(0)$ are both real. 
These correspond to solutions which start on the horizontal axis in \cref{eq:dimerdyn}.
Using \cref{eq:pu12}, the initial conditions for $q$, $r$, and $s$ are 
\[
q(0) = 0, \qquad r(0) = \sqrt{P^2 - p_0^2}, \qquad s(0) = -(1+2d)\sqrt{P^2 - p_0^2},
\]
from which it follows from \cref{eq:dimerICs} that $\dot{p}_0=0$. This in turn implies that $\phi = 0$ and $C = p_0$ in \cref{eq:pform}. Equation \cref{eq:pdoteq2} then becomes 
\begin{equation}\label{eq:pdoteqreal}
\ddot{y} = -\frac{2(1+2d)(2d P^2 - p_0^2)}{T^2} y + \frac{2d(4d^2-1) p_0^2}{T^2} y^3.
\end{equation}
If the two sites have identical initial intensity, i.e.,~$p_0 = p(0) = 0$, then $p(t) = 0$ for all $t$, which corresponds to the equilibrium at the origin in \cref{eq:dimerdyn}. Otherwise, the solution is given by
\begin{equation}\label{eq:Jreal}
p(t) = \begin{cases}
p_0\,\text{dn}\left( \sqrt{1-4d^2}\,p_0t \, ; 
m= m_0 \right)
& 0 < d < 1/2, \left(\frac{p_0}{P}\right)^2 > \frac{4d}{1+2d} \\
p_0\,\text{cn}\left( \sqrt{4d(1+2d)(P^2 - p_0^2)}t \, ; 
m= \frac{1}{m_0} \right)
& 0 < d < 1/2, \left(\frac{p_0}{P}\right)^2 < \frac{4d}{1+2d} \\
p_0\,\text{cd}\left( 
\sqrt{ (1+2d)\left(4dP^2 - (1+2d)p_0^2 \right)} t \,;
m= m_1 \right) & d > 1/2,
\end{cases}
\end{equation}
where
\[
m_0 = \dfrac{4d}{1-2d}\left[ \left(\dfrac{P}{p_0}\right)^2 -1 \right] \qquad
m_1 = \dfrac{(2d-1)p_0^2}{(2d-1)p_0^2 + 4d(P^2 - p_0^2)}.
\]
The functions $\text{cn}(t; m)$ and $\text{dn}(t; m)$ are the Jacobi elliptic functions with elliptic parameter $m$, and $\text{cd}(t; m) = \text{cn}(t; m)/\text{dn}(t;m)$. 
When $d < 1/2$, solutions which start with a large difference in initial intensities ($p_0/P$ close to 1) are in terms of the Jacobi dn function. This function has small amplitude oscillations that do not cross through 0, which leads to self-trapping behavior of the dimer (\cref{fig:dimersols}, top left). By contrast, solutions which start with a small difference in initial intensities ($p_0/P$ close to 0) are in terms of the Jacobi cn function. This function has large amplitude oscillations that cross through 0, which leads to oscillatory behavior of the dimer (\cref{fig:dimersols}, top right).
The boundary between these two regions of qualitatively distinct behavior occurs when $\left(\frac{p_0}{P}\right)^2 = \frac{4d}{1+2d}$. At this point, 
$m=1$ in the first and second lines of \cref{eq:Jreal}, and the limiting solution is given by
\begin{equation}\label{eq:sechlimit}
p(t) = p_0 \sech\left( \sqrt{1-4d^2}\,p_0 t\right),
\end{equation}
which are the heteroclinic orbits in the left panel of \cref{fig:dimerdynamic} and the middle solution in the top row of \cref{fig:dimersols}.

When $d = 1/2$, $m = 0$ in the second and third line of \cref{eq:Jreal}, and the limiting solution is
\begin{align*}
p(t) = p_0 \cos\left( 2 \sqrt{P^2 - p_0^2}\,t \right).
\end{align*}
Finally, when $d>1/2$, all solutions starting with real initial conditions are in terms of the Jacobi cd function, which exhibits large-amplitude oscillations.

\subsection{Equal intensity initial conditions}

We now consider the case where the initial intensities $|u_1(0)|^2$ and $|u_2(0)|^2$ are equal, i.e.,~$p_0 = 0$. This corresponds to solutions which start on the vertical axis in \cref{eq:dimerdyn}.
If $p_0 = 0$, then the behavior of the system depends on $\dot{p}(0)$. If $\dot{p}_0 \neq 0$, the solution $p(t)$ will not be 0 for all $t>0$. Let $u_1(0) = a$ and $u_2(0) = a e^{i\theta}$, where $a = \sqrt{P/2} > 0$ and $\theta \in (-\pi, \pi)$. Due to the gauge symmetry, we can, without loss of generality, assume that $a$ is real and positive. The initial density matrix elements are given by
\[
\rho_{11}(0) = \rho_{22}(0) = a^2, \qquad \rho_{12}(0) = a^2 e^{-i \theta}, \qquad \rho_{21}(0) = a^2 e^{i \theta},
\]
and the initial conditions for $p$, $q$, $r$, and $s$ are $p(0) = 0$ and 
\begin{align*}
q(0) = P \sin \theta, \qquad 
r(0) = P \cos \theta, \qquad
s(0) = -P^2(\cos 2\theta + 2 d),
\end{align*}
where we used the formulas from \cref{eq:qr2} and \cref{eq:s}. It follows from \cref{eq:pqrsderiv} that
\begin{equation}\label{eq:degenp0dot}
\dot{p}(0) = 4 d q(0)r(0) = 4 d P^2 \sin \theta \cos \theta = 2 d P^2 \sin 2 \theta.
\end{equation}
If $\theta = 0$ or $\theta = \pm \pi/2$, equation \cref{eq:degenp0dot} implies that $\dot{p}(0) = 0$, thus in those cases we will have $p(t)=0$ for all $t>0$. 
These are equilibrium points of \cref{eq:dimerdyn}, which correspond to the real and the staggered compacton, respectively. From here on, we will assume that $\theta \notin \{ 0, \pm \pi/2\}$.

Next, we define the constant $A_0$ by
\begin{equation}\label{A_0}
A_0 = 1 + 2 d \cos 2\theta.
\end{equation}
In terms of $A_0$, the constant $A$ is given by
\[
A = -4 d P^2(\cos 2\theta + 2 d) = -2 P^2 \left[ A_0 + (4d^2-1) \right],
\]
thus \cref{eq:pdoteq} becomes
\[
\ddot{p} = -2 P^2 \left[ A_0 + (4d^2-1) \right] p + 2(4d^2-1) p^3.
\]
The solution depends on $d$ and the sign of $A_0$. We note if $d < 1/2$, then we will always have $A_0 > 0$. The solution is given by 

\begin{equation}\label{eq:Jequal}
p(t) = \begin{cases}
\frac{dP\sin 2\theta}{\sqrt{d(1+2d\cos 2\theta)}}\,\text{sd}\left( 2P\sqrt{d(1+2d\cos 2\theta)}\,t \, ; 
m= \frac{(1-2d)\sin^2 \theta}{1+2d \cos 2\theta} \right)
& 0 < d < 1/2\\
2 P \sin \theta \sqrt{\frac{d}{2d+1}}\,\text{sn}\left( 2 P \cos \theta \sqrt{d(2d+1)}\, t \, ; 
m= \frac{2d-1}{2d+1}\tan^2\theta \right)
& d > 1/2, A_0 > 0 \\
2 P \cos \theta \sqrt{\frac{d}{2d-1}}\,\text{sn}\left( 2 P \sin \theta \sqrt{d(2d-1)}\, t \, ; 
m= \frac{2d+1}{2d-1}\cot^2\theta \right)
& d > 1/2, A_0 < 0,
\end{cases}
\end{equation}
where $\text{sd}(t;m) = \text{sn}(t;m)/\text{dn}(t;m)$. 

When $d<1/2$, all solutions starting with equal intensity conditions are in terms of the Jacobi sd function, which exhibits large-amplitude oscillations. (We can see from the left panel of \cref{fig:dimerdynamic} that self-trapping behavior is not possible for these initial conditions).
When $d = 1/2$, $m = 0$ in the first and second line of \cref{eq:Jequal}, and the limiting solution is
\begin{align*}
p(t) = P \sin \theta \sin \left((2 P \cos \theta) t \right).
\end{align*}
When $d > 1/2$, all periodic solutions are in terms of the Jacobi sn function, which also exhibits large-amplitude oscillations (left and right solutions in the bottom row of \cref{fig:dimersols}). In terms of the phase portrait in \cref{fig:dimerdynamic}, the solutions for $A_0>0$ and $A_0<0$ correspond to periodic orbits about the equilibria at $(0,0)$ and $(0, \pi/2)$, respectively.
The limiting solution for $d>1/2$ is the boundary between the two families of periodic orbits in the right panel of \cref{fig:dimerdynamic}. This occurs when $A_0 = 0$, from which it follows that $m=1$ in the second and third line of \cref{eq:Jequal}. The limiting solution is given by
\begin{equation}\label{eq:tanhlimit}
p(t) = P \tanh \left( P \sqrt{4d^2-1}\,t \right),
\end{equation}
which corresponds to the heteroclinic orbits (top and bottom of red box) in the right panel of \cref{fig:dimerdynamic}.
Using \cref{eq:pu12}, the intensities at the two lattice sites are 
\begin{equation}\label{eq:sliderint}
|u_1(t)|^2 = \frac{P}{1 + e^{2P\sqrt{4d^2-1}t} }, \qquad |u_2(t)|^2 = \frac{P}{1 + e^{-2P\sqrt{4d^2-1}t} },
\end{equation}
which can be seen in the middle solution in the bottom row of \cref{fig:dimersols}. Taking $P=1$ and $d=1$, these are the slider solutions in \cite{Colliander2010}*{(3.7)}.  

Finally, we note that all trajectories in \cref{fig:dimerdynamic} cross at least one of the two axes; since this implies that they fall into either the real initial conditions case or the equal intensity initial conditions case, we do not need to consider any other cases.

\section{Lattice traveling solutions}\label{sec:moving}

\begin{figure}
\begin{center}
\includegraphics[width=16cm]{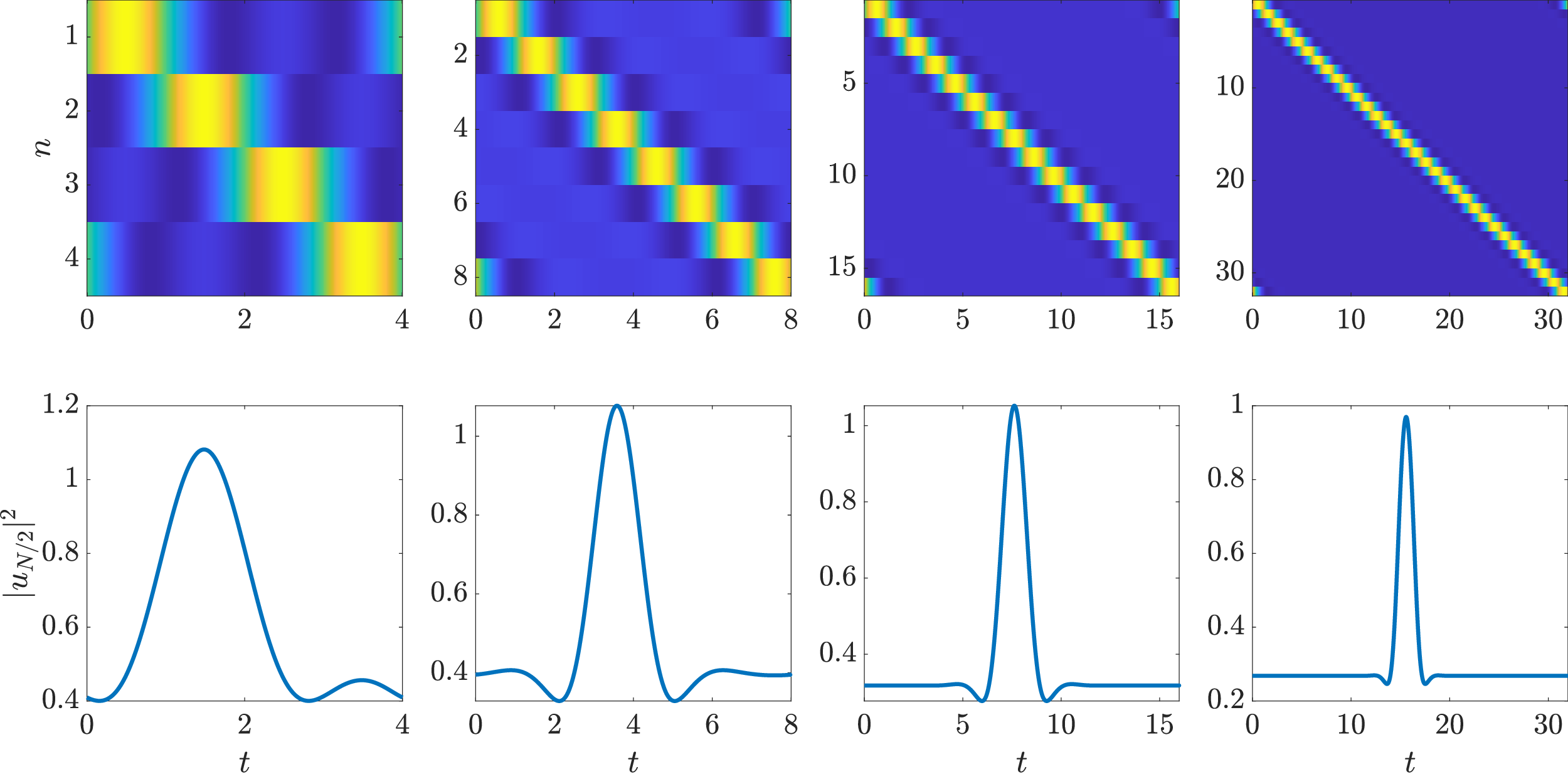}
\end{center}
\caption{Colormap showing lattice site intensity (top) and intensity of central lattice site vs. $t$ (bottom) for rightward moving solutions on a periodic lattice for lattice sizes $N=4, 8, 16, 32$. Coupling parameter $d = 0.6$. The time evolution is performed using the Dormand-Prince integrator, implemented in Matlab by means of \texttt{ode45} function.}
\label{fig:moving1}
\end{figure}

Motivated by the results we obtained from the dimer that showed the existence of solutions in which intensity is transferred between the two sites, we look for solutions in larger lattices in which the intensity flows unidirectionally along the lattice. In particular, we consider a lattice of $N$ nodes with periodic boundary conditions, i.e.,~the lattice is effectively a ring of $N$ nodes. We seek a solution in which the bulk of the intensity starts at the first lattice site at $t=0$, and then the entire solution reproduces itself exactly, except shifted one site to the right, at $t=1$. The choice of $t=1$ is arbitrary, but can be made without loss of generality due to the time-amplitude scaling from \cref{sec:model}. By symmetry, we can equivalently look for leftward-moving solutions. Thus we look to solve the boundary value problem
\begin{equation}\label{eq:moveBVP}
\begin{aligned}
\dot{u}_j &= i \left[ d (u_{j-1}^2 + u_{j+1}^2) \overline{u_j} - |u_j|^2 u_j \right] && \qquad j = 1, \dots, N \\
u_{j+1}(1) &= u_j(0),
\end{aligned}
\end{equation}
where the subscripts are taken $\Mod N$ due to the periodic lattice. In addition, due to the gauge symmetry, we can without loss of generality take $\Im u_1(0) = 0$.
To solve \cref{eq:moveBVP} numerically, we use a shooting method, which we describe in \cref{sec:shootapp}. See \cref{fig:moving1} for rightward moving solutions of varying $N$ computed numerically using this method.
The solutions at each site are identical, except shifted by an integer time, thus they all satisfy the advance-delay equation 
\begin{equation}\label{eq:advdelay}
i \dot{u}(t) + d( u(t+1)^2 + u(t-1)^2 ) \overline{u(t)} - |u(t)|^2 u(t) = 0
\end{equation}
for $t \in [0,N]$ with periodic boundary conditions. Once a solution has been obtained via a shooting method, equation \cref{eq:advdelay} is useful for parameter continuation. 

\begin{figure}
\begin{center}
\includegraphics[width=7cm]{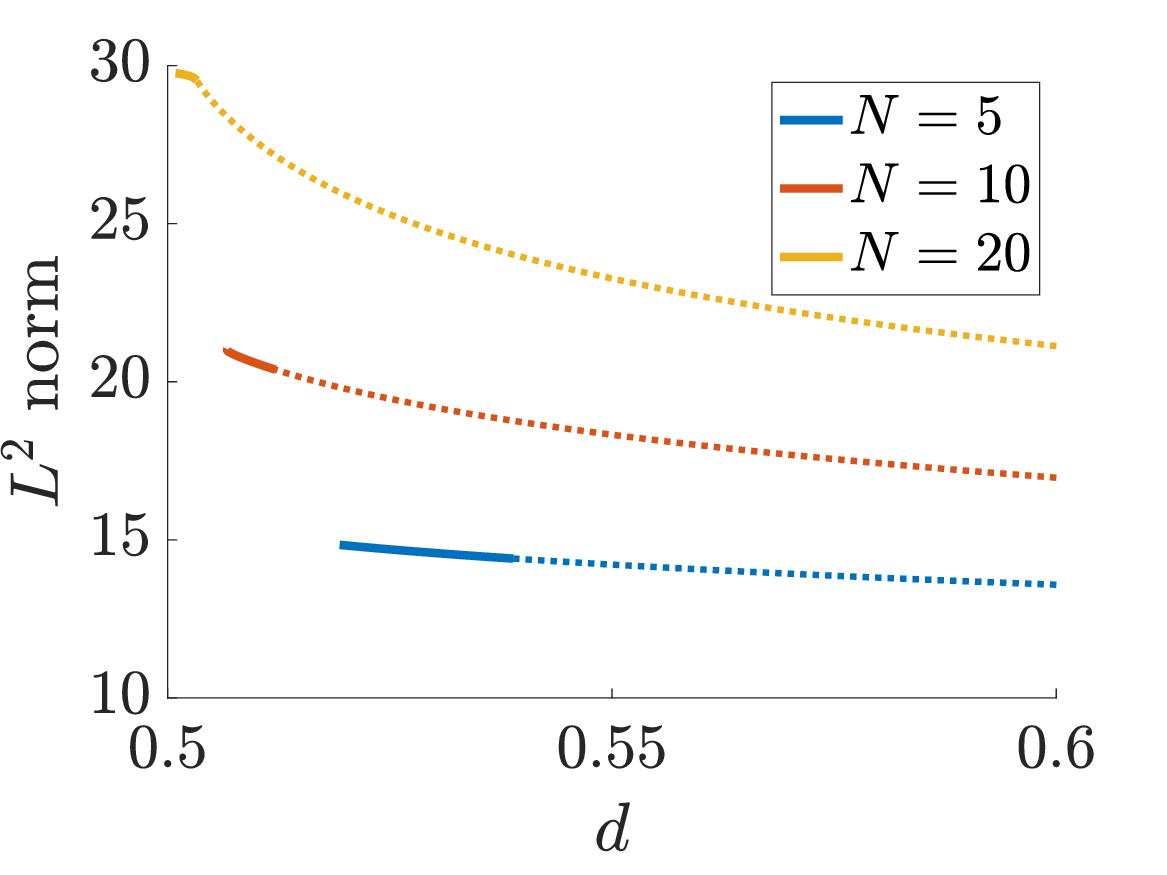}
\includegraphics[width=7cm]{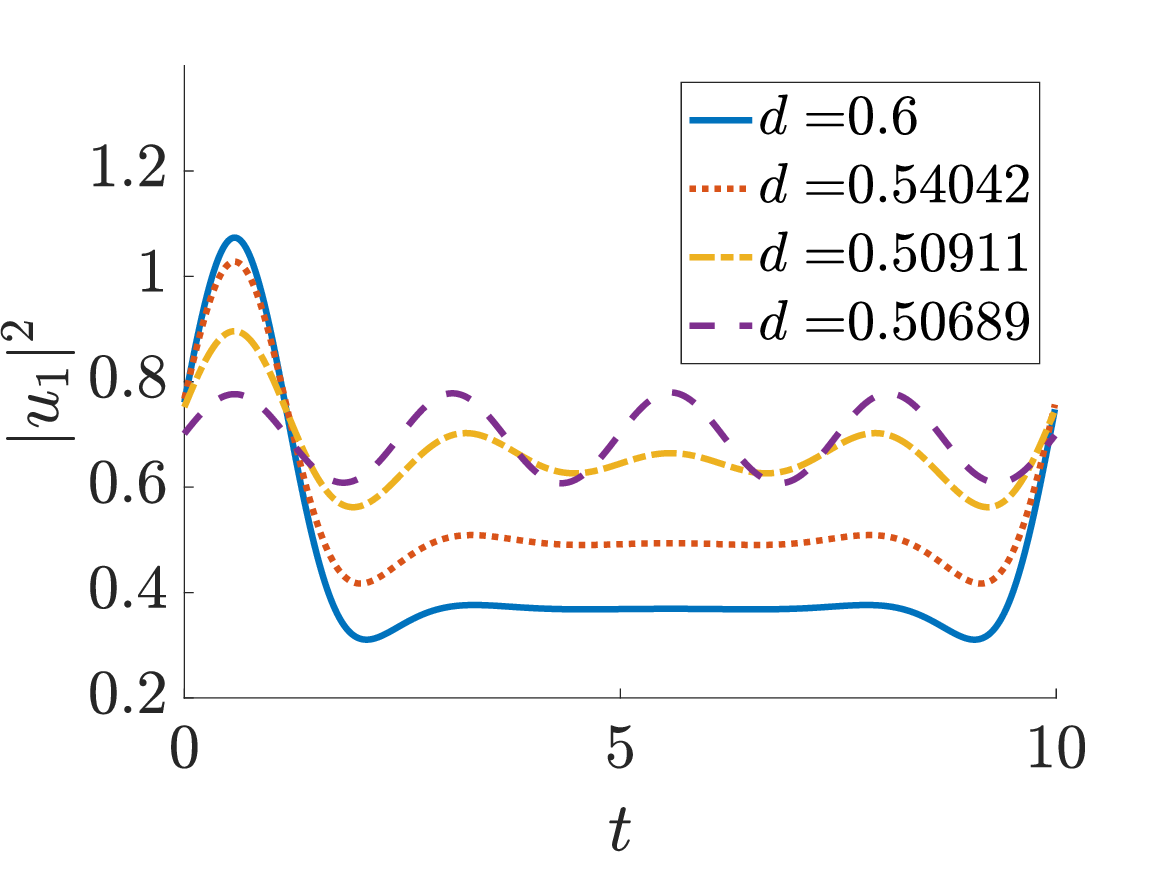}
\includegraphics[width=7cm]{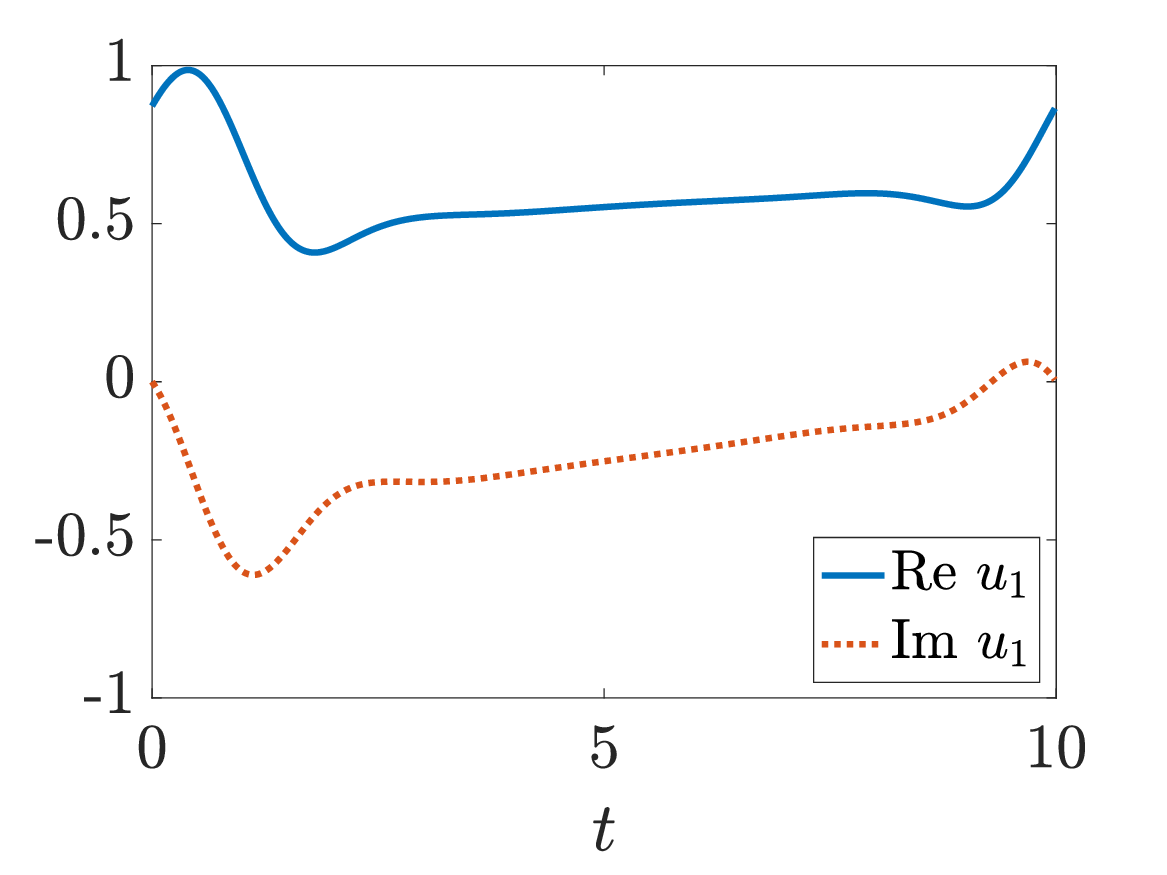}
\includegraphics[width=7cm]{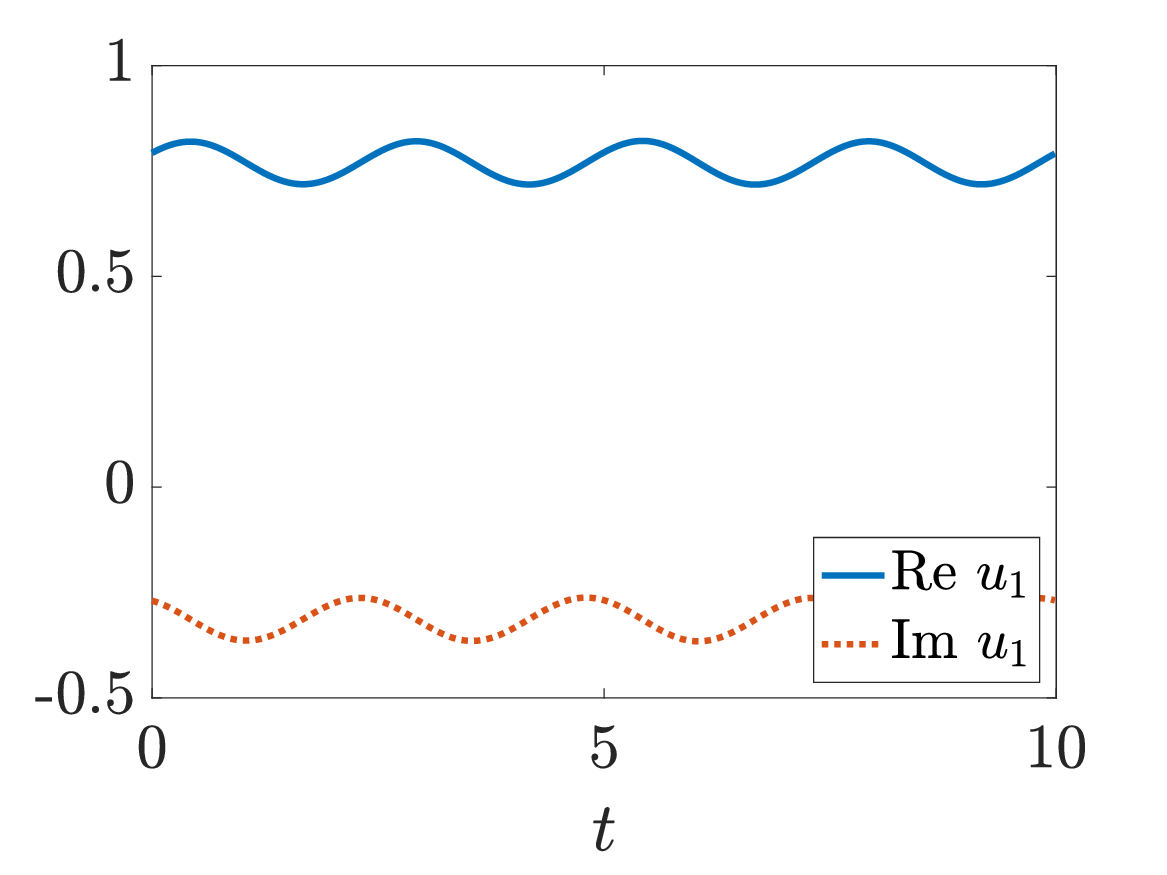}
\includegraphics[width=7cm]{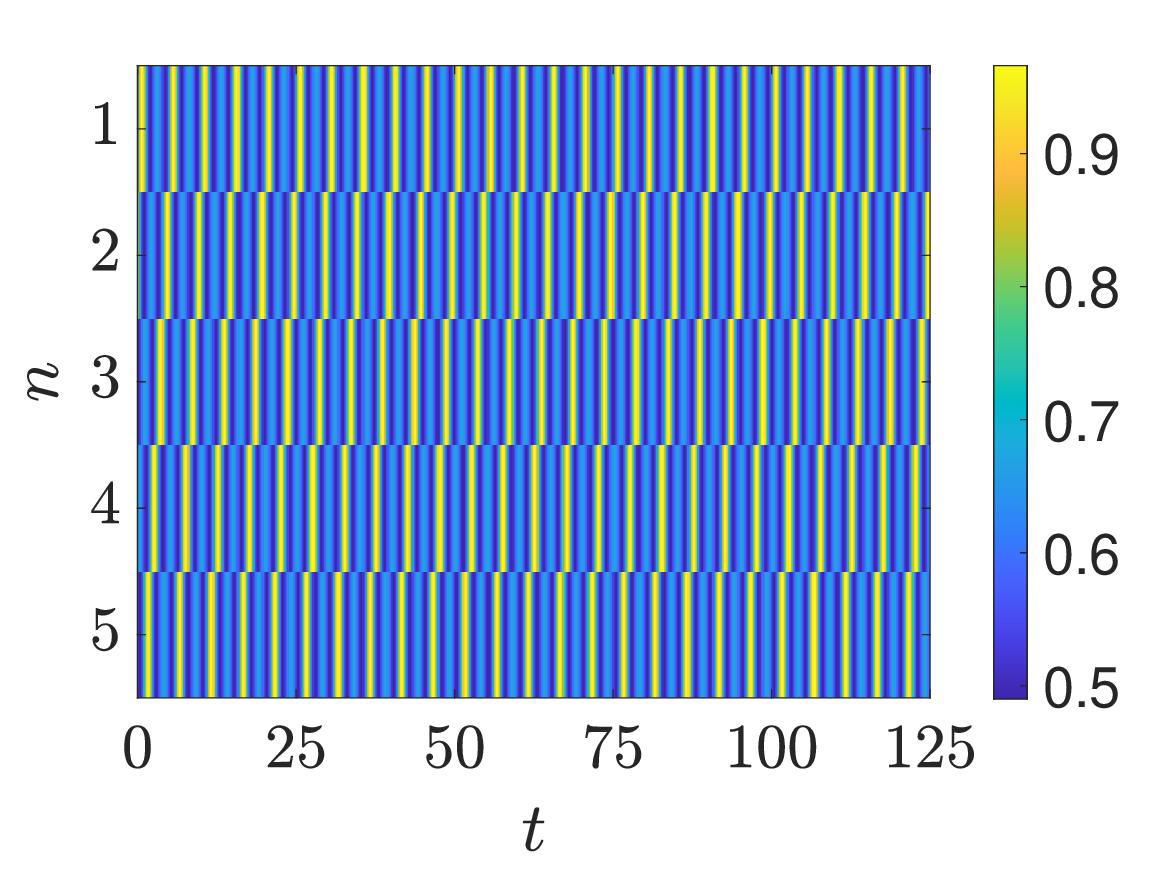}
\includegraphics[width=7cm]{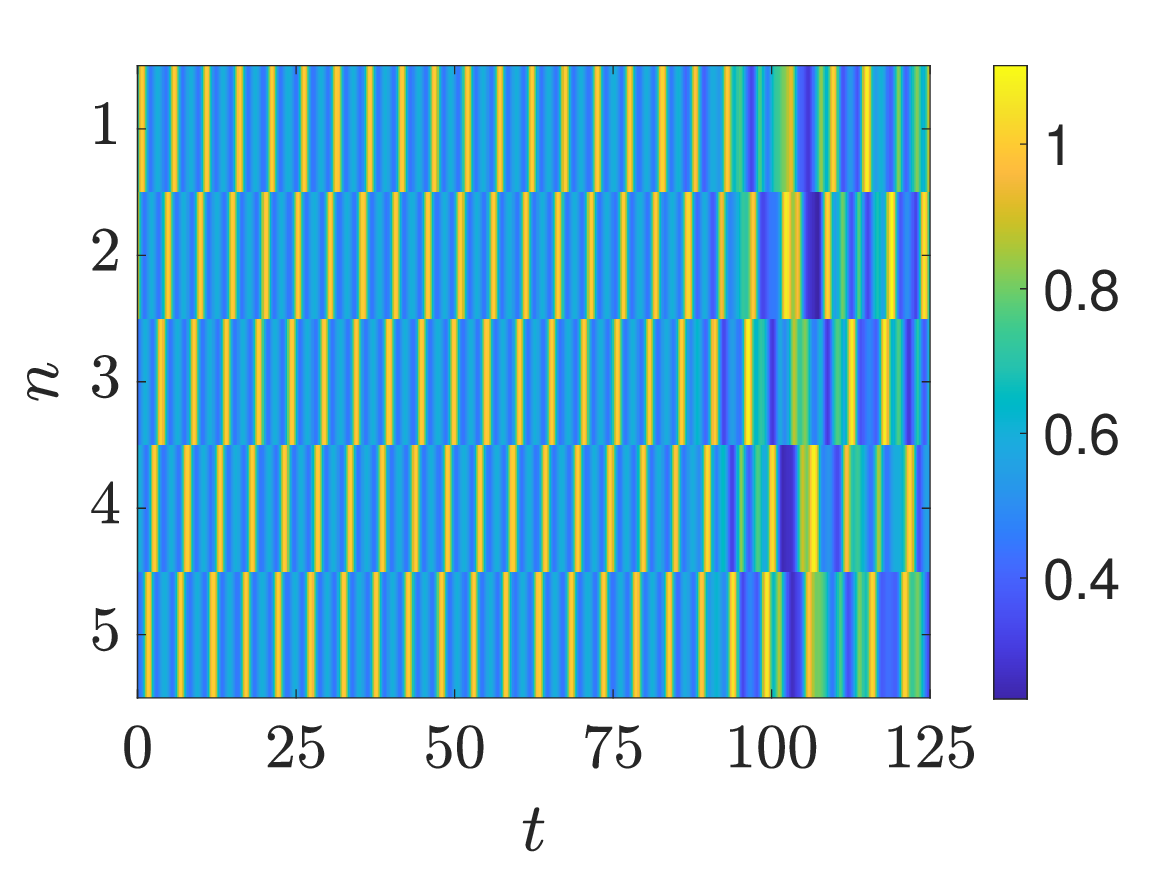}
\end{center}
\caption{Top left: $L^2$ norm of $u_1(t)$ vs. $d$ for periodic lattice with $N=5, 10$ and $20$ sites; spectrally stable (unstable) solutions denoted by solid (dotted) lines. Top right: square intensities $|u_1(t)|^2$ of periodic moving solutions for $N=10$ and varying $d$.
Middle: real and imaginary parts of $u_1(t)$ of periodic moving solutions from top left panel for $d=0.6$ (left) and $d=0.50689$ (right).
Bottom: colormap of square intensity $|u_n|^2$ of evolution of unperturbed moving solutions on a periodic lattice with $N=5$, $d=0.537$ (left) and $d=0.555$ (right). 
The integrator used is once
again the Dormand-Prince one.}
\label{fig:moving2}
\end{figure}

Representative moving solutions for four values of $N$ are shown in the top panel of \cref{fig:moving1}. Numerical experiments strongly suggest that these solutions only exist for $d > 1/2$ (\cref{fig:moving2}, top left). 
Although the intensity profile of these moving solutions for sufficiently large $d$ (and $N$) are indicative of a localized solution on a constant background (solid blue line in \cref{fig:moving2}, top right), a plot of the real and imaginary parts (\cref{fig:moving2}, middle left) shows that the background is, in fact, not constant. As $d$ decreases towards 1/2, the difference between the minimum and maximum intensity decreases (\cref{fig:moving2}, top right), and the solution takes the form of oscillations on a constant background (\cref{fig:moving2}, middle right). Nevertheless, the 
relevant oscillation is only expected to disappear in the
limit and is clearly found to persist in \cref{fig:moving2}
even close to that limit.

For exactly $d=1/2$, any constant state $u_j = C$, $j = 1,\dots,N$, is a solution to \cref{eq:moveBVP}. The parameter continuation in the top left of \cref{fig:moving2} never reaches this constant limit at $d=1/2$, but it does come closer to it for larger lattice sizes. 
Furthermore, these solutions appear to only be spectrally stable (as defined by the Floquet multipliers being confined to the unit circle) for $d$ close to 1/2, i.e.,~for $1/2 < d < d^*(N)$, where $d^*(N)$ approaches 1/2 as $N$ becomes large. The dynamical consequences of this can be observed in the lower panels of \cref{fig:moving2} for $N=5$. For $d < d^*(5)$ (bottom left of \cref{fig:moving2}), the solution remains coherent for the 25 full periods shown in the figure (one period has a length in $t$ of 5, after which the system has returned exactly to its starting condition); in fact, numerical experiments show that it remains coherent for at least 1000 periods. By contrast, for $d > d^*(5)$, the solution breaks down after approximately 20 periods (\cref{fig:moving2}, bottom right).

These findings constitute, in our view, a significant
addition to our understanding of such nonlinearly 
dispersive models. This is not only since, to our understanding,
they have not been presented previously, but also because
they appear to be central to segments of the dynamics that emerge in both our ramp (\cref{fig:evol1}) and in our modulationally
unstable (\cref{fig:MI2}) dynamical evolutions.

\section{Conclusions and Future Challenges}\label{sec:conclusions}

In the present work, we have revisited an intriguing minimal
model characterized by the interplay of nonlinear dispersion
and cubic nonlinearity. The motivation of the model
stems from its derivation as a minimal description 
for the study of cascades across (groups of) Fourier
modes in the defocusing NLS equation, which constitute the effective nodes of this lattice,
that was initiated in the work of~\cite{Colliander2010}.
This study adds to the wealth of earlier numerical~\cites{jeremy1,gideon2}
and analytical~\cites{jeremy1,Her} explorations of this model
by considering the prototypical 
nonlinear excitations thereof
and their spectral stability properties, as well as their associated
nonlinear dynamics. We found that the model exhibits 
different types of compactly supported nonlinear states,
showcased their ranges of existence, and identified the
termination and bifurcation points of the relevant structures. Both regular (monotonic)
and staggered (non-monotonic) states were explored; for $d<1/2$, it was found
that the former are spectrally stable, while the latter
are spectrally unstable. We then turned
to dynamical considerations and were able to analytically solve the simplest
scenario thereof, namely the two-site (dimer); in addition, we were able to convert this problem into one involving only two 
degrees of freedom by using the relevant conservation laws.
The phase portrait of this two-dimensional system captures the full dynamics of the dimer, explains the relevant bifurcations, and also sheds light on the subtle non-robustness
of the so-called slider solutions discussed 
in~\cite{Colliander2010}. This, in turn, prompted
us to search for generalizations of moving solutions 
in lattices with larger numbers of nodes, which we were
able to identify. The somewhat unexpected (yet, a posteriori,
justified) feature of such solutions was their apparent (for large
lattice sizes) anti-dark nature, i.e., their density profiles that
asymptoted to a constant nonzero value. 
While the states themselves
are found to be unstable for large lattices, numerical
simulations clearly illustrate their transient role in cascade
dynamics and indeed motivate their direct numerical identification. Moreover we illustrated their 
potential stability near the parameter $d=1/2$.

This study motivates a wide range of additional questions 
worth examining. It might be useful to examine if analytical
results can be extended beyond the dimer setting, e.g., into the
trimer case of $N=3$, also potentially addressing the question of
whether variants of slider states may be found therein.
A deeper understanding of the transient role of the obtained traveling
states in the dynamics, and perhaps even more importantly in the 
thermodynamics and long time asymptotics~\cite{cretegny}, of such nonlinear dispersive models would be particularly interesting to elucidate. 
While the relevance of this class of models as minimal models
for turbulence is less evident in higher dimensions, their
potential nonlinear wave patterns in the latter setting would
be quite interesting to explore in their own right, motivated
by the wealth of states accessible to higher dimensional
linearly dispersive models~\cite{LEDERER20081}. Lastly, the
implications of the present findings for continuum models
of turbulence, while perhaps more removed from the current work,
are certainly relevant to future thought and exploration.

\subsection*{Acknowledgments}

This material is based upon work supported by the U.S. National Science Foundation under RTG Grants No. DMS-1840260 (R.P. and A.A.), No. PHY-2110030, No. DMS-2204702 (P.G.K.), and No. DMS-1909559 (A.A.). J.C.-M. acknowledges support from EU (FEDER Program No. 2014-2020) through MCIN/AEI/10.13039/501100011033 under Project No. PID2020-112620GB-I00.
P.G. and P.G.K. gratefully acknowledge 
Professor Jeremy Marzuola for numerous 
informative discussions on the models considered herein.

\appendix

\section{The continuous limit}\label{app:contlimit}

The continuous or long wave limit, first considered in~\cite{jeremy1}, is an important regime for the equation under consideration, and would deserve an investigation in its own right. We will not carry it out here, but will simply sketch some of its features. 
An important connection with the earlier developments in the present article has to do with the key value $d=1/2$, which arose repeatedly as a turning point.
Namely, if one considers features such as variational properties, modulational stability, and stability of compactons, it makes sense to think of the cases $d<1/2$ (resp. $d>1/2$) as defocusing (resp. focusing).
It is an interesting coincidence that $d=1/2$ also corresponds to the only value of $d$ for which the continuous limit asymptotically makes sense, as we will see below (compare to ~\cites{jeremy1,GHGM, jeremy2}, which focus on the value $d=2$).

To investigate the continuous limit, we choose the ansatz
$$
u_j(t) = u(t,hj),
$$
where $u$ is a smooth function, and $h>0$ (this is a slight abuse of notation; from now on, $u$ is a function on the real line instead of the lattice).
Expanding in a Taylor series in $h$, one finds
$$
i \partial_t u + (2d-1) |u|^2 u + 2 h^2 d \overline{u} \partial_x (u \partial_x u) + O(h^3) = 0.
$$
The continuous limit corresponds to the case where
$$
h \to 0, \quad \frac{2d-1}{2h^2} \to \alpha, 
$$
where $\alpha$ is a real constant (this implies in particular $d \to 1/2$). Upon rescaling time, the limiting equation is then
$$
i \partial_t u  + \alpha |u|^2 u + \overline{u} \partial_x (u \partial_x u) = 0.
$$
A further rescaling enables one to restrict the value of $\alpha$ to $\alpha \in \{-1 , 0 ,1\}$. The Hamiltonian is now
$$
H(u) = \frac{1}{2} \int |u \partial_x u|^2 \,dx - \frac{\alpha}{4} \int |u|^4 \,dx.
$$
We now follow~\cite{GHGM} and seek solitary waves of the form
$$
u(t,x) = Q(x-vt) e^{-ict},
$$
where the wave profile $Q$ solves the ODE
$$
cQ - iv Q' + \overline{Q} (QQ')' + \alpha |Q|^2 Q = 0.
$$
Multiplying by $\overline{Q}$ and taking the imaginary part, or multiplying by $\overline{Q'}$ and taking the real part, one finds the two conservation laws
\begin{align*}
& \frac{v}{2} |Q|^2 + \text{Im} (|Q|^2 \overline{Q} Q') = \eta \\
& -c |Q|^2 + |Q|^2 |Q'|^2 + \frac{\alpha}{4} |Q|^4 = \kappa,
\end{align*}
for constants $\eta$ and $\kappa$. In the particular case where $\eta = \kappa = 0$, 
 which corresponds to localized waves, one finds that $Q = \psi e^{i\theta}$, where $\psi$ and $\theta$ solve the system of equations
\begin{align}
& c \psi = \psi (\psi \psi')' + \alpha \psi^3 \\
& \theta' = - \frac{v}{2} \frac{1}{\psi^2}.
\end{align}
The equation for $\psi$ can be integrated to give
$$
c \psi^2 = (\psi \psi')^2 + \frac{\alpha}{2} \psi^4 + C,
$$
for an integration constant $C$. If $C=0$, this can be integrated to give, up to translation,
$$
\psi = \sqrt{ \frac {2c} \alpha} \sin \left( \sqrt{\frac \alpha 2} (x-vt) \right),
$$
\noindent
which is valid for $\alpha =1$ or $d\rightarrow 1/2^+$. As one can see from the equation for the phase above, the solution fails to exist when $z=x-vt = n\pi$, as it creates a singularity in the phase, unless $v=0$, or the solution is stationary. 

This result can be generalized for when $C \neq 0$. In that case, let $U=\psi^2$. Then $U$ satisfies the equation
$$
2c U =  (U')^2 + \alpha U^2 + A 
$$
for a constant $A$. Completing squares and defining $W=U-\frac{c}{\alpha}$ we arrive at
$$
(W')^2 = -\alpha W^2  + \left[\left(\frac{c}{\alpha}\right)^2-A\right].
$$
A solution to this is
\[
U = \frac{c}{\alpha} \pm \sqrt{\frac{B}{\alpha}}\sin(\sqrt{\alpha}z + C),
\]
for a constant $C$, where $B = \left(\frac{c}{\alpha}\right)^2-A$, thus 
$A < \left(\frac{c}{\alpha}\right)^2$.
From the definition of $U$, only solutions that are non-negative are valid, restricting the choice of $B$ so that $\sqrt{\frac{B}{\alpha}} < \frac{c}{\alpha}$. 
Exploring the potential of constructing weak
solutions out of a single period of these 
sinusoidal (static and traveling) waveforms,
appropriately glued to a constant background
(in the spirit of the compactons of~\cite{rosenau})
would constitute an interesting direction for future study.


\section{Shooting method}\label{sec:shootapp} 

Shooting methods are highly useful techniques for solving boundary value problems numerically (for a good reference, see Chapter 18.1 of \cite{NumericalRecipes}). 
The shooting method reduces a boundary value problem to an initial value problem. We then systematically solve the corresponding initial value problem for different initial conditions until we obtain a solution which also satisfies the desired boundary conditions.

We use a shooting method to find traveling solutions in the lattice by solving the boundary value problem \cref{eq:moveBVP}. First, we define $u(t; u_0)$ to be the solution to the corresponding initial value problem
\begin{equation}\label{eq:shootIVP}
\begin{aligned}
\dot{u}_j &= i \left[ d (u_{j-1}^2 + u_{j+1}^2) \overline{u_j} - |u_j|^2 u_j \right] && \qquad j = 1, \dots, N
\end{aligned}
\end{equation}
with initial condition $u_0$ at $t=0$. The solution $u(t; u_0)$ solves the boundary value problem \cref{eq:moveBVP} on $t \in [0, 1]$ if $F(u_0) = 0$, where
\begin{equation}
F(u_0) = u(1; u_0) - R u_0, \qquad
R = \begin{pmatrix}
0 & & & 1 \\
1 & 0 \\
& \ddots & \ddots \\
& & 1 & 0 \\
\end{pmatrix}.
\end{equation}
The matrix $R$ is responsible for the rightward direction of travel on the lattice (we can obtain leftward motion by using $R^\top$ in place of $R$). We use a standard root-finding method (e.g., the trust-region-dogleg algorithm, implemented by means of \texttt{fsolve} function in Matlab) to solve $F(u_0) = 0$ for a suitable initial seed. For the initial guess, we use the vector $(1, \epsilon, \dots, \epsilon)^\top$, where $\epsilon$ is small but nonzero. We cannot take $\epsilon = 0$, since sites which start at intensity of 0 will remain at 0 for all $t$.

\bibliographystyle{amsplain}
\bibliography{main}

\end{document}